\documentclass{aa}

\usepackage{graphicx}
\usepackage{txfonts}
\usepackage{color}
\usepackage{enumitem}
\usepackage{subfigure}
\usepackage{longtable}

\begin{document}

   \title{A matched filter approach for blind joint detection of galaxy clusters in X-ray and SZ surveys}

   \author{P. Tarr\'io
		   	\inst{1}
          \and
          J.-B. Melin
          \inst{2}
          \and
          M. Arnaud
          \inst{1}
          }

   \institute{IRFU, CEA, Université Paris-Saclay, F-91191 Gif-sur-Yvette, France \\
   	Université Paris Diderot, AIM, Sorbonne Paris Cité, CEA, CNRS, F-91191, Gif-sur-Yvette, France\\
   	\email{paula.tarrio-alonso@cea.fr}
   	\and
   	IRFU, CEA, Université Paris-Saclay, F-91191 Gif-sur-Yvette, France\\
   }

   \abstract{The combination of X-ray and SZ observations can potentially improve the cluster detection efficiency when compared to using only one of these probes, since both probe the same medium: the hot ionized gas of the intra-cluster medium. 
   	We present a method based on matched multifrequency filters (MMF) for detecting galaxy clusters from SZ and X-ray surveys. 
   	This method builds on a previously proposed joint X-ray-SZ extraction method \citep{Tarrio2016} and allows to blindly detect clusters, that is finding new clusters without knowing their position, size or redshift, by searching on SZ and X-ray maps simultaneously. 
   	The proposed method is tested using data from the ROSAT all-sky survey and from the \emph{Planck} survey. The evaluation is done by comparison with existing cluster catalogues in the area of the sky covered by the deep SPT survey. Thanks to the addition of the X-ray information, the joint detection method is able to achieve simultaneously better purity, better detection efficiency and better position accuracy than its predecessor \emph{Planck} MMF, which is based on SZ maps only. For a purity of 85\%, the X-ray-SZ method detects 141 confirmed clusters in the SPT region, whereas to detect the same number of confirmed clusters with \emph{Planck} MMF, we would need to decrease its purity to 70\%. We provide a catalogue of 225 sources selected by the proposed method in the SPT footprint, with masses ranging between 0.7 and 14.5~$\cdot~10^{14}~M_{\sun}$ and redshifts between 0.01 and 1.2. 
   	}

   \keywords{Methods: data analysis --
                Techniques: image processing --
                Galaxies: clusters: general --
                large-scale structure of Universe --
                X-rays: galaxies: clusters
               }

   \authorrunning{P. Tarr\'io et al.}
   \maketitle

\section{Introduction}\label{sec:intro}

Galaxy clusters can be detected from observations at different bands of the electromagnetic spectrum, each of them probing a different component of the cluster. In optical observations we can see the individual galaxies inside the cluster, which contribute to around 1\% of the total cluster mass. Clusters are identified in these images as overdensities of galaxies. Clusters can also be detected in X-ray observations, where they appear as bright sources with extended emission. In these images we observe the emission of the hot gas of the intra cluster medium (ICM), which accounts for 10\%-15\% of the cluster mass. Over the last decade, this gas has also begun to be detected thanks to the characteristic spectral distortion it produces on the cosmic microwave background (CMB) due to Compton scattering of the CMB photons by the ICM electrons. This effect is known as the Sunyaev-Zeldovich (SZ) effect \citep{Sunyaev1970,Sunyaev1972}. 

State-of-the-art galaxy cluster detection techniques usually rely on the analysis of single-survey observations. 
However, combining information from different surveys at different wavelengths can potentially improve the detection performance, allowing to find more distant or less massive clusters. Although multi-wavelength, multi-survey detection of clusters has been theoretically conceived some years ago \citep{Maturi2007,Pace2008}, it is a very complex task and, up to now, it has only been attempted in practice in the pilot study of \citet{Schuecker2004} on X-ray data from the ROSAT All-Sky Survey (RASS) \citep{Truemper1993,Voges1999} and optical data from the Sloan Digital Sky Survey (SDSS) \citep{York2000}. In our previous work \citep{Tarrio2016}, we proposed a new analysis tool based on matched multifrequency filters (MMF) for extracting clusters from SZ and X-ray maps. The method was based on the combination of the classical SZ MMF \citep{Herranz2002, Melin2006, Melin2012} and an analogous single-frequency matched filter developed for X-ray maps. It was shown that combining these two complementary sources of information improved the  signal-to-noise (S/N) ratio with respect to SZ-only or X-ray-only cluster extraction, and also provided correct photometry as long as the physical relation between X-ray and SZ emission of the clusters, namely the expected $F_{\rm X}/Y_{500}$ relation, was known. The filter was used as an extraction tool to estimate some properties of already detected clusters, but not to detect new clusters in a blind manner, since the position, the size and the redshift of the clusters were assumed to be known.

In this paper we propose a blind detection method based on the X-ray-SZ filter studied in \citet{Tarrio2016}. The goal is to adapt this filter to use it as a blind cluster detection tool, dealing with the fact that we do not know the position, the size and the redshift of the clusters. As demonstrated in \citet{Tarrio2016}, combining X-ray and SZ information increases the cluster S/N with respect to single-map extractions. This gain in S/N will translate directly into a higher detection probability for a given threshold in the S/N. We would also expect to obtain, in principle, a higher purity than using the classical SZ MMF, since the combined version will clean from objects whose emission is far from the expected $F_{\rm X}/Y_{500}$ relation. However, even if an object does not follow the expected relation, it could still pass the detection threshold if it has a very strong signal in the X-ray band. This is similar to the strong infrared emissions that were detected with the classical SZ MMF used by the \citet{PlanckEarlyVIII,Planck2013ResXXIX, Planck2015ResXXVII} despite the fact that their spectra did not fit the expected SZ spectrum. This may be the case of some non-cluster X-ray sources, such as Active Galactic Nuclei (AGNs). Since the X-ray filter was designed to be easily compatible with the classical SZ MMF and it is not specifically optimized for X-ray cluster detection, it does not consider the extent of the sources as other X-ray cluster detection techniques \citep{Bohringer2000,Vikhlinin1998,Pacaud2006,Ebeling1993,Scharf1997}. As a result, when we add the X-ray information, we will also add false detections, produced by non-cluster X-ray sources (mainly AGNs). Therefore, as already remarked in \citet{Tarrio2016}, the main challenge to be solved when using the proposed X-ray-SZ MMF for blind detection is to obtain a high purity.

The proposed method is applied on observations from the ROSAT All-Sky Survey (RASS) and the \textit{Planck} survey, the latest full-sky X-ray and SZ surveys available to date.  Nevertheless, the proposed joint detection technique is general and also applicable to other surveys, including those from future missions such as e-ROSITA \citep{Merloni2012}, a 4-year X-ray survey which is planned to start in 2018 and to be much deeper than RASS.  

The structure of the paper is as follows. Sect. \ref{sec:RASSandPlanck} presents RASS and \textit{Planck} observations. In Sect. \ref{sec:jointdetection} we describe the joint X-ray-SZ detection algorithm. Sect. \ref{sec:evaluation} presents an evaluation of its performance using RASS and \textit{Planck} maps by comparing its results with other cluster catalogues in the SPT region. Finally, we conclude the paper and discuss ongoing and future research directions in Sect. \ref{sec:conclusions}.  

Throughout, we adopt a flat $\Lambda$CDM cosmological model with $H_0 = 70$ km s$^{-1}$ Mpc$^{-1}$ and $\Omega_{\rm M} = 1-\Omega_{\Lambda} = 0.3$. We define $R_{500}$ as the radius within which the average density of the cluster is 500 times the critical density of the universe, $\theta_{500}$ as the corresponding angular radius and $M_{500}$ as the mass enclosed within $R_{500}$.

\section{Description of the observations}\label{sec:RASSandPlanck}

Although the joint algorithm proposed in this paper is a general technique that can be applied, in principle, to any X-ray and SZ surveys of the sky, we have tested it in this paper using all-sky maps from \textit{Planck} and RASS surveys. In this section, we briefly describe these observations.

\subsection{RASS data description}

The ROSAT All-Sky Survey (RASS) is, so far, the only full-sky X-ray survey conducted with an X-ray telescope \citep{Truemper1993,Voges1999}. The survey data release\footnote{ftp://legacy.gsfc.nasa.gov/rosat/data/pspc/processed\_data/rass/release, or http://www.xray.mpe.mpg.de/rosat/survey/rass-3/main/help.html\#ftp} contains 1378 individual RASS fields in three different bands: TOTAL (0.1-2.4 keV), HARD (0.5-2.0 keV) and SOFT (0.1-0.4 keV). Each field covers an area of 6.4 $\deg$ x 6.4 $\deg$ (512 x 512 pixels) and has a resolution of 0.75 arcmin/pixel. 

In this paper, we use an X-ray all-sky HEALPix map that we built from the HARD band information. Similarly to other cluster detection surveys based on RASS data, such as REFLEX \citep{Bohringer2001,Bohringer2013}, we chose to use the HARD band because it provides a better S/N for the clusters, due to the fact that the SOFT band is dominated by the diffuse X-ray background of the local bubble. This map has a resolution of 0.86 arcmin/pixel (HEALPix resolution closest to the RASS resolution). The details of its construction can be found in Appendix B of \cite{Tarrio2016}. 

\subsection{Planck data description}

\textit{Planck} is the most recent space mission that was launched to measure the anisotropy of the CMB. It observed the sky in nine frequency bands from 30 to 857 GHz with high sensitivity and angular resolution. The Low Frequency Instrument (LFI) covered the 30, 44, and 70 GHz bands, while the High Frequency Instrument (HFI) covered the 100, 143, 217, 353, 545, and 857 GHz bands.

In this paper, we use only the six temperature channel maps of HFI, which are the same channels used by the Planck Collaboration to produce their cluster catalogues \citep{PlanckEarlyVIII,Planck2013ResXXIX, Planck2015ResXXVII}. In particular, we used the latest version of these maps, whose description can be found in \cite{Planck2015ResVIII}. The published full resolution maps have a resolution of 1.72 arcmin/pixel. However, to make them directly compatible with the all-sky X-ray map mentioned before, we up-sampled them to a resolution of 0.86 arcmin/pixel by zero-padding in the spherical harmonics domain.

\section{Joint detection of galaxy clusters on X-ray and SZ maps}\label{sec:jointdetection}

In this section, we describe the proposed algorithm for blind detection of galaxy clusters using X-ray and sub-mm maps. The algorithm is based on the X-ray-SZ extraction method proposed in \cite{Tarrio2016}, which aimed at extracting the characteristics of a cluster given its known position, size and redshift. In this paper, we adapted this extraction method to perform a blind detection of clusters, i.e., to discover clusters in the maps without knowing their positions, sizes or redshifts. 

\subsection{X-ray-SZ MMF}\label{ssec:jointalgorithm}

Let us first briefly recall the joint X-ray-SZ extraction method proposed in \cite{Tarrio2016}. This method is based on a matched filter approach and was designed to be compatible with the SZ MMF known as MMF3, described by \cite{Melin2012} and used by the \cite{PlanckEarlyVIII,Planck2013ResXXIX, Planck2015ResXXVII} to construct their SZ cluster catalogues. 

The main idea of the joint extraction algorithm is to consider the X-ray map as an additional SZ map at a given frequency and to introduce it, together with the other SZ maps, in the classical SZ-MMF. In order to do so, the X-ray map needs to be converted into an equivalent SZ map at a reference frequency $ \nu_{\rm ref} $, leveraging the expected $F_{\rm X}/Y_{500}$ relation.  The details of this conversion are described in Appendix B of \cite{Tarrio2016}. Once the X-ray map is expressed in the same units as the SZ maps we can apply the classical MMF to the complete set of maps (the original $N_{\nu}$ SZ maps obtained at sub-mm frequencies $\nu_1, ..., \nu_{N_\nu}$ and an additional SZ map at the reference frequency $ \nu_{\rm ref} $, obtained from the X-ray map). The reference frequency $ \nu_{\rm ref} $ is just a fiducial value with no effect on the extraction algorithm. In our case, we took $\nu_{\rm ref}$ = 1000 GHz.

 \begin{figure*}[]
 	\centering
 	\subfigure[]{\includegraphics[width=0.99\columnwidth]{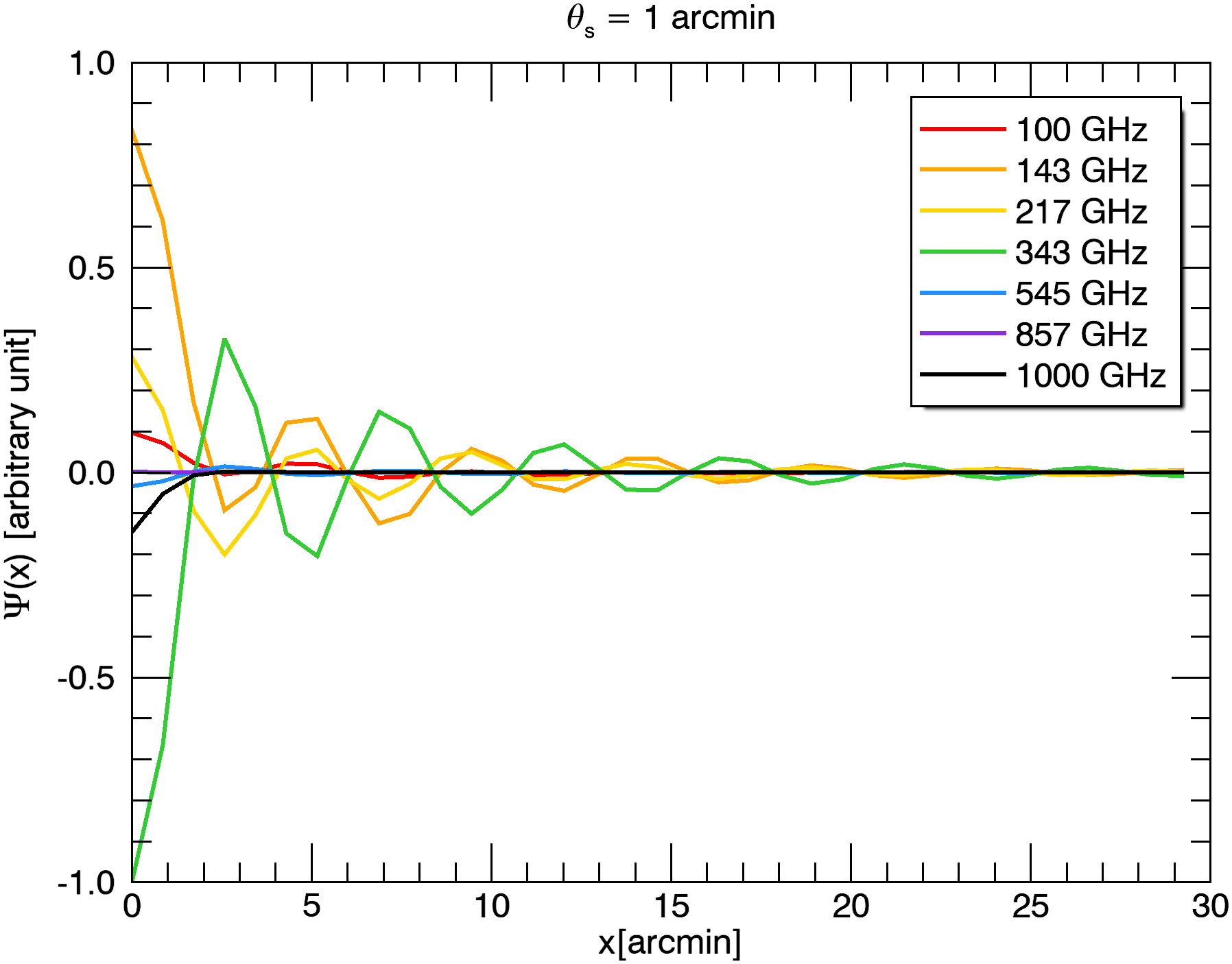}\label{fig:filter_1arcmin}}
 	\subfigure[]{\includegraphics[width=0.99\columnwidth]{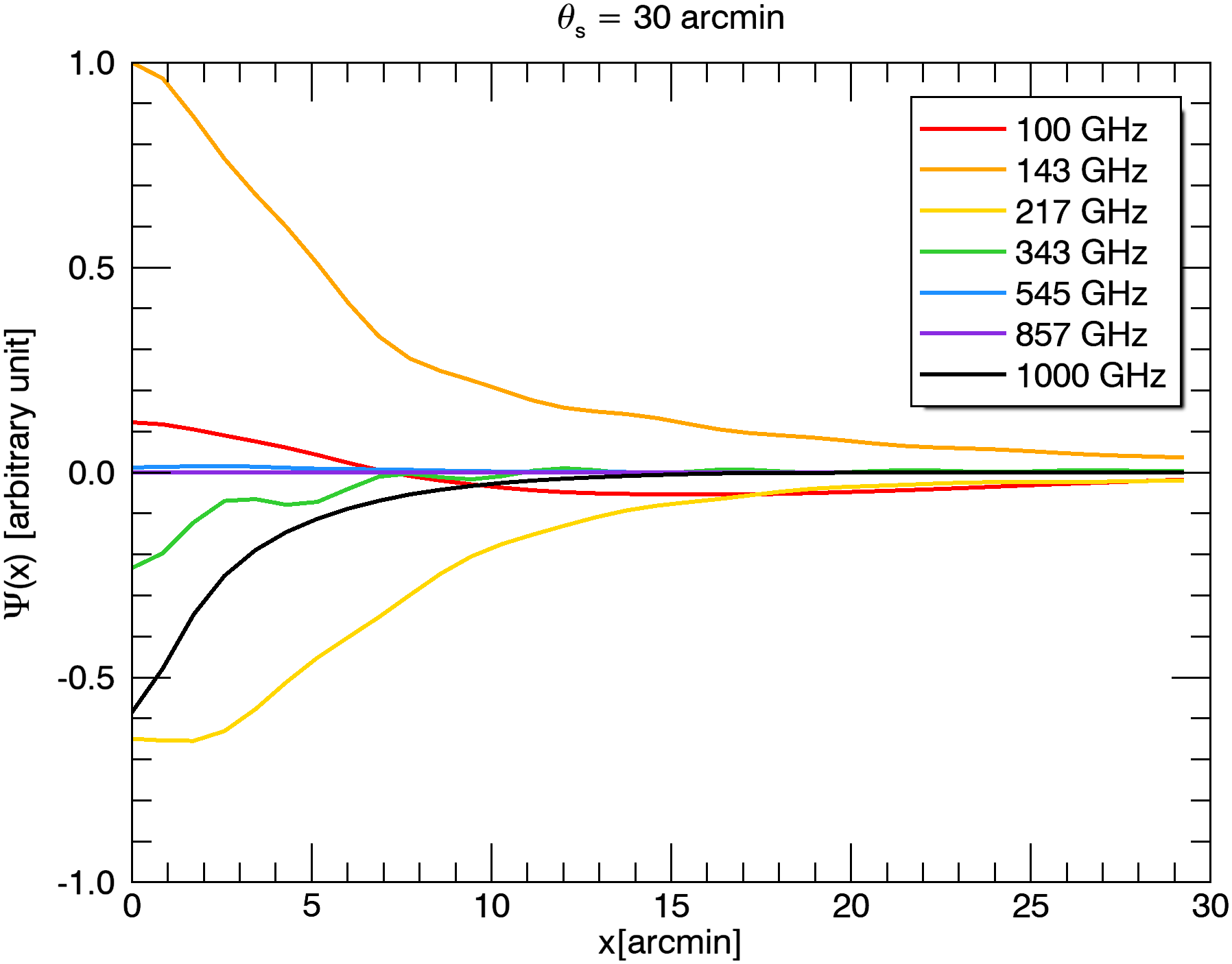}\label{fig:filter_30arcmin}}
 	\caption{Examples of the matched filter $\mathbf{\Psi}_{\theta_{\rm s}}$ for $\theta_{\rm s}=1$ arcmin (a) and $\theta_{\rm s} = 30$ arcmin (b). The curves give the radial profiles of the filters, which are symmetric because we have chosen a symmetric cluster template. The filter is normalized so that its maximum amplitude is equal to 1.}
 	\label{fig:filter}		
 \end{figure*}
 
The X-ray-SZ MMF presented in \cite{Tarrio2016} is given, in Fourier space\footnote{We use $\mathbf{k}$ to denote the 2-dimensional spatial frequency, corresponding to the 2D-position $\mathbf{x}$ in the Fourier space. All the variables expressed as a function of $\mathbf{k}$ are then to be understood as variables in the Fourier space.}, by
\begin{equation}\label{eq:filter_sz}
\mathbf{\Psi}_{\theta_{\rm s}}(\mathbf{k}) = \sigma_{\theta_{\rm s}}^2 \mathbf{P}^{-1}(\mathbf{k})  \mathbf{F}_{\theta_{\rm s}}(\mathbf{k})
\end{equation}
with
\begin{equation}\label{eq:sigma_sz}
\sigma_{\theta_{\rm s}}^2 =  \left[ \sum_{\mathbf{k}}   \mathbf{F}_{\theta_{\rm s}}^{\rm T}(\mathbf{k})  \mathbf{P}^{-1}(\mathbf{k})  \mathbf{F}_{\theta_{\rm s}}(\mathbf{k}) \right] ^{-1}
\end{equation}
$\mathbf{\Psi}_{\theta_{\rm s}}$ is a $(N_\nu+1) \times 1 $ column vector whose $i$th component will filter the map at observation frequency $\nu_i$.  $\sigma_{\theta_{\rm s}}^2$ is, approximately, the background noise variance after filtering. $\mathbf{P}(\mathbf{k})$ is the noise power spectrum, a $(N_\nu+1) \times (N_\nu+1)$ matrix whose $ij$ component is given by $\left\langle N_i(\mathbf{k})N_j^\ast(\mathbf{k}')\right\rangle  = P_{ij}(\mathbf{k}) \delta(\mathbf{k}-\mathbf{k}')$, where $N_i(\mathbf{k})$ is the noise map at observation frequency $\nu_i$, which includes instrumental noise and astrophysical sources different from the cluster signal (extragalactic point sources, diffuse Galactic emission and the primary CMB anisotropy, for the SZ maps and X-ray background for the X-ray map). Finally, $\mathbf{F}_{\theta_{\rm s}}$ is a $(N_\nu+1) \times 1 $ column vector defined as
\begin{equation} \label{eq:F_joint}
\mathbf{F}_{\theta_{\rm s}}(\mathbf{x})  = 
[j(\nu_{1}) T_{1}(\mathbf{x}), ..., j(\nu_{N_{\nu}}) T_{N_{\nu}}(\mathbf{x}), C j(\nu_{\rm ref}) T^{\rm{x}}_{\theta_{\rm s}}(\mathbf{x})]^{\rm T}
\end{equation} 
where $j(\nu_{i})$ is the SZ spectral function at frequency $\nu_{i}$ and $T_{i}(\mathbf{x}) = \tilde{T}_{\theta_{\rm s}}(\mathbf{x}) \ast B_{\nu_i}(\mathbf{x})$ and $ T^{\rm{x}}_{\theta_{\rm s}}(\mathbf{x}) = \tilde{T}^{\rm{x}}_{\theta_{\rm s}}(\mathbf{x}) \ast B_{\rm xray}(\mathbf{x})$ are the convolutions of the cluster 2D spatial profiles ($ \tilde{T}_{\theta_{\rm s}}(\mathbf{x}) $ for the SZ profile and $ \tilde{T}^{\rm{x}}_{\theta_{\rm s}}(\mathbf{x}) $ for the X-ray profile) with the point spread function (PSF) of the instruments at the different frequencies.
The 2D cluster profiles $ \tilde{T}_{\theta_{\rm s}}(\mathbf{x}) $  and $ \tilde{T}^{\rm{x}}_{\theta_{\rm s}}(\mathbf{x}) $ are normalized so that their central value is 1. Finally, the constant $C$ is a geometrical factor that accounts for the different shapes of the SZ and X-ray 3D profiles and it is defined in Eq. 25 of \cite{Tarrio2016} as the ratio of the integrated fluxes of the normalized SZ and X-ray 3D profiles up to $R_{500}$.
As we can see, the filter is determined by the shape of the cluster signal and by the power spectrum of the noise, hence, the name of matched filter.

This matched filter approach relies on the knowledge of the normalized cluster profile. This profile is not known in practice, so we need to approximate it by the theoretical profile that best represents the clusters we want to detect. As in \cite{Tarrio2016}, we assume the generalized Navarro-Frenk-White (GNFW) profile \citep{Nagai2007} given by \begin{equation}\label{eq:pressure_prof}
p(x) \propto \frac{1}{\left( c_{500}x\right) ^{\gamma} \left[ 1+\left( c_{500}x\right) ^{\alpha}\right]^{(\beta-\gamma)/\alpha}  }
\end{equation}
with parameters given by
\begin{equation}\label{eq:sz_param}
\left[ \alpha, \beta, \gamma, c_{500}\right]   = \left[ 1.0510, 5.4905, 0.3081, 1.177\right]
\end{equation}
and
\begin{equation}\label{eq:xray_param}
\left[ \alpha, \beta, \gamma, c_{500}\right] = \left[ 2.0, 4.608, 1.05, 1/0.303\right] 
\end{equation}
for the components corresponding to the original SZ maps and the additional X-ray map, respectively. These parameters come from assuming the 3D pressure profile of \citet{Arnaud2010} and the average gas density profile from \citet{Piffaretti2011}, respectively. Note that $x=\theta/\theta_{500}$ represents here the 3D-distance to the center of the cluster in $\theta_{500}$ units, and $\theta_{500}$ relates to the characteristic cluster scale $\theta_{\rm s}$ through the concentration parameter $c_{500}$ ($\theta_{\rm s} = \theta_{500}/c_{500}$). The cluster profile is then obtained by numerically integrating these 3D GNFW profiles along the line-of-sight. 

Finally, this cluster profile needs to be convolved by the instrument beams. As in \cite{Tarrio2016}, in this paper we will use the six highest frequency \textit{Planck} maps and the X-ray maps of the ROSAT All-Sky Survey. Therefore, we will use the same instrument beams as in \cite{Tarrio2016}, namely, a Gaussian PSF for the SZ components, with FWHM depending on the frequency, as shown in Table 6 of \citet{Planck2015ResVIII}, and a PSF for the X-ray component that was estimated numerically by stacking observations of X-ray point sources from the Bright Source Catalogue \citep{Voges1999}.

Figure \ref{fig:filter} shows two examples of the radial profiles of the filter $\mathbf{\Psi}_{\theta_{\rm s}}$ for $\theta_{\rm s}=1$ and $\theta_{\rm s} = 30$ arcmin, where we can see the spectral and the spatial weighting introduced by the filter. The filter was computed at a random position, with galactic coordinates 260.356$\degr$, -20.332$\degr$. We remark that the last component of the filter corresponds to the X-ray band, and that its relative amplitude depends on the chosen reference frequency $\nu_{\rm ref}$.

\subsection{Blind procedure}\label{ssec:blind}

In \cite{Tarrio2016}, the above-described filter was proposed as an extraction tool, i.e., to estimate the flux of a cluster once we know that there is a cluster at a given position $\mathbf{x}_0$, and we know its size $\theta_{\rm s}$ and redshift (necessary to convert the X-ray into an equivalent SZ map through the $F_{\rm X}/Y_{500}$ relation). In this section we describe how this method is adapted to become a detection tool. 

Given a set of $N_\nu+1$ maps (SZ + X-ray) of a given region of the sky $\mathbf{M}(\mathbf{x})=[M_{1}(\mathbf{x}), ..., M_{N_\nu}(\mathbf{x}), M_{\rm ref}(\mathbf{x})]^{\rm T}$, where $M_{\rm ref}$ is the X-ray map already converted into SZ units, the first step to detect new clusters consists of filtering the maps with the filter defined in Eq. \ref{eq:filter_sz} as follows 
	\begin{equation}\label{eq:y0_estim}
	\hat{y}(\mathbf{x}) = \sum_{\mathbf{x'}} \mathbf{\Psi}_{\theta_{\rm s}}^{\rm T}(\mathbf{x'}-\mathbf{x}) {\mathbf{M}}(\mathbf{x'})
	\end{equation}
In this way, we obtain a $\hat{y}$-map (filtered map) and a signal-to-noise (S/N) map ($\hat{y}(\mathbf{x})$/$\sigma_{\theta_{\rm s}}$) with the same size as the observed maps. 

We note that to calculate the filter $\mathbf{\Psi}_{\theta_{\rm s}}$, we first need to estimate the noise power spectrum $\mathbf{P}(\mathbf{k})$. This is done in practice from the X-ray and SZ images themselves, assuming that they contain mostly noise. In the case of the X-ray images, this assumption may not be true due to bright X-ray sources with strong signals. Therefore, to minimize this effect, we masked some regions of the X-ray images for the calculation of $\mathbf{P}(\mathbf{k})$. In particular, we masked the areas defined in Table 1 of \citet{Bohringer2001} corresponding to the Large and Small Magellanic Clouds, and we also masked the X-ray sources of the ROSAT bright source catalogue \citep{Voges1999} that have a countrate greater than 0.3 counts/s. 

Since the size of the clusters is unknown, we repeat the filtering process using a set of $N_{\rm s}$ filters with different sizes, covering the expected range of radii. In our case, we vary $\theta_{500}$ from 0.94 to 35.31 arcmin, in $N_{\rm s}=32$ steps equally spaced in logarithmic scale. For each size, we obtain a filtered map and a S/N map.
The clusters are then detected as peaks in these S/N maps, down to a given threshold.

Finally, for the conversion of the X-ray map into an equivalent SZ map we need to assume a $F_{\rm X}/Y_{500}$ relation, which depends on the redshift. As studied in \cite{Tarrio2016}, the assumed $F_{\rm X}/Y_{500}$ relation does not have a big impact in the estimated S/N, which makes the detection robust against possible errors in the assumed relation. For this reason, we have fixed the redshift to a reference value of $z_{\rm ref}=0.8$ and assumed the relation found by the \citet{PlanckIntI2012}:
\begin{equation}\label{eq:FxY500relation}
\frac{F_X \left[ \mbox{erg s$^{-1}$ cm$^{-2}$}\right] }{Y_{500} \left[ \mbox{arcmin}^{2}\right] } = 4.95 \cdot 10^{-9} \cdot E(z)^{5/3} (1+z)^{-4} K(z),
\end{equation}
where the K-correction can be obtained by interpolating in table 2 of \cite{Tarrio2016}. 

To implement the detection procedure in practice, we proceed in two passes. 
\begin{enumerate}
	\item Produce a preliminary list of candidates. In this first pass, we project the all-sky maps into 504 small $10\degr \times 10\degr$ tangential patches, as done in MMF3 \citep{PlanckEarlyVIII,Planck2013ResXXIX,Planck2015ResXXVII}. Each patch is filtered by the X-ray-SZ filter $\mathbf{\Psi}_{\theta_{\rm s}}$ (eq. \ref{eq:filter_sz}) using $N_{\rm s}=32$ different sizes, which produces $N_{\rm s}$ S/N maps. Then, we construct a list with the peaks in these maps that are above a specified S/N threshold $q$. The procedure is as follows:
	\begin{enumerate}
		\item We look for the highest peak among all the $N_{\rm s}$ S/N maps. 
		\item If it is above the specified threshold $q$, we include its position in a preliminary candidate list and mask it in the $N_{\rm s}$ S/N maps. The size of the mask is defined as the radius at which the value of the filtered template is 1\% of its maximum value. The filtered template is the 2D cluster profile corresponding to the size at which the highest peak was found, convolved by the PSF, and filtered by the X-ray-SZ filter $\mathbf{\Psi}_{\theta_{\rm s}}$ with the $N_{\rm s}=32$ different sizes. Thus, the size of the mask is different in each of the $N_{\rm s}$ S/N maps.
		\item Then we repeat the search until there are no more peaks above the specified threshold.
	\end{enumerate}
	Finally, we merge the 504 lists into a single preliminary all-sky list of candidates by merging peaks that are close to each other by less than 10 arcmin, as done in MMF3 \citep{PlanckEarlyVIII,Planck2013ResXXIX,Planck2015ResXXVII}. 
	\item Refine the list of candidates. 
	In this second pass, we re-analyze each candidate in the preliminary candidate list. This second pass is necessary to better estimate the candidate properties and S/N, since the results from the first pass may not be accurate. This is especially true if the candidate is situated close to a border of the map, since the estimated noise in this case may not be representative of the noise around the candidate. For each candidate we follow the next procedure:
	\begin{enumerate}
		\item We produce a set of $N_\nu+1$ (SZ + X-ray) $10\degr \times 10\degr$ tangential maps centered at the candidate position. 
		\item We filter these maps with the different filter sizes, obtaining $N_{\rm s}$ S/N maps.
		\item We estimate the S/N of the detection by selecting a small circular region around the center in each of the $N_{\rm s}$ S/N maps and searching for local maxima inside this volume. This is necessary because the position of the peak may have changed slightly when centering the tangential maps. Among all the local maxima, we select the one with highest S/N that is not on the border of the circles (to avoid tails of nearby objects).
		\item If this S/N is above a specified threshold $q$, we add the detection, with its new position and corresponding size, to the final candidate list, otherwise we discard it. 
	\end{enumerate}
\end{enumerate}

\subsection{Determination of the threshold on the joint S/N}\label{ssec:jointthreshold}

An important point of the blind joint detection algorithm is the selection of the threshold $q$ that is applied to the peaks found in the first and second pass. The goal of this threshold is to discard false detections (noise peaks) with a given confidence. This can be achieved by setting the probability that a detection is due to a random fluctuation to a sufficiently low value. In the MMF3 method, a fixed threshold is used under the assumption that the noise distribution is Gaussian, so that a fixed S/N threshold leads to a fixed number of noise peak detections. In the joint X-ray-SZ detection, the Gaussian assumption is no longer valid, as explained next, so the threshold must be selected differently.

The probability density function (PDF) of the S/N in the joint filtered maps depends on the noise properties of the observed maps. Due to the Poisson nature of the noise in the X-ray maps, the final PDF of the joint S/N is not Gaussian. Its shape depends on the exposure time of the X-ray map and also on the filter size. In particular, it becomes more long-tailed when the exposure time is low, especially for small filter sizes. This is due to the fact that in these cases, most of the pixels of the X-ray map contain zero photons, and just a few pixels contain one photon. As a consequence, the average background is very low and the S/N of the filtered map, defined as $\hat{y}(\mathbf{x})$/$\sigma_{\theta_{\rm s}}$, at the few  pixels with one photon can be easily quite high.

Since the PDF of the joint S/N will have different shapes in different regions of the sky, using a fixed S/N threshold to detect cluster candidates everywhere in the sky will produce a different number of false detections, e.g. more detections will appear in low exposure time regions due to single noise pixels with high S/N. To have an approximately constant number of false detections over the whole sky, we need to establish an adaptive threshold that depends on the noise characteristics of each region. Since the PDF of the joint S/N cannot be calculated analytically, we have determined this adaptive threshold numerically by means of Monte Carlo simulations.

In particular, we performed an experiment in which we simulated a set of $N_\nu+1=7$ maps emulating in a simple manner the noise properties of \textit{Planck} and RASS maps. 
\begin{itemize}
	\item The RASS noise map was simulated as an homogeneous Poisson random field, characterized by a given mean value $\lambda$ (in counts/pixel). This noise represents the instrumental noise and the astrophysical X-ray background (mainly due to diffuse galactic emission and non-resolved point sources). To express this map into X-ray flux units, we assumed an exposure time of 400 s and a $N_{\rm H}$ of $2 \cdot 10^{20}$ cm$^{-2}$ (average values in the SPT region). We remark that the simulation results obtained for these values can be converted to the ones that would be obtained for any other values of exposure time and $N_{\rm H}$, as explained in Appendix \ref{app:conversion}, so this choice does not have any implications. Finally, as done with the real RASS maps, this X-ray flux map is converted into an equivalent SZ map at the reference frequency $\nu_{\rm ref}$  by following the  procedure detailed in Appendix B of \cite{Tarrio2016}.
	\item The $N_\nu=6$ \textit{Planck} noise maps were simulated as the sum of two independent components: primary anisotropies and white Gaussian noise. First, we used the \textit{Planck} Sky Model \citep{Delabrouille2013} to obtain a realization of the CMB for the $N_\nu=6$ \textit{Planck} frequencies (100, 143, 217, 353, 545, 857 GHz). Second, for each frequency, we added zero-mean Gaussian random noise with a frequency-dependent variance. In particular, the variance at frequency $\nu$ was fixed to the following value: $\sigma_\nu = \sigma_{217}*[1.66, 0.70, 1.00, 3.12, 19.50, 649.87]$\footnote{These ratios correspond to the ones of the real \textit{Planck} maps in the SPT region.}, where $\sigma_{217}$ is the standard deviation of the Gaussian noise in the 217 GHz map. Therefore, the simulated \textit{Planck} noise maps are characterized by this single parameter $\sigma_{217}$.
\end{itemize}

We repeated the experiment for different values of the mean Poisson level $\lambda$ and the Gaussian noise level $\sigma_{217}$, and for each pair of values $\lambda$-$\sigma_{217}$ we used 450 different realizations of the noise maps. At each realization, we changed the Poisson and the Gaussian noises (maintaining their levels), as well as the CMB realization.

Each set of $N_\nu+1=7$ maps was then filtered using the proposed joint filter with the $N_{\rm s}=32$ different sizes, yielding the variance of the X-ray filtered map $\sigma_{\theta_{\rm s}}^{\rm x}$, the variance of the SZ filtered maps $\sigma_{\theta_{\rm s}}^{\rm sz}$ and the S/N map for each filter size. 

 From these results, we calculated the average values of $\sigma_{\theta_{\rm s}}^{\rm x}$ and $\sigma_{\theta_{\rm s}}^{\rm sz}$ corresponding to each noise level and filter size, which are reported in tables \ref{table:filteredsigmasxr} and \ref{table:filteredsigmassz}. Finally, we established the joint S/N threshold $q_{\rm J}$ for a given $\lambda$-$\sigma_{217}$-$\theta_{\rm s}$ triplet as the S/N value for which the fraction of pixels (considering the 450 realizations) with S/N>$q_{\rm J}$ does not exceed a given false alarm probability $P_{\rm{FA}}$. We used the number of pixels with S/N>$q_{\rm J}$ as an approximation to the number of detections with S/N>$q_{\rm J}$. Due to the iterative blind detection procedure, where each S/N peak is masked after detection (step 1b of the blind procedure described in Sect. \ref{ssec:blind}), one detection spans more than one pixel. However, the approximation allows a much faster computation and it is also accurate enough, especially for small filter sizes, which is the regime in which the Poisson noise peaks become more important and which we need to characterize better.

The value of $P_{\rm{FA}}$ serves to select the operational point of the detection method. The higher the $P_{\rm{FA}}$, the lower the threshold $q_{\rm J}$ and the more candidates we keep, resulting in a catalog with higher completeness and lower purity. On the contrary, if we want a very pure catalog at the expense of being less complete, we will choose a small value of $P_{\rm{FA}}$. Table \ref{table:snthreshold} summarizes the S/N thresholds for each combination of noise and some filter sizes calculated for false alarm rate of $P_{\rm{FA}}=3.4 \cdot 10^{-6}$, which corresponds to a cut at $4.5\sigma$ in a zero-mean Gaussian distribution.

	\begin{table*}
		\caption{Average standard deviation of the X-ray filtered map $\sigma_{\theta_{\rm s}}^{\rm x}$ (in $\Delta T/T$  units) for different values of mean Poisson noise $\lambda$ (in counts/pixel) and several filter sizes (in arcmin), corresponding to a map with exposure time of 400 s and $N_{\rm H}=2 \cdot 10^{20}$ cm$^{-2}$.}
		\label{table:filteredsigmasxr}
		\centering 
		\begin{tabular}{c | c c c c c c c c c}
			\hline
			\noalign{\smallskip}
			 & $\lambda=0.03$ &  $\lambda=0.06$ & $\lambda=0.09$ & $\lambda=0.15$ & $\lambda=0.25$ & $\lambda=0.40$ & $\lambda=0.70$ & $\lambda=2.00$ & $\lambda=7.50$\\
			\noalign{\smallskip}
			\hline
			\noalign{\smallskip}
$\theta_{\rm s}=$0.80 & 5.00e-5 & 7.08e-5 & 8.68e-5 & 1.12e-4 & 1.45e-4 & 1.84e-4 & 2.43e-4 & 4.10e-4 & 7.93e-4 \\ 
$\theta_{\rm s}=$1.28 & 3.07e-5 & 4.34e-5 & 5.33e-5 & 6.90e-5 & 8.93e-5 & 1.13e-4 & 1.49e-4 & 2.52e-4 & 4.87e-4 \\ 
$\theta_{\rm s}=$2.04 & 1.76e-5 & 2.49e-5 & 3.06e-5 & 3.96e-5 & 5.12e-5 & 6.49e-5 & 8.56e-5 & 1.44e-4 & 2.79e-4 \\ 
$\theta_{\rm s}=$3.25 & 9.40e-6 & 1.33e-5 & 1.64e-5 & 2.12e-5 & 2.75e-5 & 3.48e-5 & 4.59e-5 & 7.73e-5 & 1.49e-4 \\ 
$\theta_{\rm s}=$5.19 & 4.96e-6 & 7.03e-6 & 8.64e-6 & 1.12e-5 & 1.45e-5 & 1.84e-5 & 2.42e-5 & 4.08e-5 & 7.88e-5 \\ 
$\theta_{\rm s}=$8.29 & 2.71e-6 & 3.86e-6 & 4.75e-6 & 6.17e-6 & 8.01e-6 & 1.01e-5 & 1.33e-5 & 2.24e-5 & 4.32e-5 \\ 
$\theta_{\rm s}=$13.23 & 1.50e-6 & 2.14e-6 & 2.64e-6 & 3.44e-6 & 4.47e-6 & 5.67e-6 & 7.41e-6 & 1.24e-5 & 2.39e-5 \\ 
$\theta_{\rm s}=$21.12 & 8.81e-7 & 1.26e-6 & 1.56e-6 & 2.04e-6 & 2.67e-6 & 3.38e-6 & 4.38e-6 & 7.29e-6 & 1.41e-5 \\ 
$\theta_{\rm s}=$30.00 & 5.91e-7 & 8.51e-7 & 1.06e-6 & 1.39e-6 & 1.82e-6 & 2.30e-6 & 2.96e-6 & 4.90e-6 & 9.43e-6 \\ 
			\noalign{\smallskip}
			\hline
		\end{tabular}
	\end{table*}
	
\begin{table*}
			\caption{Average standard deviation of the SZ filtered maps $\sigma_{\theta_{\rm s}}^{\rm sz}$ (in $\Delta T/T$  units) for different values of mean Gaussian noise $\sigma_{217}$ (in $\Delta T/T$  units) and several filter sizes (in arcmin).}
			\label{table:filteredsigmassz}
			\centering 
			\begin{tabular}{c | c c c c c c c}
				\hline
				\noalign{\smallskip}
				& $\sigma_{217}=10^{-5}$ &  $\sigma_{217}=2 \cdot 10^{-5}$ & $\sigma_{217}=2.5 \cdot 10^{-5}$ & $\sigma_{217}=3 \cdot 10^{-5}$ & $\sigma_{217}=4 \cdot 10^{-5}$ & $\sigma_{217}=5 \cdot 10^{-5}$ & $\sigma_{217}=6 \cdot 10^{-5}$\\
				\noalign{\smallskip}
				\hline
				\noalign{\smallskip}
$\theta_{\rm s}=$0.80 & 4.77e-5 & 9.01e-5 & 1.11e-4 & 1.33e-4 & 1.76e-4 & 2.18e-4 & 2.61e-4 \\  
$\theta_{\rm s}=$1.28 & 2.95e-5 & 5.56e-5 & 6.87e-5 & 8.19e-5 & 1.08e-4 & 1.35e-4 & 1.61e-4 \\ 
$\theta_{\rm s}=$2.04 & 1.74e-5 & 3.25e-5 & 4.02e-5 & 4.79e-5 & 6.32e-5 & 7.86e-5 & 9.39e-5 \\  
$\theta_{\rm s}=$3.25 & 9.78e-6 & 1.82e-5 & 2.24e-5 & 2.67e-5 & 3.52e-5 & 4.37e-5 & 5.22e-5 \\ 
$\theta_{\rm s}=$5.19 & 5.50e-6 & 1.01e-5 & 1.25e-5 & 1.48e-5 & 1.95e-5 & 2.43e-5 & 2.90e-5 \\  
$\theta_{\rm s}=$8.29 & 3.26e-6 & 5.94e-6 & 7.30e-6 & 8.68e-6 & 1.14e-5 & 1.42e-5 & 1.70e-5 \\  
$\theta_{\rm s}=$13.23 & 1.94e-6 & 3.50e-6 & 4.30e-6 & 5.11e-6 & 6.74e-6 & 8.37e-6 & 1.00e-5 \\ 
$\theta_{\rm s}=$21.12 & 1.20e-6 & 2.16e-6 & 2.65e-6 & 3.15e-6 & 4.15e-6 & 5.16e-6 & 6.17e-6 \\  
$\theta_{\rm s}=$30.00 & 8.19e-7 & 1.48e-6 & 1.81e-6 & 2.15e-6 & 2.84e-6 & 3.53e-6 & 4.22e-6 \\ 
				\noalign{\smallskip}
				\hline
			\end{tabular}
		\end{table*}
		
	\begin{table*}
		\caption{Joint S/N threshold $q_{\rm J}$ for different values of mean Poisson noise $\lambda$ (expressed in counts/pixel), different values of mean Gaussian noise $\sigma_{217}$ (in $\Delta T/T$  units) and several filter sizes (in arcmin). These thresholds correspond to a false alarm rate of $3.4 \cdot 10^{-6}$ (equivalent to a $4.5\sigma$ cut in a Gaussian distribution). They correspond to the case where exposure time is 400 s and $N_{\rm H}=2 \cdot 10^{20}$ cm$^{-2}$.}
		\label{table:snthreshold}
		\centering 
		\begin{tabular}{c | c c c c c c c c c}
			\hline
			\noalign{\smallskip}
			$\theta_{\rm s}=0.8$ & $\lambda=0.03$ &  $\lambda=0.06$ & $\lambda=0.09$ & $\lambda=0.15$ & $\lambda=0.25$ & $\lambda=0.40$ & $\lambda=0.70$ & $\lambda=2.00$ & $\lambda=7.50$\\
			\noalign{\smallskip}
			\hline
			\noalign{\smallskip}
			$\sigma_{217}=10^{-5}$           & 7.31 & 5.67 & 5.09 & 4.71 & 4.59 & 4.52 & 4.50 & 4.51 & 4.51 \\ 
			$\sigma_{217}=2 \cdot 10^{-5}$   & 8.84 & 7.02 & 6.21 & 5.45 & 4.94 & 4.70 & 4.55 & 4.50 & 4.52 \\ 
			$\sigma_{217}=2.5 \cdot 10^{-5}$ & 9.15 & 7.40 & 6.57 & 5.77 & 5.17 & 4.84 & 4.61 & 4.53 & 4.53 \\ 
			$\sigma_{217}=3 \cdot 10^{-5}$   & 9.34 & 7.64 & 6.84 & 6.03 & 5.38 & 4.98 & 4.69 & 4.54 & 4.55 \\ 
			$\sigma_{217}=4 \cdot 10^{-5}$   & 9.55 & 7.95 & 7.17 & 6.38 & 5.69 & 5.23 & 4.86 & 4.60 & 4.56 \\ 
			$\sigma_{217}=5 \cdot 10^{-5}$   & 9.65 & 8.10 & 7.36 & 6.59 & 5.93 & 5.44 & 5.01 & 4.65 & 4.57 \\ 
			$\sigma_{217}=6 \cdot 10^{-5}$   & 9.71 & 8.20 & 7.47 & 6.72 & 6.08 & 5.58 & 5.15 & 4.70 & 4.59 \\ 
			\noalign{\smallskip}
			\hline
			\noalign{\smallskip}
			$\theta_{\rm s}=5.19$ & $\lambda=0.03$ &  $\lambda=0.06$ & $\lambda=0.09$ & $\lambda=0.15$ & $\lambda=0.25$ & $\lambda=0.40$ & $\lambda=0.70$ & $\lambda=2.00$ & $\lambda=7.50$\\
			\noalign{\smallskip}
			\hline
			\noalign{\smallskip}
$\sigma_{217}=10^{-5}$           & 6.19 & 5.14 & 4.79 & 4.60 & 4.52 & 4.43 & 4.42 & 4.42 & 4.42 \\ 
$\sigma_{217}=2 \cdot 10^{-5}$   & 7.06 & 5.96 & 5.50 & 5.07 & 4.76 & 4.55 & 4.44 & 4.41 & 4.38 \\ 
$\sigma_{217}=2.5 \cdot 10^{-5}$ & 7.23 & 6.17 & 5.72 & 5.25 & 4.92 & 4.65 & 4.52 & 4.44 & 4.41 \\ 
$\sigma_{217}=3 \cdot 10^{-5}$   & 7.34 & 6.32 & 5.87 & 5.40 & 5.05 & 4.76 & 4.58 & 4.48 & 4.43 \\ 
$\sigma_{217}=4 \cdot 10^{-5}$   & 7.47 & 6.49 & 6.06 & 5.59 & 5.23 & 4.92 & 4.71 & 4.53 & 4.46 \\ 
$\sigma_{217}=5 \cdot 10^{-5}$   & 7.52 & 6.58 & 6.17 & 5.70 & 5.35 & 5.04 & 4.81 & 4.58 & 4.49 \\ 
$\sigma_{217}=6 \cdot 10^{-5}$   & 7.56 & 6.64 & 6.23 & 5.76 & 5.43 & 5.14 & 4.90 & 4.63 & 4.51 \\ 
			\noalign{\smallskip}
			\hline
			\noalign{\smallskip}
			$\theta_{\rm s}=30.0$ & $\lambda=0.03$ &  $\lambda=0.06$ & $\lambda=0.09$ & $\lambda=0.15$ & $\lambda=0.25$ & $\lambda=0.40$ & $\lambda=0.70$ & $\lambda=2.00$ & $\lambda=7.50$\\
			\noalign{\smallskip}
			\hline
			\noalign{\smallskip}
			$\sigma_{217}=10^{-5}$           & 5.11 & 4.54 & 4.40 & 4.18 & 4.21 & 4.17 & 4.01 & 4.10 & 4.10 \\ 
			$\sigma_{217}=2 \cdot 10^{-5}$   & 5.51 & 4.97 & 4.67 & 4.43 & 4.50 & 4.39 & 4.23 & 4.28 & 4.22 \\ 
			$\sigma_{217}=2.5 \cdot 10^{-5}$ & 5.59 & 5.06 & 4.74 & 4.52 & 4.60 & 4.44 & 4.29 & 4.36 & 4.26 \\ 
			$\sigma_{217}=3 \cdot 10^{-5}$   & 5.63 & 5.11 & 4.80 & 4.58 & 4.64 & 4.48 & 4.34 & 4.43 & 4.28 \\ 
			$\sigma_{217}=4 \cdot 10^{-5}$   & 5.66 & 5.16 & 4.88 & 4.65 & 4.69 & 4.51 & 4.40 & 4.51 & 4.28 \\ 
			$\sigma_{217}=5 \cdot 10^{-5}$   & 5.68 & 5.17 & 4.91 & 4.69 & 4.70 & 4.52 & 4.45 & 4.56 & 4.28 \\ 
			$\sigma_{217}=6 \cdot 10^{-5}$   & 5.69 & 5.18 & 4.94 & 4.71 & 4.70 & 4.54 & 4.49 & 4.58 & 4.26 \\ 
			\noalign{\smallskip}
			\hline
		\end{tabular}
	\end{table*}

For simplicity reasons, we apply this adaptive threshold $q_{\rm J}$ after we have obtained the final candidate list from the second pass of the blind procedure. The threshold $q$ to be applied in the first and second pass is established to a sufficiently low value so that it does not introduce any different selection effect, i.e. it does not discard any candidate above $q_{\rm J}$. In our case, we selected $q = 4$ for the first and second pass, which is lower than any of the adaptive thresholds shown in table \ref{table:snthreshold}. 
The adaptive threshold $q_{\rm J}$ is then used to discard noise-detections in the following way. 
\begin{enumerate}
	\item For each detection in the final candidate list provided at the second pass of the joint blind algorithm, we save the joint S/N of the detection, the corresponding filter size $\theta_{\rm s}$, the standard deviation of the X-ray filtered map $\sigma_{\theta_{\rm s}}^{\rm x}$ and the standard deviation of the SZ filtered maps $\sigma_{\theta_{\rm s}}^{\rm sz}$. 
	\item We then calculate the mean Gaussian level $\hat{\sigma}_{217}$ that corresponds to the measured $\sigma_{\theta_{\rm s}}^{\rm sz}$ by interpolating in the simulation results (table \ref{table:filteredsigmassz}).
	\item Then, we calculate the mean Poisson level $\hat{\lambda}$ that corresponds to the measured $\sigma_{\theta_{\rm s}}^{\rm x}$. To this end, we need to take into account that the measured $\sigma_{\theta_{\rm s}}^{\rm x}$ corresponds to a map with an exposure time and a $N_{\rm H}$ that are different from the ones used in the simulations ($t_{\rm exp} = 400$ s and $N_{\rm H}=2 \cdot 10^{20}$ cm$^{-2}$). Therefore, we will first convert the measured $\sigma_{\theta_{\rm s}}^{\rm x}$ into the value that we would have obtained with the values of exposure time and $N_{\rm H}$ used in the simulations, and then use table \ref{table:filteredsigmasxr} to interpolate the value of $\lambda$. The conversion from the measured $\sigma_{\theta_{\rm s}}^{\rm x}$ into its simulation-equivalent counterpart is done using eq. \ref{eq:sigmareal_sigmasimu}. A detailed description of this conversion can be found in Appendix \ref{app:conversion}.
	\item We choose the two simulated values of $\sigma_{217}$ that are closer to $\hat{\sigma}_{217}$  (one above: $\sigma_{217}^+$, one below: $\sigma_{217}^-$) and the two simulated values of $\lambda$ that are closer to $\hat{\lambda}$ ($\lambda^+$ and $\lambda^-$). Then, we select the four simulations corresponding to the filter size $\theta_{\rm s}$ and the four possible combinations of $\sigma_{217}^+$,  $\sigma_{217}^-$, $\lambda^+$ and $\lambda^-$. 
	\item Then we calculate the S/N threshold for each of the four selected simulations. We cannot take directly the thresholds in table \ref{table:snthreshold} because they correspond to an X-ray map with $t_{\rm exp} = 400$ s and $N_{\rm H}=2 \cdot 10^{20}$ cm$^{-2}$ and the real map where we have detected the candidate will have, in general, different characteristics. To correct for this effect, we need to convert the S/N maps obtained in the simulations into equivalent S/N maps corresponding to the exposure time and $N_{\rm H}$ of the real map. This is done using eq. \ref{eq:SNRreal_SNRsimu}, as explained in Appendix \ref{app:conversion}. Then, for each of the 4 selected simulations, we will calculate the S/N threshold  as the value $q_{\rm Ji}$ for which the fraction of pixels on the transformed S/N maps with S/N $>q_{\rm Ji}$ is at most $P_{\rm{FA}}$. 
	\item Finally, the threshold $q_{\rm J}$ to be applied to our detection is obtained via a 2-dimensional interpolation between the four values $q_{\rm Ji}$. If the S/N of the detection is above this threshold (S/N $> q_{\rm J}$), we will keep it in the list since it is not likely to be a noise detection (with confidence 1-$P_{\rm{FA}}$); otherwise we will discard it.
\end{enumerate}
	
\subsection{Catalogue preparation}\label{ssec:cuts}

The blind detection outputs an all-sky catalogue of joint X-ray-SZ detections that may still contain non-cluster objects or detections caused by noise or contamination. As demonstrated in \cite{Tarrio2016}, adding the X-ray information to the SZ maps increases the cluster detection probability, allowing us to detect fainter or more distant clusters with respect to the catalogues constructed from purely SZ information.  
However, introducing the X-ray maps also increases the number of false detections, produced by non-cluster X-ray sources (mainly AGNs) and Poisson noise.
Furthermore, the SZ observations contain regions contaminated with infrared emission that may also produce false detections. Thus, a main challenge of the proposed X-ray-SZ blind detection is to obtain a high purity. To achieve it, the catalogue produced by the blind detection method needs to be cleaned to discard detections in contaminated regions of the sky, in regions with poor statistics, or that correspond to non-cluster objects. In the rest of this section, we introduce two masking procedures to avoid detections in regions with infrared contamination (Sect. \ref{ssec:cleaning}) or X-ray poor statistics (Sect. \ref{ssec:xraymask}), and a method to discard real detections corresponding to non-cluster X-ray objects (Sect. \ref{ssec:classif_pointsources}). 

\subsubsection{SZ mask}\label{ssec:cleaning}  

To avoid SZ contaminated regions, we follow the same procedure used to build the second \textit{Planck} cluster catalogue (PSZ2) described in \citet{Planck2015ResXXVII}, i.e. we discard all the detections inside the PSZ2 survey mask, we reject also detections within 5$\sigma_{\rm beam}$ of any SZ compact source of the second \textit{Planck} catalogue of compact sources (PCCS2) \citep{Planck2015ResXXVI} with S/N>10 in any of the six HFI \textit{Planck} channels, and we remove 7 arcmin matches with the \textit{Planck} cold-clump catalogue (C3PO), or with PCCS2 detections at both 545GHz and 857GHz to eliminate infra-red spurious detections.
	
\subsubsection{X-ray mask}\label{ssec:xraymask} 
	
In the regions of the sky where the X-ray exposure time is very low, the X-ray count-rate map contains few noise pixels with very high count-rate value (typically, one count divided by a very low exposure time) compared to adjacent pixels (with zero counts).  
These bright pixels may introduce false detections, which could be discarded using the adaptive threshold calculated by numerical simulations. However, since the amount of simulation time required to properly simulate these regions is significant, and given that the X-ray information provided by these regions is very limited, we decided to just set a threshold in the exposure time to mask the low-exposure regions.
In the case of RASS, we decided to use a threshold of 100 seconds, which masks only 5.5\% of the sky, i.e. 2300 $\deg^2$, 910 $\deg^2$ having a non-zero exposure time and 1390 $\deg^2$ having no RASS observations. The overlap of this masked area with the SPT footprint, where the proposed method will be evaluated (see Sect. \ref{sec:evaluation}), is 209 $\deg^2$, 80 $\deg^2$ with non-zero exposure time and 129 $\deg^2$ with no RASS observations.

\subsubsection{Classification to distinguish clusters from point sources}\label{ssec:classif_pointsources}

  \begin{figure*}[]
  	\centering
  	\includegraphics[width=1.99\columnwidth]{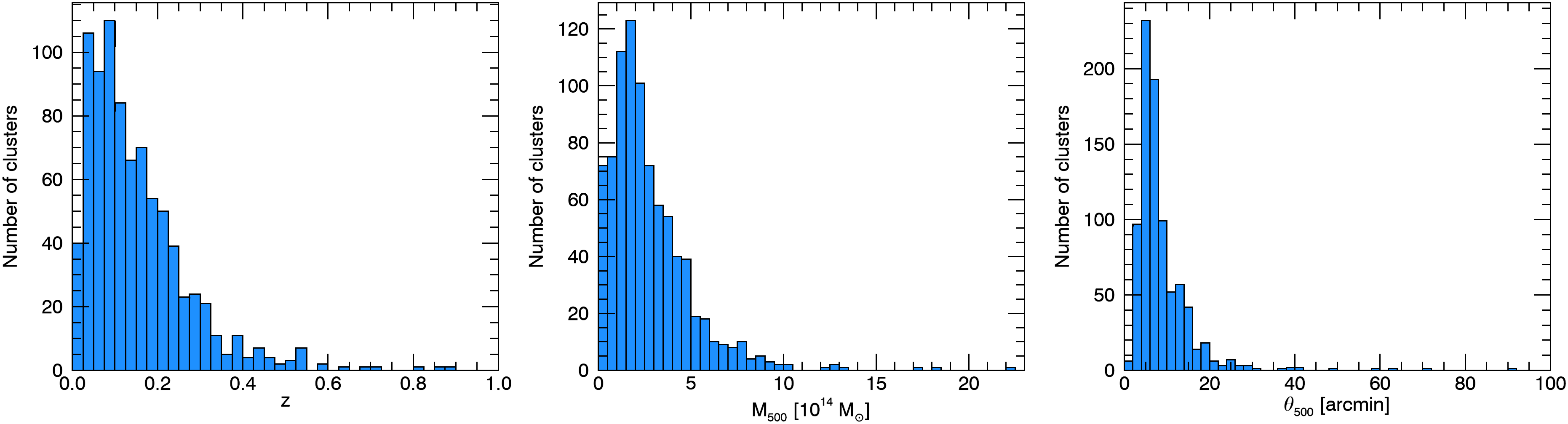}
  	 \caption{Distributions of the redshift (left panel), mass (middle panel) and size (right panel) of the clusters used for the training of the SVM classifier.}
  	 \label{fig:SVM_hist}
  \end{figure*}

Some of the objects detected with the blind joint detection method correspond to point sources in the X-ray maps that coincide with an SZ noise peak. Although the estimated size can be used as a criteria to distinguish between a real cluster and a point source, it is difficult to distinguish between a cluster with a small apparent size and a point source. 

Our ideal aim would be to recognize if a detection is a real cluster or a point source given the parameters extracted during the filtering process. To check if this was possible, we cross-matched some joint detections with a list of known clusters and known point sources (see details below), and we labeled each of our matching candidates as belonging to one or the other class. 
Then, we characterized each sample of this labeled list using five features: the estimated S/N, size and flux of the blind joint detection, and the X-ray and SZ components of the S/N: (S/N)$_{\rm XR}$ and (S/N)$_{\rm SZ}$. 
Finally, in order to get an upper bound on the best classification accuracy that can be obtained with a linear classifier we trained a support vector machine (SVM) classifier with this labeled list.

\begin{figure}[]
 	\centering
 	\includegraphics[width=0.99\columnwidth]{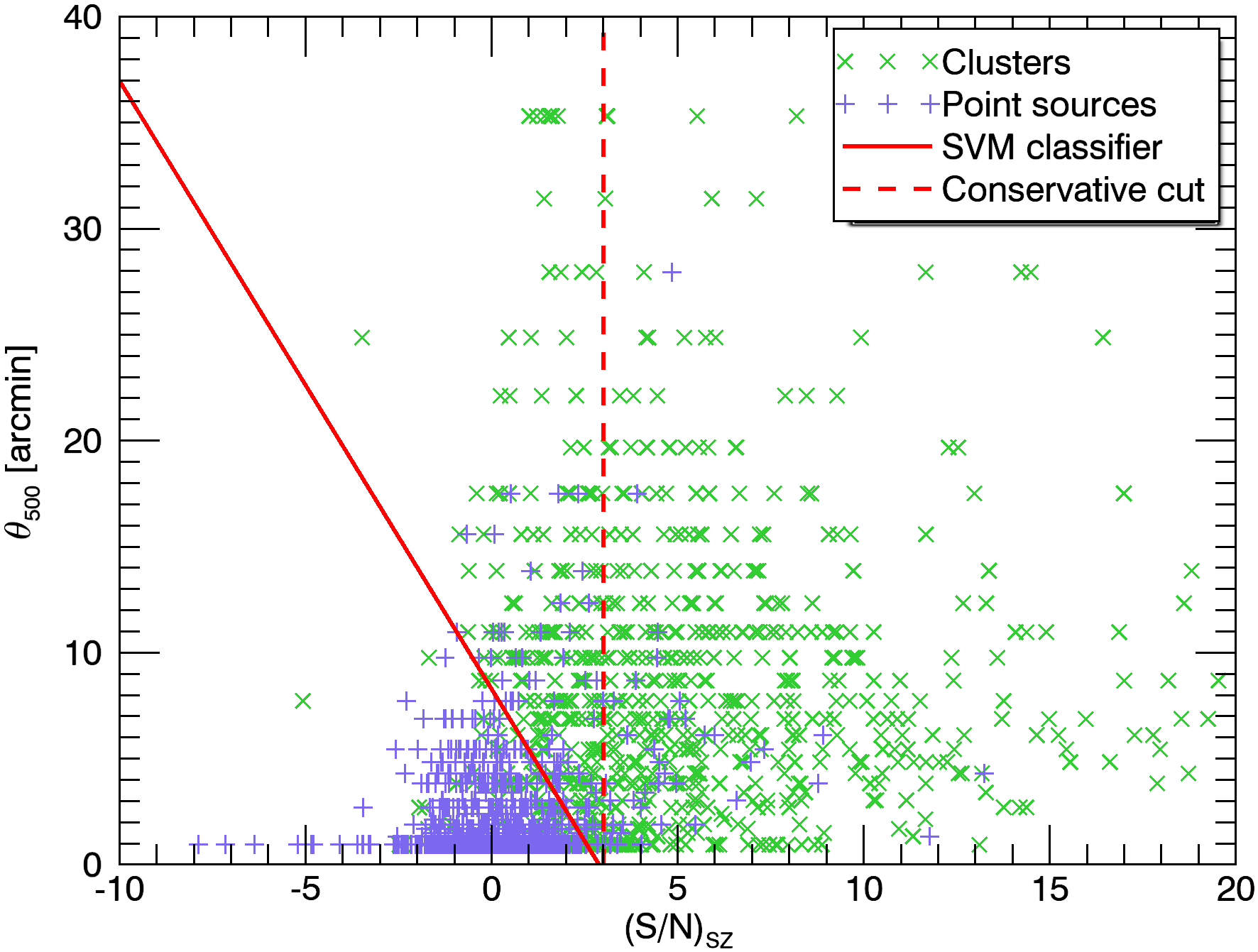}
 	\caption{Estimated (S/N)$_{\rm SZ}$ and size $\theta_{500}$ of the joint detections in the Northern hemisphere that match a known cluster or a known X-ray point source. The continuous red line shows the classification boundary provided by a SVM classifier trained with this dataset. The dashed red line shows the conservative cut that we adopted for discarding point sources.}
 	\label{fig:SVMclassifier}
 \end{figure}

Using 10-fold cross-validation\footnote{The dataset is randomly divided into 10 subsets. Then, 9 subsets are used for training the SVM and 1 is used for testing the classifier. This validation process is repeated 10 times, with each subset used only once as test set, and the 10 results are averaged.}, we obtained a classification accuracy of 88\%, with 5\% of the mis-classifications being point sources classified as clusters and 7\% being clusters classified as point sources. 
We noticed that the parameter that plays the most important role in the classification is (S/N)$_{\rm SZ}$, followed by the estimated size $\theta_{500}$. This is logical, because we do not expect to find an SZ signal at the position of an AGN, and we expect them to have a small size (they are point-like sources). A classification considering only these two parameters gives the same performance as the one obtained with the five parameters. Figure \ref{fig:SVMclassifier} shows the detections used for this experiment in the (S/N)$_{\rm SZ}$ - $\theta_{500}$ plane, color-coded according to the type of object to which they are associated.  
A red line indicates the best linear classification boundary determined by the trained SVM.
A simple classification boundary of (S/N)$_{\rm SZ}$ = 2 provides almost the same performance as the complete SVM classifier: 85\% correct classifications, with 4\% of the point sources classified as clusters and 11\% of the clusters classified as point sources. So, for simplicity reasons, we decided to use only (S/N)$_{\rm SZ}$ for the cluster/point source classification. Finally, since for our purity purposes we prefer to have less false clusters at the expense of a lower classification accuracy (and thus, lower completeness), we decided to modify the classification threshold to (S/N)$_{\rm SZ}$ = 3, which provides 82\% correct classifications, with only 2\% of the point sources classified as clusters (and 16\% of the clusters classified as point sources).
 
As mentioned before, the classification is based on the labels obtained by cross-matching some joint detections with a list of known clusters and known point sources. In particular, for constructing the list of known clusters, we used the MCXC \citep{Piffaretti2011}, ESZ \citep{PlanckEarlyVIII}, PSZ1 \citep{Planck2013ResXXIX}, PSZ2 \citep{Planck2015ResXXVII}, SPT \citep{Bleem2015} and ACT \citep{Hasselfield2013} cluster catalogues, and considered only confirmed clusters. Figure \ref{fig:SVM_hist} shows the distribution of redshift, mass and size of these clusters. For constructing the list of known point sources, we took the ROSAT bright source catalogue \citep{Voges1999} and we applied the selection criteria of MACS \citep{Ebeling2001}. All the objects in the resulting list have been followed up for confirmation by MACS, so by eliminating the objects that match with a known cluster, we are left with a list of X-ray point sources. Since all these point sources belong to the Northern hemisphere, the list of joint detections that we used for this test was obtained by running the joint detection algorithm on the Northern hemisphere. Then, we then cross-matched our list of detections with the two lists to label our detections as clusters or point sources. This cross-match was done based on distance, with a matching radius of 2 arcmin. It is worth mentioning that this labeling may not be completely accurate, first because the catalogues of known point sources and clusters may not be completely correct, and second because the cross-matching done according to distance may introduce some incorrect matches. Therefore, the classification results reported before just provide a good idea of the real classification performance (with respect to the unknown ground truth). Finally, we want to emphasize that these classification results are based on the selected training dataset, so they cannot be generalized to the problem of distinguishing point sources from any kind of cluster.

\subsection{Output parameters and mass estimation}\label{ssec:mass_estimation} 

For each detection, the joint algorithm provides its position, the size $\theta_{500}$ of the filter that gives the best joint S/N, the corresponding flux $Y_{500}$ and joint S/N, and the SZ and X-ray components of this S/N: (S/N)$_{\rm SZ}$ and (S/N)$_{\rm XR}$. 

Additionally, the joint method also provides a value for the significance of each detection. This value is calculated from the simulation results described in Sect. \ref{ssec:jointthreshold} in the following way. 
\begin{enumerate}
	\item For each detection, we select the four simulations corresponding to the filter size  that are closer to the mean Poisson level $\lambda$ and mean Gaussian level $\sigma_{217}$ of the analyzed map,  and we convert the S/N maps obtained in the simulations into equivalent S/N maps corresponding to the exposure time and $N_{\rm H}$ of the real map, as done in steps 1 to 3 of Sect. \ref{ssec:jointthreshold}.  
	\item Then, we calculate the fraction of pixels in the simulations with a S/N on these transformed S/N maps greater that the joint S/N of the detection. This measures the probability of a false detection. 
	\item Finally, we perform a 2-dimensional interpolation using these four values to get the probability that the detection is due to noise. From this probability, we calculate the value of significance corresponding to a Gaussian distribution.  
	\item We remark that if the joint S/N is very  high, there are no pixels in the simulations with a higher S/N. In these cases, it is not possible to calculate the significance directly, and we will use the following expression to estimate it: $\rm{significance} = 4.5 + 0.68\cdot((\rm{S/N})_{\rm J}-q_{\rm J}$). Appendix \ref{app:significance} explains how this expression was obtained. 
\end{enumerate}

Finally, since the size estimation is not very accurate, as occurred for PSZ2 catalogue, the blind detection provides the degeneracy curves $Y_{500}(\theta_{500})$ and (S/N)$_{\rm J}(\theta_{500})$ for the assumed reference redshift $z_{\rm ref}$, which allow to determine more precisely the size and flux of the cluster given some a priori information (e.g. the redshift) about the cluster. 
	
Apart from the degeneracy curve $Y_{500}(\theta_{500})$ corresponding to the reference redshift $z_{\rm ref}$, we can as well re-extract the degeneracy curves for different redshifts at the position given by the blind detection. Then, if the detection matches a cluster with known redshift, we can interpolate between these degeneracy curves to obtain the curve corresponding to the real redshift of the cluster. This size–flux degeneracy can be further broken using the $M_{500} - D^2_A Y_{500}$ relation, that relates $\theta_{500}$ and $Y_{500}$ when z is known, as explained in Sect. 7.2.2 of \cite{Planck2013ResXXIX}. In this way, we can obtain an estimate of the mass $M_{500}$ of the candidate. In Sect. \ref{ssec:mass_estimation} we will compare the mass estimated following this approach with the published mass for some of the joint detections that match known clusters.

\section{Evaluation in SPT area}\label{sec:evaluation}

In this section, we present an evaluation of the proposed blind detection method in the region of the sky covered by the SPT survey.  
This region was selected because it is a wide-area region of the sky (2500 deg$^2$) where we can assume that (almost) all the massive clusters ($M_{500} > 7 \cdot 10^{14} M_{\sun}$) up to redshift 1.5 are already known. On one hand, the SPT survey, which is deeper than the \textit{Planck} survey, is almost 100\% complete at z>0.25 for clusters with mass $M_{500} > 7 \cdot 10^{14} M_{\sun}$ \citep{Bleem2015} ($\sim$ 90\% complete for $M_{500} > 6 \cdot 10^{14} M_{\sun}$). On the other hand, the X-ray MCXC catalog \citep{Piffaretti2011} should include almost all the clusters with mass $M_{500} > 5 \cdot 10^{14} M_{\sun}$ at z<0.25, since it contains the REFLEX sample, which is highly complete (> 90\%) for that redshift-mass range (see Fig. \ref{fig:clusters_in_MZ_plane}). There is however a small redshift range around 0.2-0.3 where some massive clusters could be still unknown. Less massive clusters could also be unknown in a broader redshift range.
A comparison of the blind detection results with these catalogues allows us to determine if the detected candidates are real clusters (purity) and which fraction of real clusters we do detect (detection efficiency). Nevertheless, we should keep in mind that there could be some clusters in the transition region that are neither in SPT nor in MCXC. Also, other clusters could be missing due to masked regions in the surveys. 

We run the blind joint detection algorithm on the SPT footprint\footnote{Defined as  (R.A. < 104.8$\degr$ or R.A. > 301.3$\degr$) and -65.4$\degr$ < decl. < -39.8$\degr$.} and obtained 2767 detections in the second pass (using $q=4$). Then, we applied the cuts in the \textit{Planck} S/N (S/N$_{\rm SZ} > 3$), the RASS exposure time ($t_{\rm exp}>100$ s) and the joint S/N (S/N$_{\rm J}>q_{\rm J}$), and we applied the SZ cleaning procedure described in Sect. \ref{ssec:cleaning}. If we choose a false alarm rate of $P_{\rm FA}=3.40 \cdot 10^{-6}$ to calculate the joint S/N threshold $q_{\rm J}$ to apply to the 2767 detections, we are left with 225 candidates. Table \ref{table:allcandidates1} summarizes the properties of these candidates. If we decrease this false alarm probability we get fewer candidates, for example, for $P_{\rm FA}=2.04 \cdot 10^{-7}$ (equivalent to a 5$\sigma$ cut in a Gaussian distribution) we get 185 candidates and for $P_{\rm FA}=1.90 \cdot 10^{-8}$ (equivalent to a 5.5$\sigma$ cut in a Gaussian distribution) we get 165 candidates.

\begin{figure}[]
	\centering
	\includegraphics[width=0.99\columnwidth]{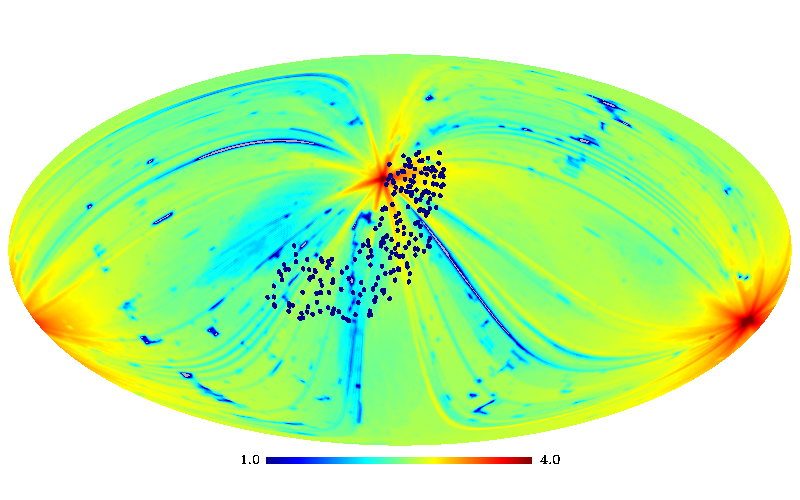}
	\caption{Positions of the joint detections with respect to the RASS exposure map. Blue dots represent the 225 candidates corresponding to a false alarm rate of $P_{\rm FA}=3.40 \cdot 10^{-6}$. The sky map is color-coded according to the logarithm of the RASS exposure time.}
	\label{fig:detections_in_rass_texp_map}
\end{figure}

\begin{figure}[]
	\centering
	\includegraphics[width=0.99\columnwidth]{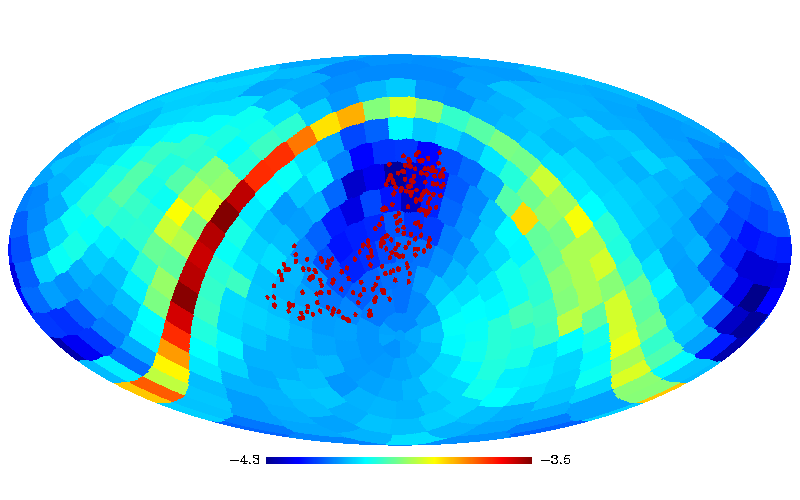}
	\caption{Positions of the joint detections with respect to the \textit{Planck} noise map. Red dots represent the 225 candidates corresponding to a false alarm rate of $P_{\rm FA}=3.40 \cdot 10^{-6}$. The sky map is color-coded according to the logarithm of the \textit{Planck} noise standard deviation map.}
	\label{fig:detections_in_planck_noise_map}
\end{figure}

The detection area is not covered homogeneously: there are slightly more candidates in the regions where the RASS exposure time is higher and where the \textit{Planck} noise is lower. This is expected, since in those regions both surveys are deeper. This effect is shown in Figs. \ref{fig:detections_in_rass_texp_map} and \ref{fig:detections_in_planck_noise_map}. 


\subsection{Crossmatch with other cluster catalogues}\label{ssec:crossmatch}

\begin{table*}
	\caption{Number of previously known clusters or cluster candidates in the considered region that match with our detections for $P_{\rm FA}=3.40 \cdot 10^{-6}$, $P_{\rm FA}=2.04 \cdot 10^{-7}$ and $P_{\rm FA}=1.90 \cdot 10^{-8}$ within a distance of 10 arcmin. Planck refers to the combination of the three \textit{Planck} catalogues \citep{PlanckEarlyVIII,Planck2013ResXXIX,Planck2015ResXXVII}, whereas PSZ2 refers only to the last one. MMF3 is the subsample of objects in the PSZ2 catalogue that were detected using the MMF3 detection algorithm. RASS refers to the subsample of objects in the MCXC catalogue that were detected from RASS observations. SZ refers to the combination of all the SZ catalogues (Planck, SPT and ACT).}
	\label{table:detectedclusters}
	\centering 
	\begin{tabular}{c c | c c | c c | c c}
		\hline
		\noalign{\smallskip}
		&                     & \multicolumn{2}{c|}{$P_{\rm FA}=3.40 \cdot 10^{-6}$}   & \multicolumn{2}{c|}{$P_{\rm FA}=2.04 \cdot 10^{-7}$} & \multicolumn{2}{c}{$P_{\rm FA}=1.90 \cdot 10^{-8}$}\\
		\noalign{\smallskip}
		Cluster      &   Clusters in the   & Clusters  & Percentage & Clusters  & Percentage & Clusters  & Percentage\\
		catalogue    &   considered region & detected  &  (\%)      & detected  &  (\%)      & detected  &  (\%)\\
		\noalign{\smallskip}
		\hline
		\noalign{\smallskip}
		all MMF3          & 126   &  118  & 93.7   &   111 &  88.1  & 106   & 84.1\\   
		confirmed MMF3    & 113   &  110  & 97.3   &   106 &  93.8  & 102   & 90.3\\
		MMF3 candidates   &  13   &    8  & 61.5   &     5 &  38.5  &   4   & 30.8\\      
		\noalign{\smallskip}
		all PSZ2          & 154   &  133  & 86.4   &   124 &  80.5  & 117   & 76.0\\         
		confirmed PSZ2    & 131   &  125  & 95.4   &   120 &  91.6  & 114   & 87.0\\
		PSZ2 candidates   &  23   &    8  & 34.8   &     4 &  17.4  &  3    & 13.0\\    
		\noalign{\smallskip}
		all Planck        & 174   &  134  & 77.0   &   125 &  71.8  & 118   & 67.8\\   
		confirmed Planck  & 137   &  126  & 92.0   &   121 &  88.3  & 115   & 83.9\\
		Planck candidates &  37   &    8  & 21.6   &     4 &  10.8  &   3  &  8.1\\         
		\noalign{\smallskip} 
		confirmed SPT     & 494   &  111  & 22.5   &    98 &  19.8  &  89   & 18.0\\
		\noalign{\smallskip}
		ACT               &  22   &   16  & 72.7   &    14 &  63.6  &  14   & 63.6\\
		\noalign{\smallskip}
		MCXC              & 138   &   72  & 52.2   &    72 &  52.2  &  72   & 52.2\\
		RASS              & 103   &   71  & 68.9   &    71 &  68.9  &  71   & 68.9\\
		\noalign{\smallskip}
		Abell             & 896   &  104  & 11.6   &    96 &  10.7  &  91   & 10.2\\
		\hline
		\noalign{\smallskip}
		MCXC not SZ                 &    70 &     9 &  12.9  &     9 &  12.9 &   9   & 12.9\\
		\noalign{\smallskip}        
		confirmed SPT not Planck    &   405 &    32 &   7.9  &    23 &   5.7 &  18   &  4.4\\
		\noalign{\smallskip}
		all SZ                      &   734 &   163 &  22.2  &   146 &  19.9 & 134   & 18.3\\        
		confirmed SZ                &   544 &   155 &  28.5  &   142 &  26.1 & 131   & 24.1\\
		SZ candidates               &   190 &     7 &   3.7  &     3 &   1.6 & 2     & 1.1\\    
		\noalign{\smallskip}
		all SZ + MCXC               &   804 &   172 &  21.4  &   155 &  19.3  & 143  & 17.8\\          
		confirmed SZ + MCXC         &   614 &   164 &  26.7  &   151 &  24.6  & 140  & 22.8\\
		SZ + MCXC candidates        &   190 &    10 &   3.7  &     3 &   1.6  & 2    & 1.1\\  
		\noalign{\smallskip}
		\hline
		\noalign{\smallskip}
		\multicolumn{2}{c}{Total number of detections}            & \multicolumn{2}{|c}{225} & \multicolumn{2}{|c}{185} &      \multicolumn{2}{|c}{165}  \\
		\multicolumn{2}{c}{Detections matching any confirmed cluster}     & \multicolumn{2}{|c}{187} & \multicolumn{2}{|c}{166} &     \multicolumn{2}{|c}{151}  \\ 
		\multicolumn{2}{c}{Purity (w.r.t. confirmed clusters)}    & \multicolumn{2}{|c}{>83.1 \%} & \multicolumn{2}{|c}{>89.7 \%} &  \multicolumn{2}{|c}{>91.5 \%} \\ 
		
		\hline
	\end{tabular}
\end{table*}

To estimate the purity and the detection efficiency of these catalogues, we cross-matched all the candidates with various published catalogues of clusters. In particular, we took several SZ-selected catalogues covering the considered region, namely the three \textit{Planck} catalogues: ESZ \citep{PlanckEarlyVIII}, PSZ1 \citep{Planck2013ResXXIX} and PSZ2 \citep{Planck2015ResXXVII}, the SPT catalogue \citep{Bleem2015} and the ACT catalogue \citep{Hasselfield2013}. It is worth mentioning that a subsample of the PSZ2 catalogue, namely the MMF3 sub-catalogue, is especially interesting for us, since the proposed joint detection method is based on the MMF3 detection method. The SPT and ACT surveys are deeper than the \textit{Planck} survey, so these catalogues contain additional clusters that were not detected by \textit{Planck}. We also took as reference the X-ray selected MCXC catalogue \citep{Piffaretti2011}. This is a metacatalog of X-ray detected clusters that was constructed from publicly available cluster catalogues of two kinds: RASS-based catalogues, obtained from the RASS survey data, and serendipitous catalogues, based on deeper pointed X-ray observations. Finally, we also considered the optically-selected Abell catalogue \citep{Abell1989}. We did not use Zwicky and redMaPPer catalogues since they do not contain clusters in the considered region.

\begin{figure}[]
	\centering
	\includegraphics[width=0.99\columnwidth]{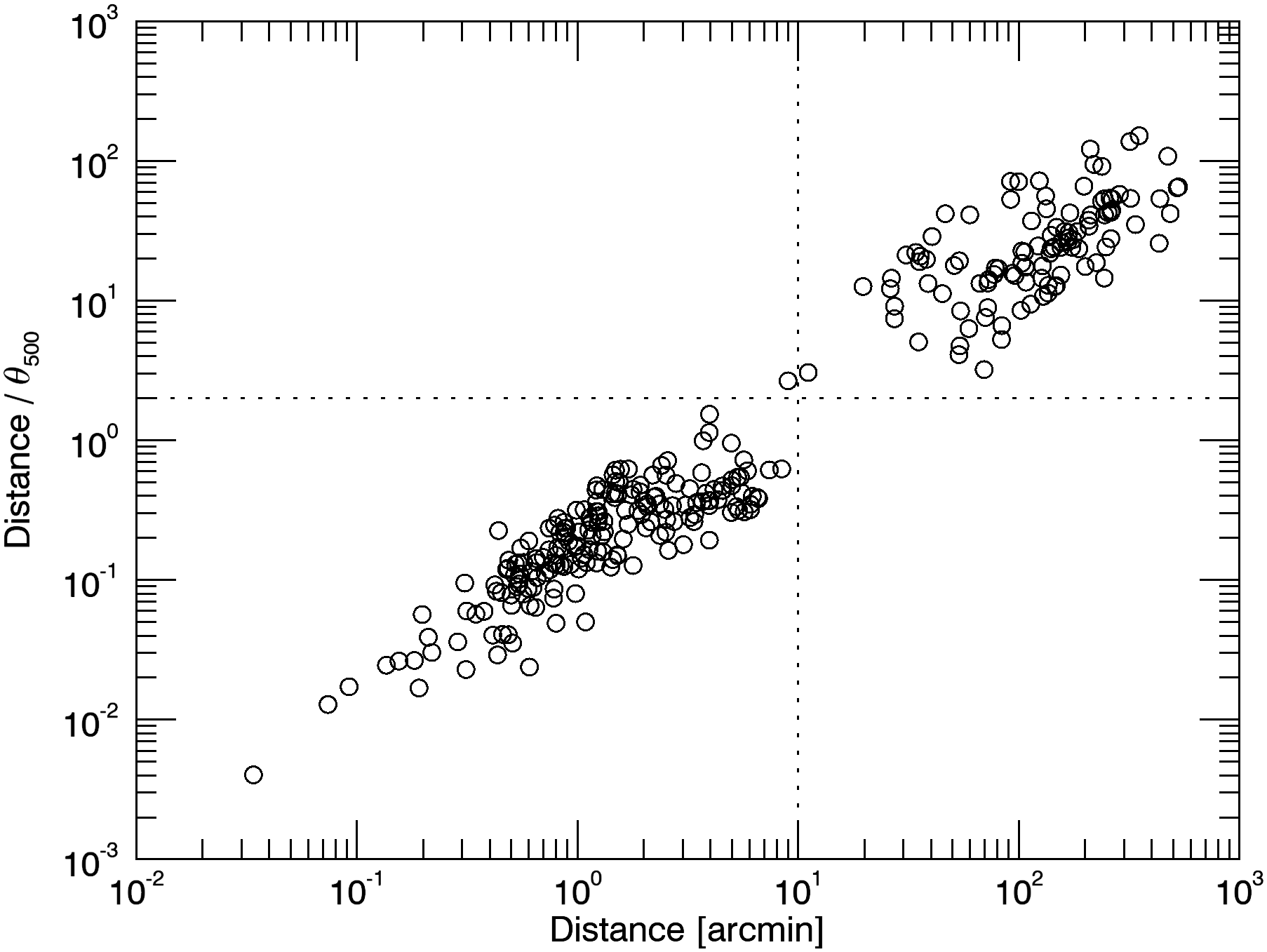}
	\caption{Distance from the joint position to the position of the closest cluster versus the distance normalized to the cluster size. Only the objects with known redshift and mass in the considered SZ and X-ray catalogues were taken as clusters.}
	\label{fig:crossmatch_distance}
\end{figure}

To decide whether our detections match or not these previously-known candidates, we first determined the closest cluster to each of our detections. To this end, we selected only the objects in the considered SZ and X-ray catalogues with known redshift and mass (i.e. confirmed clusters). Figure \ref{fig:crossmatch_distance} shows a scatter plot of the absolute distance versus the relative distance (in terms of $\theta_{500}$) between the associated objects. We observe two types of associations: those with a small distance both in absolute and relative terms, and those with a long distance both in absolute and relative terms. The first group of points represents true detections of clusters, whereas the second group corresponds to the detections that are randomly distributed with respect to the considered known clusters. From this observation, we decided to use the following association rule: if the distance is less than 10 arcmin the detection is considered as associated with a known cluster, otherwise the detection is considered as not associated with a known cluster. We show in Sect. \ref{ssec:mass_comparison} that the resulting associations are valid, since the masses of the detected objects and the associated clusters agree. Furthermore, given that the considered catalogues contain also objects without redshift and mass informations, we decided not to introduce an additional criteria based on the relative distance, which can be only calculated if the $\theta_{500}$ of the object is known. This association rule is very simple, but has the advantage that it can be applied to all the candidates in the considered catalogues. 

After the cross-match of our candidates with these published catalogues, we found that 187 of the 225 detections corresponding to $P_{\rm FA}=3.40 \cdot 10^{-6}$ match with a previously-known confirmed cluster within a distance of 10 arcmin. This corresponds to a purity larger than 83.1\%. For the case of $P_{\rm FA}=2.04 \cdot 10^{-7}$ we found that 166 of the 185 detections match with a previously-known confirmed cluster, whereas for $P_{\rm FA}=1.90 \cdot 10^{-8}$ there are 151 matches out of 165 detections. This corresponds to a purity larger than 89.7\% and 91.5\% respectively, which is higher than before, as expected, since we have decreased the false detection probability. 
Table \ref{table:detectedclusters} shows more details about the number of candidates matching the different cluster catalogues that we considered.

In this context, and for the rest of this section, we have defined "purity" as the percentage of joint detections that are associated with a confirmed cluster of these published catalogues. It is important to keep in mind that these values of purity are just rough estimations, since our simple association rule could introduce a few wrong associations. Furthermore, the candidates without a match may also be real clusters that were not detected or included in the published catalogues (for example, objects in masked regions, objects in a mass-redshift region where the considered surveys are not complete, etc.). Therefore, the values of purity with respect to previously-known confirmed clusters can be considered as an approximate lower limit. 

\begin{figure*}[]
	\centering
	\subfigure[]{\includegraphics[width=0.99\columnwidth]{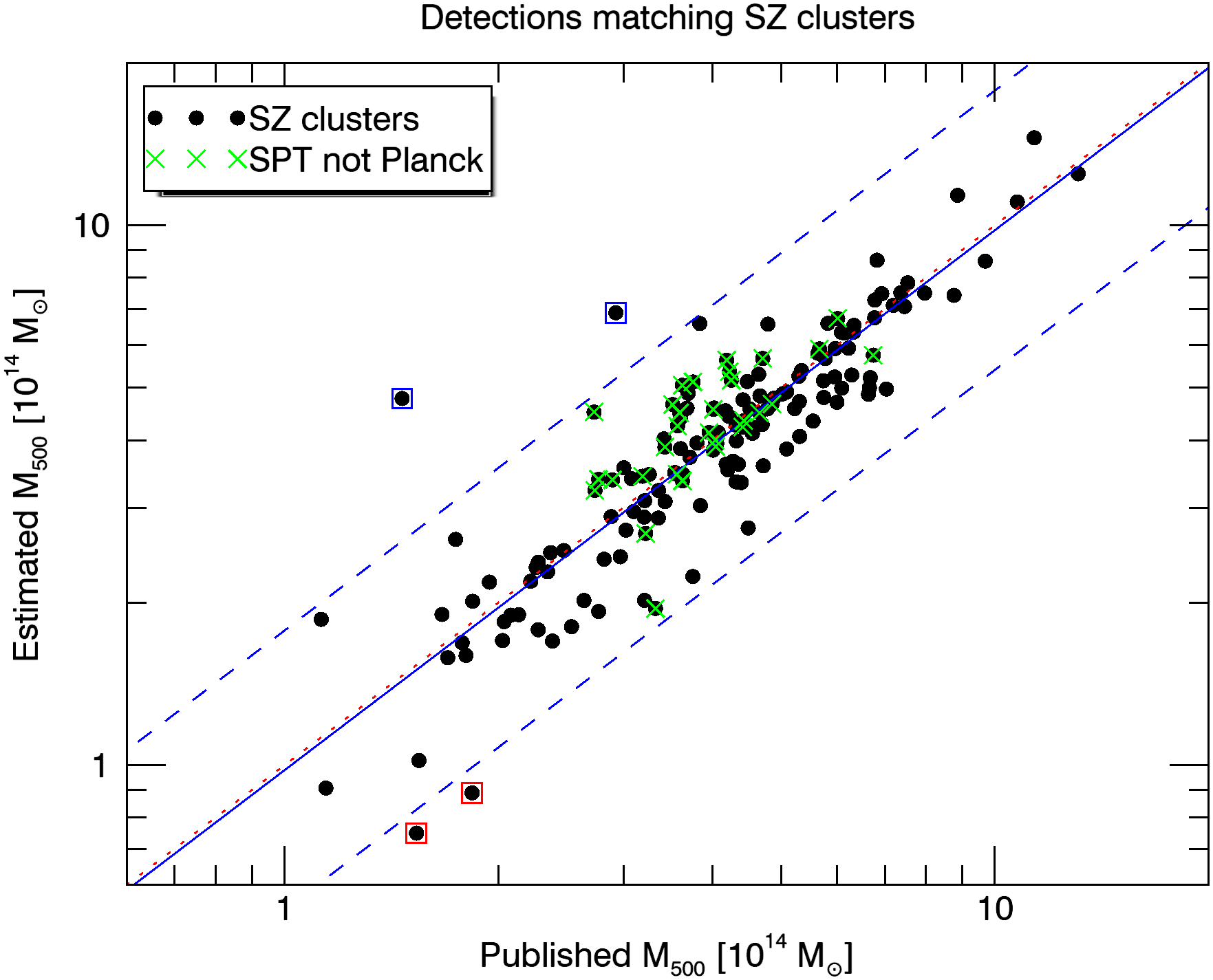}\label{fig:mass_xsz-mass_sz}}
	\subfigure[]{\includegraphics[width=0.99\columnwidth]{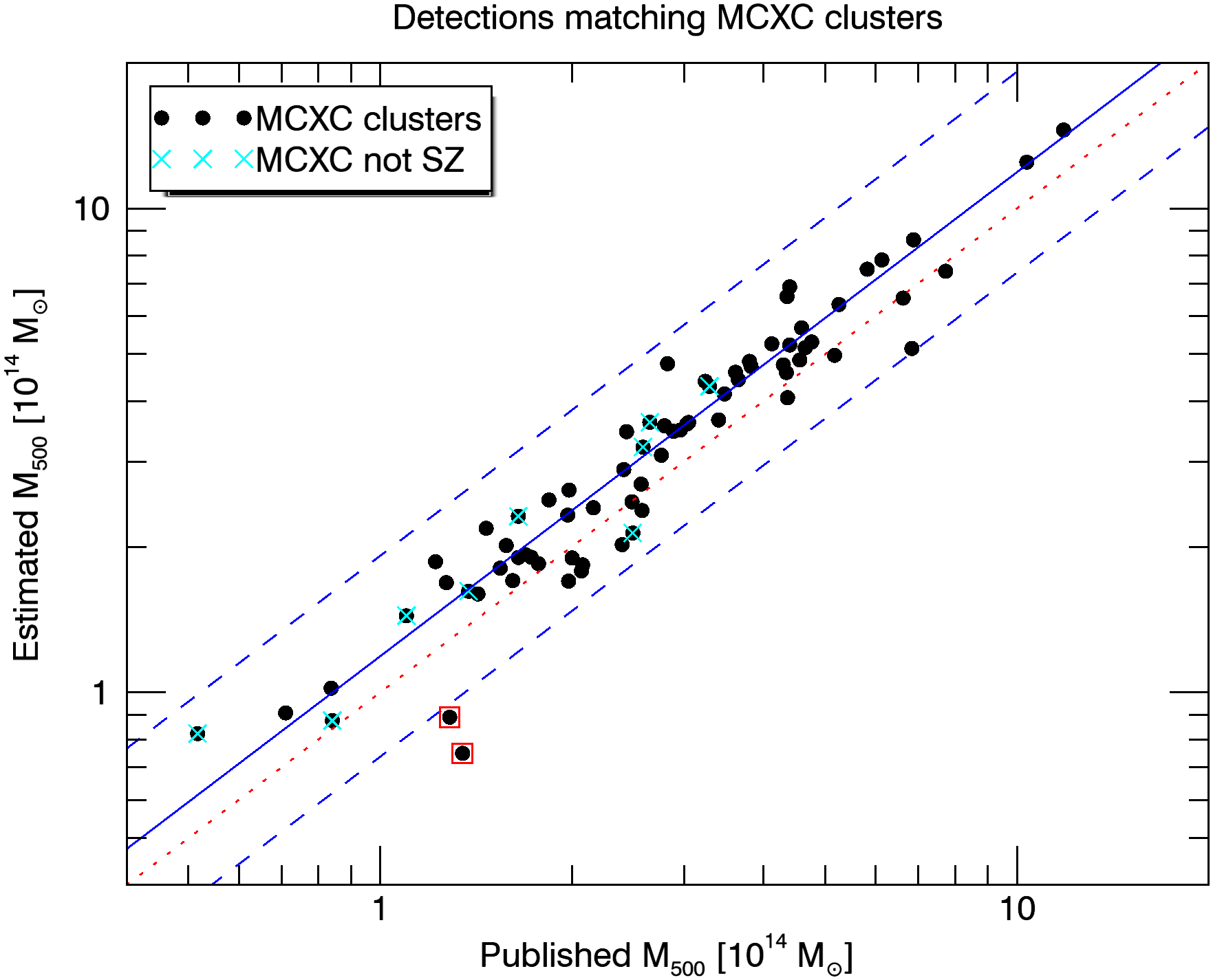}\label{fig:mass_xsz-mass_x}}
	\caption{$M_{500}$ estimated from the joint detection for the candidates matching a confirmed cluster versus the published $M_{500}$ of the corresponding cluster. (a) compares the estimated mass to the published SZ mass for the 155 candidates matching an SZ cluster with known mass. (b) compares the estimated mass to the published MCXC mass for the 72 candidates matching an MCXC cluster. The dotted red line indicates the line of zero intercept and unity slope. The solid blue line indicates the median ratio. The dashed blue lines indicate the interval of $\pm 2.5\sigma$ around the median ratio. Outliers are highlighted with a blue (high estimated mass) or red (low estimated mass) square and they are discussed in the text. Green crosses indicate SPT clusters that are not in \textit{Planck} catalogues and cyan crosses indicate MCXC clusters that are not in \textit{Planck} or SPT.  
	}
	\label{fig:mass_xsz-mass}
\end{figure*}

On the other hand, we define the "detection efficiency" as the percentage of candidates of the considered published catalogues that are detected by our joint algorithm. This magnitude is related to the completeness. To calculate it, we have cross-matched all the previously-known clusters in the considered region (SPT region with RASS exposure time greater than 100 s, outside the PSZ2 masked region) with our detections. Table \ref{table:detectedclusters} shows the results from this cross-match for different values of $P_{\rm FA}$. A higher $P_{\rm FA}$ allows to recover more clusters, but at the expense of a lower purity, 
as seen at the bottom line of Table \ref{table:detectedclusters}. As expected, the proposed method is able to recover most of the MMF3 objects, given that it is based on the same data and a similar approach. A more detailed comparison with respect to MMF3 can be found in Sect. \ref{ssec:purity-completeness}. The recovery rate of PSZ2 and \textit{Planck} clusters is also high. On the contrary, the method recovers only a small fraction of SPT clusters not detected by \textit{Planck}, which was foreseen, since the SPT survey data is deeper. Finally, 68.9\% of the RASS clusters situated in the considered region are recovered. Given that the proposed method also uses RASS observations, this value may seem low, however it is not due to the detection algorithm, but to the additional cut we imposed to discard possible X-ray point sources. A more detailed comparison with respect to RASS clusters can be found in Sect. \ref{ssec:comparison_rass}.
 
\subsection{Mass comparison}\label{ssec:mass_comparison} 

Following the procedure described in Sect. \ref{ssec:mass_estimation} and using the $M_{500} - D^2_A Y_{500}$ relation proposed in \cite{Planck2013ResXX}, we estimated the mass $M_{500}$ for the detections matching confirmed SZ or MCXC clusters. There are a total of 163 detections matching a confirmed SZ cluster (155 with known mass)  
and 72 detections matching MCXC clusters. Figure \ref{fig:mass_xsz-mass} shows the relation between the estimated mass and the published mass for the corresponding clusters. 

The comparison with respect to the published SZ mass (Fig. \ref{fig:mass_xsz-mass_sz}) shows that the estimated mass follows well the published mass, with a median ratio of 0.98, very close to 1.  
We identified four outliers that are at more than 2.5$\sigma$ from the median ratio, two with overestimated masses and two with underestimated masses. Figure \ref{fig:mass_xsz-mass_sz} also shows that the SPT clusters that were not detected by \textit{Planck} have, in average, a higher mass ratio than the ones detected by \textit{Planck}. This behaviour can be explained due to the Malmquist bias.

The two outliers with overestimated mass correspond to clusters PSZ2 G252.99-56.09 (also RXC J0317.9-4414, ABELL 3112) and PSZ2 G348.46-64.83 (also SPT-CLJ2313-4243, RXC J2313.9-4244 and ACO S 1101). Both clusters are known strong cool-cores according to the classification of \cite{Hudson2010}, so the assumed $F_{\rm X}/Y_{500}$ relation used in the detection does not represent these clusters accurately. Since our mass estimation is obtained from the combination of X-ray and SZ information, its value, compared to the SZ mass, is boosted due to the high X-ray luminosity. We checked that the estimated mass using only the SZ information agrees with the published mass, which supports this explanation. 

\begin{figure*}[]
	\centering
	\includegraphics[width=1.5\columnwidth]{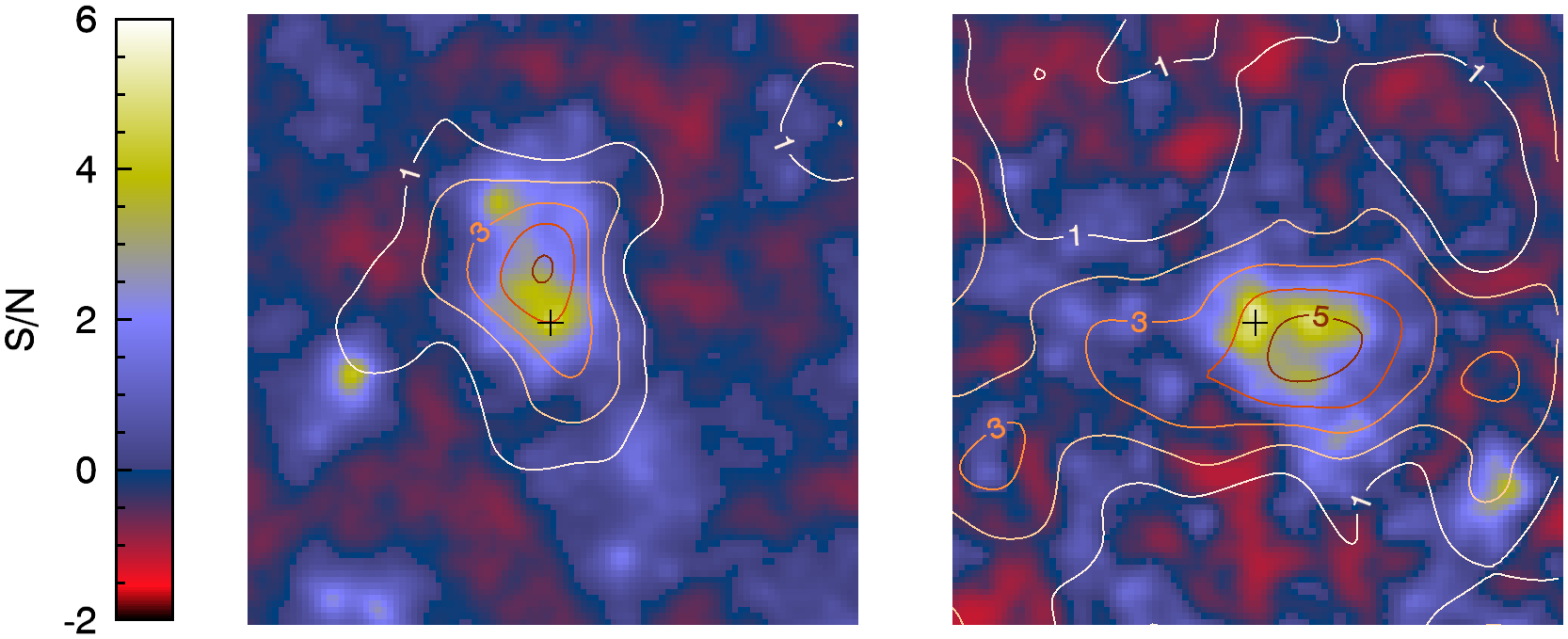}
	\caption{S/N maps for the two outliers with underestimated mass with respect to the published SZ mass and X-ray mass: PSZ2 G265.21-24.83 (left) and PSZ2 G269.36-47.20 (right). The colour images show the S/N corresponding to the X-ray map and the contours indicate the S/N corresponding to the SZ maps. The black crosses indicate the position of the joint detection. The angular size of the images is 86.7 arcmin. The distance between the joint detection and the published SZ position is 8.4 and 7.4 arcmin, respectively.
	}
	\label{fig:outliers_sn_contours}
\end{figure*}

The two outliers with underestimated masses correspond to clusters PSZ2 G265.21-24.83 (also RXC J0631.3-5610) and PSZ2 G269.36-47.20 (also RXC J0346.1-5702 and ABELL 3164). They can be justified due to the high distance between the joint detection and the published position, which is 8.4 and 7.4 arcmin respectively. This implies that the SZ signal at the detected position is not at its peak value, which explains the low value obtained for the mass. The reason for this distance is that the detection is centered in the X-ray peak while the X-ray emission is not coincident with the SZ emission, as it can be seen in Fig. \ref{fig:outliers_sn_contours}. We note that in both cases the distance normalized by the cluster size is less than one, so the association can be still considered to be correct.

The comparison with respect to the published MCXC mass (Fig. \ref{fig:mass_xsz-mass_x}) shows that the ratio between the estimated mass and the published mass is greater than one, with a median value of 1.19. The same value is found for the ratio between the published SZ mass and the published MCXC mass for the same clusters. This kind of behaviour was also observed by the \cite{Planck2015ResXXVII} when they compared the SZ mass and the X-ray luminosity of common PSZ2-MCXC objects. We identified two outliers that are at more than 2.5$\sigma$ from the median ratio. They coincide with the two outliers with underestimated mass with respect to published SZ mass (Fig. \ref{fig:mass_xsz-mass_sz}).

This mass comparison indicates that the joint extraction provides in general a good mass proxy when the redshift is known. The main sources of bias in the mass estimation are the presence of a cool core, that will tend to overestimate the mass, and an offset between the X-ray and SZ peaks, that will tend to give underestimated masses.

The good agreement between the estimated and the published masses also indicate that the 10-arcmin association rule is appropriate.

\subsection{Position accuracy}\label{ssec:position_comparison}
\begin{figure}[]
	\centering
	\includegraphics[width=0.99\columnwidth]{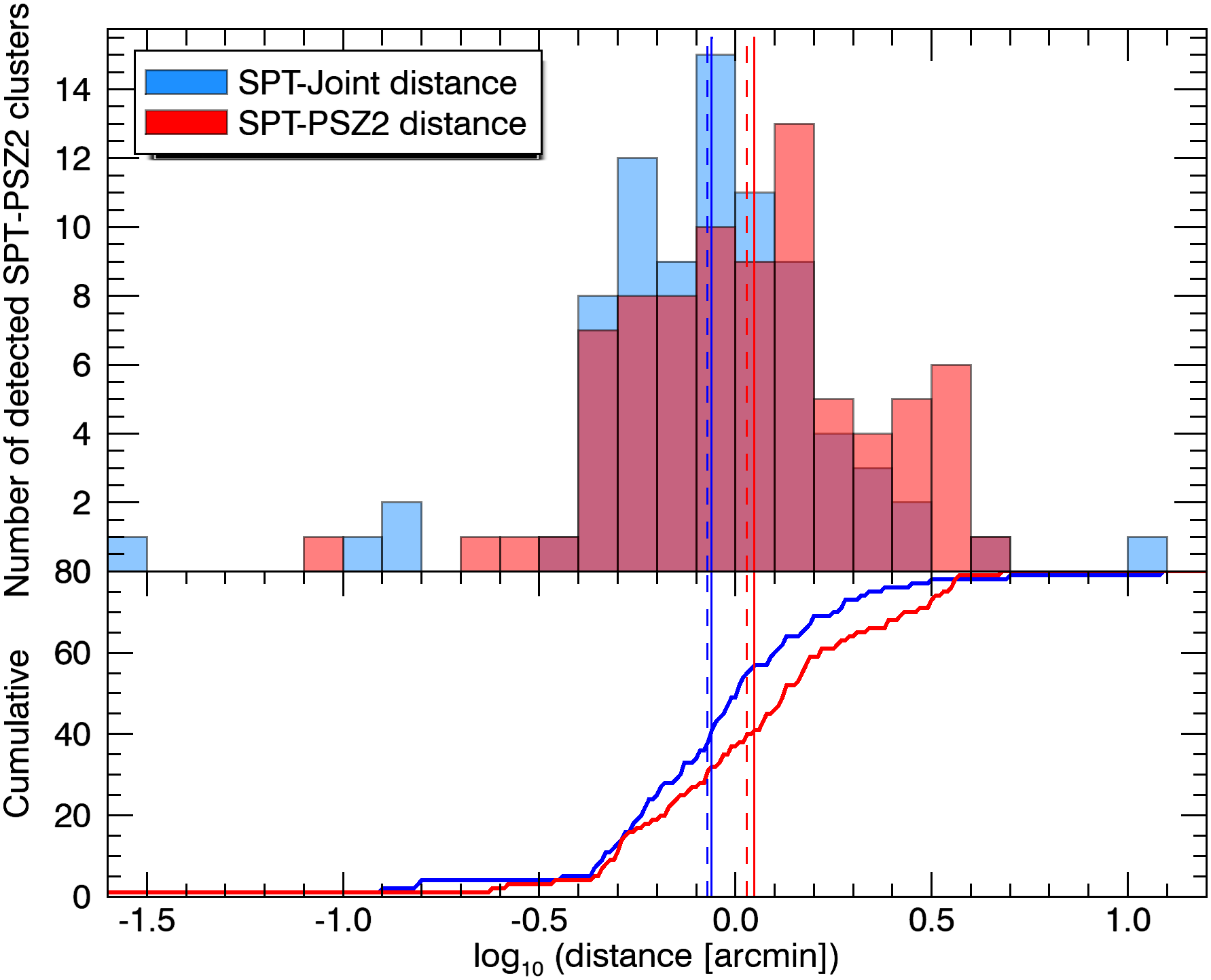}
	\caption{Histograms of the distance between the SPT and joint positions (blue) and of the distance between the SPT and PSZ2 positions (red). The median and mean values of each set of distances are represented by solid and dashed lines, respectively. The corresponding cumulative distributions are shown in the bottom panel.}
	\label{fig:distance_comparison_to_spt}
\end{figure}

Since the proposed method combines \textit{Planck} maps with RASS observations, which have better position accuracy due to the smaller beam, we expect the positions provided by the joint detection method to be more accurate than those provided by \textit{Planck}. To assess this accuracy, we took as a reference the positions given in the SPT catalogue, which are more accurate than \textit{Planck} positions. Then, we selected the joint detections that match clusters detected both by SPT and PSZ2 and calculated the distance between the joint position and the SPT position. Finally, we compared this distance to the distance between the SPT and the PSZ2 position for the same clusters. Figure \ref{fig:distance_comparison_to_spt} summarizes this comparison. On average, the joint position is closer to the SPT position than the PSZ2 position is. Therefore, we can conclude that the joint detection method introduces a gain in the position determination with respect to \textit{Planck}, thanks to the use of the X-ray information.

\subsection{Performance comparison with MMF3 method}\label{ssec:purity-completeness}

\begin{figure*}[]
	\centering
	\subfigure[]{\includegraphics[width=1.8\columnwidth]{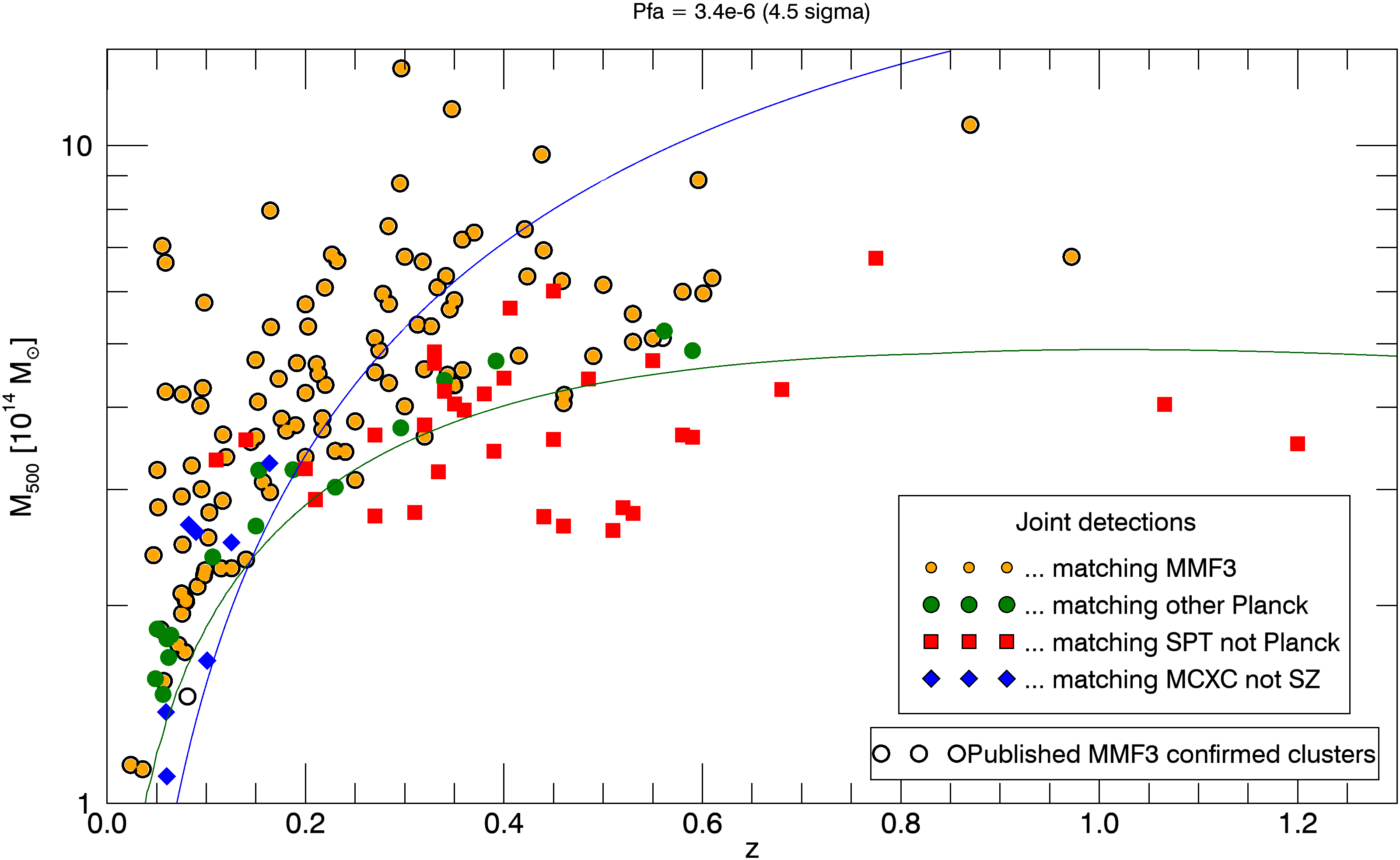}\label{fig:clusters_in_MZ_plane_3.4e-6}}
	\subfigure[]{\includegraphics[width=1.8\columnwidth]{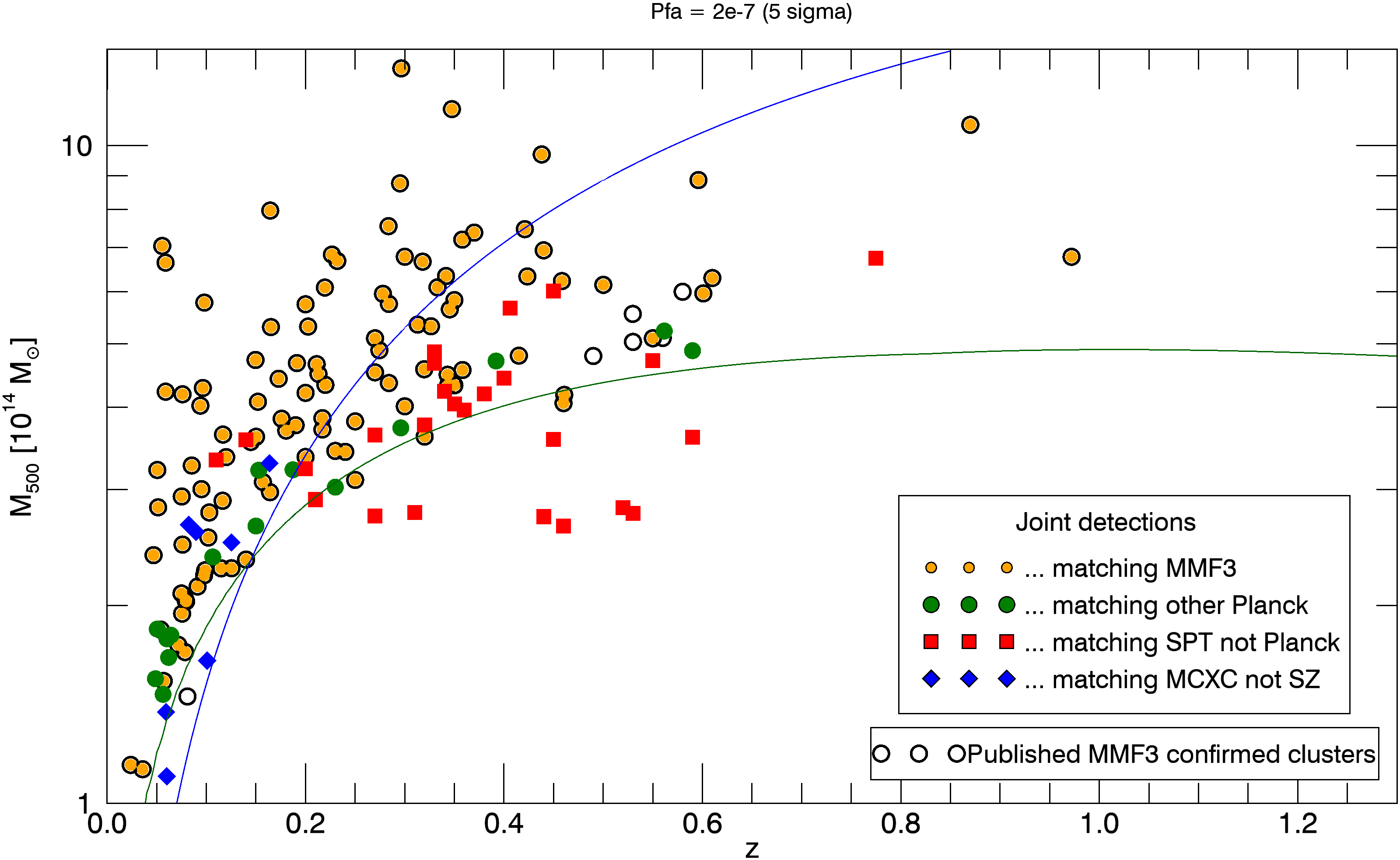}\label{fig:clusters_in_MZ_plane_2e-7}}

	\caption{Mass and redshift of the clusters in the MMF3 catalogue and of the clusters detected with the proposed method for (a) $P_{\rm FA}=3.40 \cdot 10^{-6}$ and (b) $P_{\rm FA}=2.04 \cdot 10^{-7}$. Open circles represent the MMF3 confirmed clusters in the considered region, while filled symbols represent the joint detections colour-coded according to the associated cluster. Yellow-filled circles represent joint detections matching confirmed MMF3 clusters, green-filled circles represent joint detections matching other confirmed \textit{Planck} clusters (not MMF3), red-filled squares represent joint detections matching confirmed SPT clusters not detected by \textit{Planck}, and blue-filled diamonds represent joint detections matching confirmed MCXC clusters that do not match any of the previously mentioned catalogues. 
		The blue solid line shows the REFLEX detection limit, calculated from the REFLEX flux limit and the $L_X-M_{500}$ relation presented in \cite{Piffaretti2011}. It corresponds to a completeness of at least 90\% \citep{Bohringer2001}. The green solid line shows the \textit{Planck} mass limit for the SPT zone at 20\% completeness.
		}
	\label{fig:clusters_in_MZ_plane}
\end{figure*}

\begin{figure*}[ht]
	\centering
	\includegraphics[width=1.5\columnwidth]{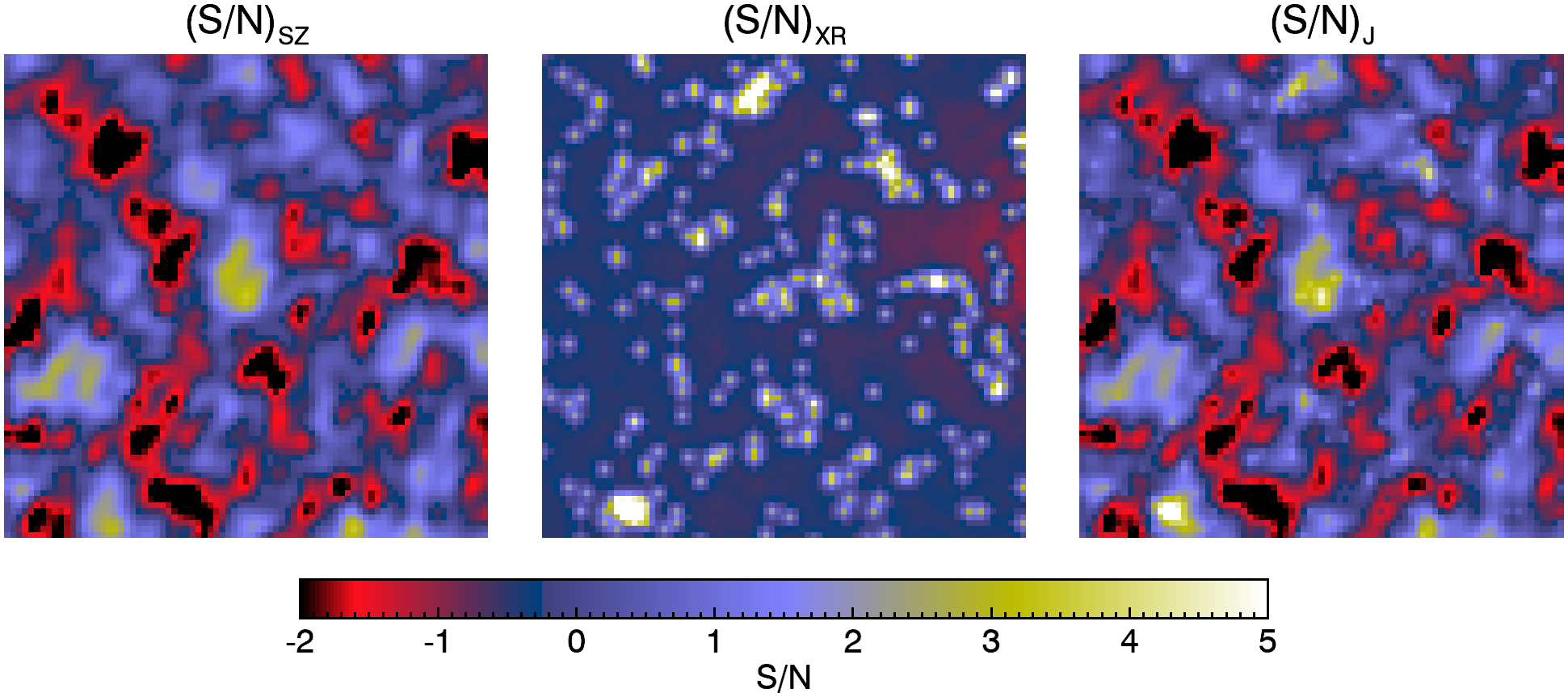}
	\caption{S/N maps for the joint detection that matches SPT-CLJ0351-4109, an SPT cluster not detected by \textit{Planck} situated at z=0.68 with $M_{500}=4.26 \cdot 10^{14} M_{\odot}$. The three colour images show the S/N corresponding to the SZ filtered maps (left), the X-ray filtered map (middle) and the joint filtered maps (right). The filter size is 0.8 arcmin, which corresponds to the one that provides the best S/N for this detection. The angular size of the images is 68.7 arcmin. The RASS exposure time at the position of the detection is 545 s. }
	\label{fig:detection75_spt}
\end{figure*}
\begin{figure*}[]
	\centering
	\includegraphics[width=1.5\columnwidth]{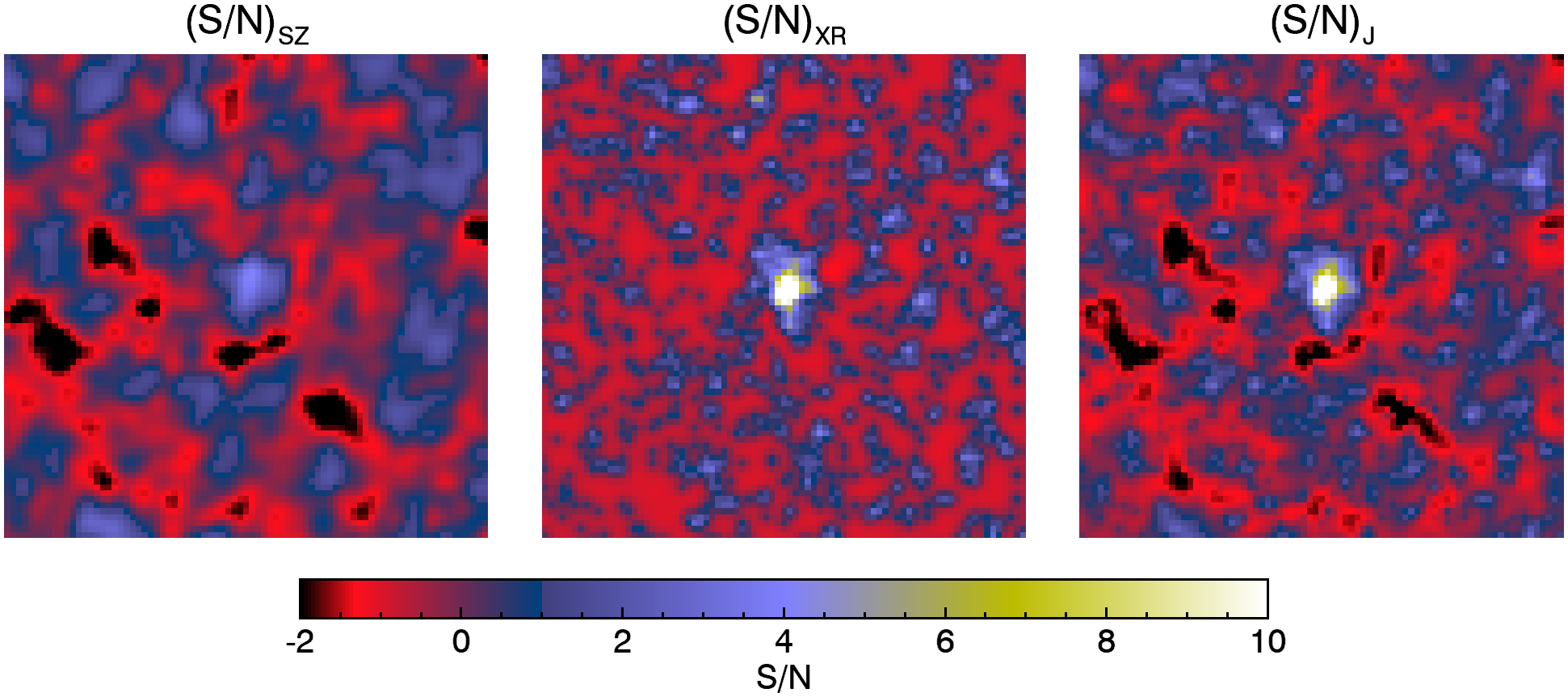}
	\caption{S/N maps for the joint detection that matches RXC J0211.4-4017, a MCXC cluster not detected by \textit{Planck} or SPT situated at z=0.1 with $M_{500}=1.65 \cdot 10^{14} M_{\odot}$. The three colour images show the S/N corresponding to the SZ filtered maps (left), the X-ray filtered map (middle) and the joint filtered maps (right). The filter size is 0.8 arcmin, which corresponds to the one that provides the best S/N for this detection. The angular size of the images is 68.7 arcmin. The RASS exposure time at the position of the detection is 946 s.}
	\label{fig:detection159_mcxc}
\end{figure*}

\begin{figure}[]
	\centering
	\includegraphics[width=0.99\columnwidth]{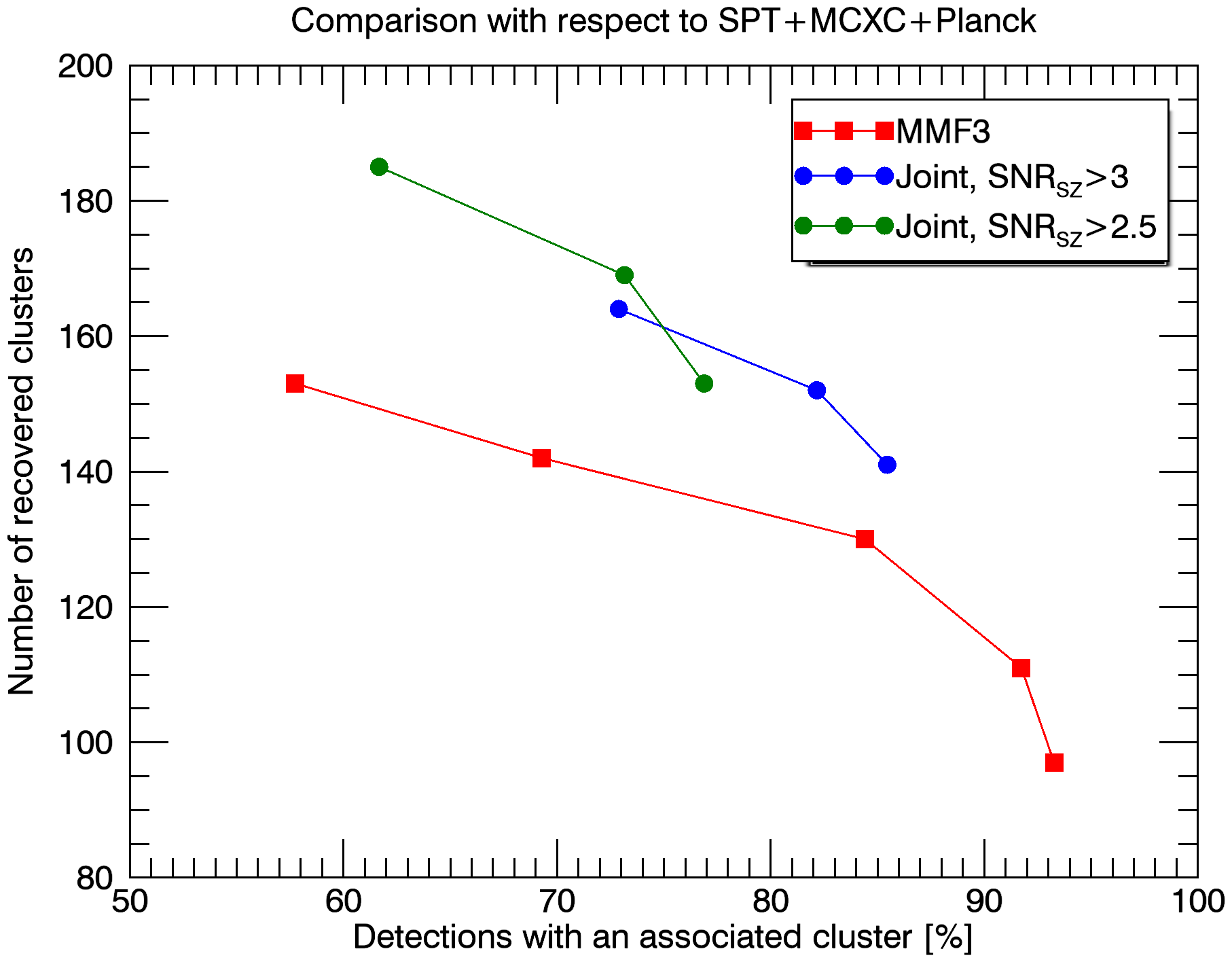}
	\caption{Comparison of the performance of the proposed method (in blue and green) and the MMF3 method (in red). The horizontal axis shows the percentage of detections which match a real cluster within a 10-arcmin radius. It is an indicator of the purity of the methods. The vertical axis shows the number of real clusters which are detected (which match a detection within a 10-arcmin radius). It is related to the detection efficiency of the methods. We assume that the real clusters in the considered region are all the clusters from SPT, MCXC and \textit{Planck} catalogues with known mass and redshift (confirmed). The different points in the curves correspond to different operational points of the detection algorithms. For the MMF3 case, we have represented the results for S/N thresholds of 4.00, 4.25, 4.50, 4.75, and 5.00, increasing from left to right. For the proposed algorithm, we have represented the results for $P_{\rm FA}=3.40 \cdot 10^{-6}$, $P_{\rm FA}=2.04 \cdot 10^{-7}$ and $P_{\rm FA}=1.90 \cdot 10^{-8}$, decreasing from left to right.}
	\label{fig:purity-completeness}
\end{figure}

Since the proposed joint detection method is build as an extension of the MMF3 detection method, we expect it to have a better performance than that of its predecessor. 

As shown in Table \ref{table:detectedclusters}, the proposed method is able to recover most of the MMF3 candidates for $P_{\rm FA} =3.40 \cdot 10^{-6}$ and only eight MMF3 candidates are missing, three of them being confirmed clusters. The three clusters that are missed were initially detected (in the second pass), but then were discarded because the joint S/N was not high enough.  
The five MMF3 unconfirmed clusters that do not appear in our candidate list are missing due to several reasons: two of them are below the initial threshold $q=4$ used to include S/N peaks in the list, another one has a joint S/N lower than the corresponding threshold $q_{\rm J}$ and two of them were discarded because they have a (S/N)$_{\rm SZ}$ < 3. These 5 objects belong to the PSZ2 catalogue, but they are not detected by SPT or ACT, and they do not have any known external counterpart. 
The fact that the recovery rate for MMF3 confirmed clusters is much higher than that of MMF3 candidates indicates that our joint detection is able to clean the MMF3 catalogue from non-cluster objects, thanks to the combination with the X-ray information.

Even though the proposed method misses a small fraction of the MMF3 clusters, it detects other previously known clusters (as shown in 
Table \ref{table:detectedclusters}) that are missed by MMF3. In particular, for $P_{\rm FA}=3.40 \cdot 10^{-6}$, it detects 16 additional \textit{Planck} clusters, 32 SPT clusters that were not detected by \textit{Planck} and 9 MCXC clusters that were not detected by \textit{Planck} or SPT. The overall effect is an improvement of the purity-detection efficiency performance with respect to the reference method MMF3. A comparison of the two methods can be seen in Fig. \ref{fig:clusters_in_MZ_plane}, which shows the MMF3 clusters and the joint detections in the mass-redshift plane for $P_{\rm FA}=3.40 \cdot 10^{-6}$ and $P_{\rm FA}=2.04 \cdot 10^{-7}$. The MMF3 clusters are represented as black open circles, whereas the joint detections are represented as colored symbols. For $P_{\rm FA}=3.40 \cdot 10^{-6}$, the proposed method is able to recover almost all the MMF3 clusters while detecting at the same time additional clusters down to a mass of $2.6\cdot 10^{14} M_{\sun}$ at redshift 0.5. For $P_{\rm FA}=2.04 \cdot 10^{-7}$, the proposed method recovers less clusters due to the increased purity.

Figures \ref{fig:detection75_spt} and \ref{fig:detection159_mcxc} show two examples of additional clusters detected by the proposed method thanks to the combination of SZ and X-ray information. Figure \ref{fig:detection75_spt} shows SPT-CLJ0351-4109, a cluster at z=0.68 with $M_{500}=4.26 \cdot 10^{14} M_{\odot}$ detected by SPT but not detected by \textit{Planck}. The SZ S/N obtained from \textit{Planck} observations is too low to pass the \textit{Planck} detection threshold. However, the presence of some X-ray photons at the same position (11 photons within a 4 arcmin-radius circle, as compared to 3.5 photons expected from the average background level) boosts the joint S/N so that the cluster is detected. The “red” smooth region on the right side of the middle panel of Fig. \ref{fig:detection75_spt} is due to a negative ripple introduced by the filter around a very strong X-ray source (X-ray S/N of 75.4) situated to the right of the cluster, at a distance of 39 arcmin (outside the region represented here). Figure \ref{fig:detection159_mcxc} shows RXC J0211.4-4017, a cluster at z=0.1 with $M_{500}=1.65 \cdot 10^{14} M_{\odot}$ included in the MCXC catalogue, but not detected by \textit{Planck} or SPT. In this case, the presence of a strong X-ray signal at the same position of a faint SZ signal allows the detection of this cluster. 

A direct comparison of the purity-detection efficiency performance of the joint detection method and MMF3 can be seen in Fig. \ref{fig:purity-completeness}. The purity and the detection efficiency are calculated with respect to all confirmed clusters from \textit{Planck}, SPT and MCXC catalogues in the considered region, thus, they are both rough estimations. Nevertheless, they serve as indicators to compare our method with the reference MMF3. The figure shows different operational points of both methods. For MMF3, the operational point is chosen through the S/N threshold. For the nominal \textit{Planck} catalogue, this threshold is set to 4.5, but different thresholds can be used, producing catalogues with different purity and detection efficiency (thus, different completeness). The proposed joint method can be tuned by changing the false alarm probability that is used to calculated the joint S/N threshold. This figure shows that our detection method outperforms MMF3 in the sense that it can simultaneously achieve higher purity and higher detection efficiency if the operational point is chosen appropriately.

\subsection{Comparison with RASS clusters}\label{ssec:comparison_rass}
\begin{figure}[]
	\centering
	\includegraphics[width=0.99\columnwidth]{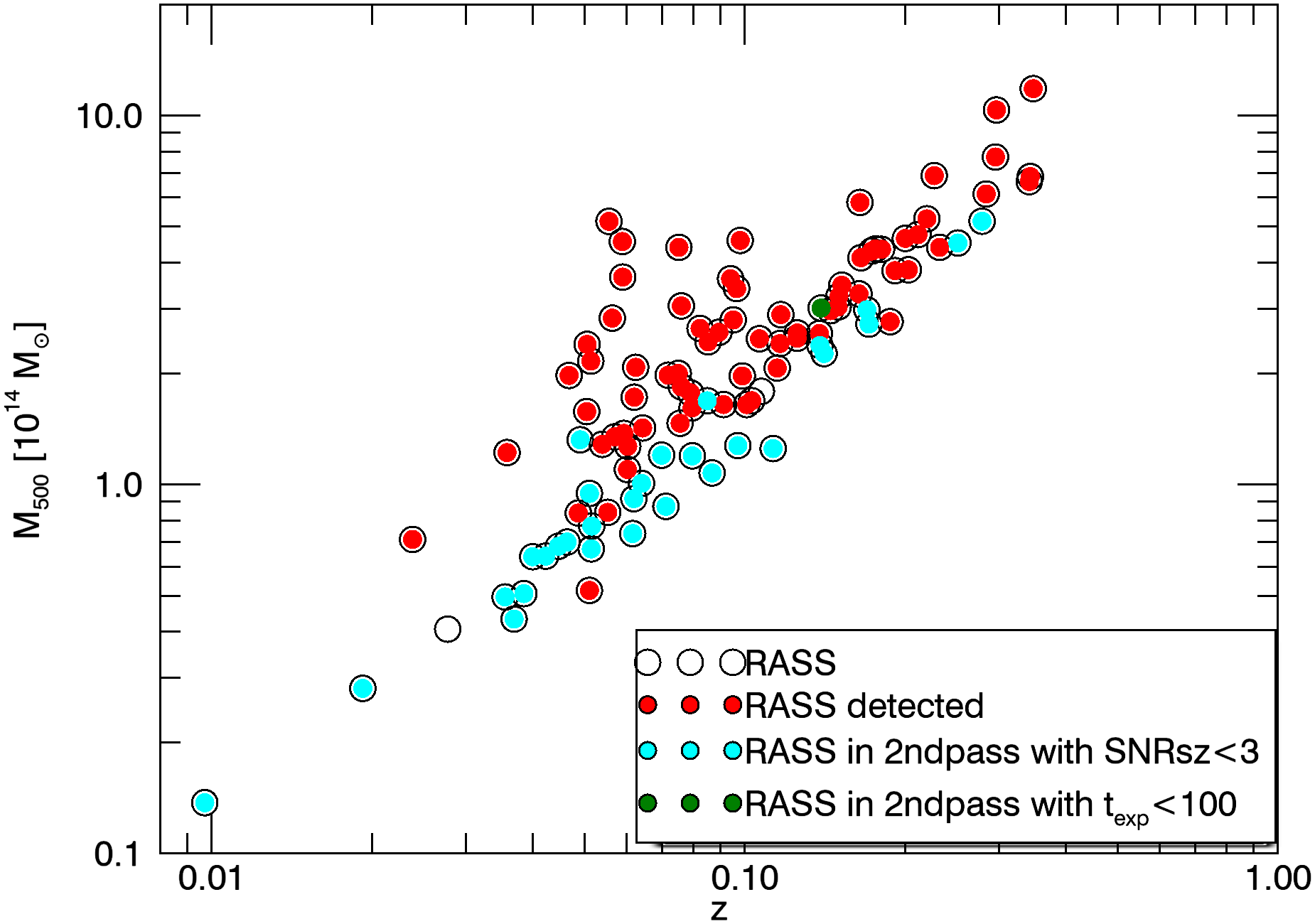}
	\caption{Mass and redshift of the clusters in the RASS catalogue. Open circles represent RASS clusters in the considered region, red-filled circles represent RASS clusters that are detected by the joint detection algorithm, cyan-filled circles represent RASS clusters that were detected in the second pass of the algorithm, but discarded due to a low (S/N)$_{\rm SZ}$ and the green-filled circle represents the RASS cluster that was detected in the second pass of the algorithm, but discarded due to a low exposure time.}
	\label{fig:RASS_clusters_in_MZ_plane}
\end{figure}

Since the proposed joint detection method uses RASS observations, it is interesting to check whether it is able to recover known clusters that have been previously detected using the same observations. Table \ref{table:detectedclusters} shows that we detect 71 of the 103 RASS clusters situated in the considered region (SPT area with RASS exposure time greater than 100 s, outside the PSZ2 masked region), which corresponds to 68.9 \%. Most of the RASS clusters that we do not recover (30 of the 32) were in fact included in the list of detections provided by the second pass of the algorithm, but 29 were discarded because their (S/N)$_{\rm SZ}$ was lower than 3 and one was discarded because the RASS exposure time at the detection's position is lower than 100 s. There are just 2 RASS clusters that were not originally detected by the joint algorithm because their joint S/N does not reach the threshold of $q=4$: RXC J0040.1-5607 and RXC J2326.7-5242, which are quite faint both in X-ray and SZ. To cross-check these results, we used the MMF3 method of \cite{Planck2015ResXXVII} to extract the S/N of RASS clusters from \textit{Planck} maps. We obtained that the 2 not-detected clusters and the 29 discarded due to a low (S/N)$_{\rm SZ}$ have a very low S/N, which supports our results. Therefore, we can conclude that the joint detection method is able to recover almost all the RASS clusters, as expected, but we discard some of them later in order to maintain a high purity by eliminating possible AGN detections with a threshold in (S/N)$_{\rm SZ}$, which has a similar effect to a mass cut at each redshift.
Figure \ref{fig:RASS_clusters_in_MZ_plane} illustrates this comparison by showing the RASS clusters and the joint detections in the mass-redshift plane.  

\subsection{New candidates}\label{ssec:candidates} 
As mentioned before, 193 of the 225 detections corresponding to $P_{\rm FA}=3.40 \cdot 10^{-6}$ match with a real cluster within a distance of 10 arcmin. This means that our catalogue contains 32 candidates that are not known clusters. They could be either false detections due to noise or X-ray point sources, or true clusters not detected before. Table \ref{table:candidates} summarizes the coordinates and some additional information of these 32 new candidates.

Since we set a false alarm probability of $P_{\rm FA}=3.40 \cdot 10^{-6}$ to calculate the joint S/N threshold, we expect to have 1.7 pixels in each filtered patch with (S/N)$_{\rm J}$ greater than the threshold. This means that in the whole SPT area we expect to have 42 pixels above the threshold, producing at most 42 false detections due to noise fluctuations. This number is close to the number of new candidates, so can we expect most of them to be false detections. 
On the other hand, for $P_{\rm FA}=2.04 \cdot 10^{-7}$ and $P_{\rm FA}=1.90 \cdot 10^{-8}$ we would expect at most 2.5 and 0.2 false detections due to noise fluctuations in the SPT footprint, respectively. However, the number of new candidates for these two false alarm probabilities is 14 and 10, respectively 
(see Table  \ref{table:detectedclusters}). We expect then that most of these candidates are real detections (either clusters or other objects like X-ray point sources). We indicate in Table \ref{table:candidates} which of the 32 candidates corresponding to $P_{\rm FA}=3.40 \cdot 10^{-6}$ are also candidates for $P_{\rm FA}=2.04 \cdot 10^{-7}$ and $P_{\rm FA}=1.90 \cdot 10^{-8}$.

\begin{table*}
	\caption{List of candidates for $P_{\rm FA}=3.40 \cdot 10^{-6}$ that do not match with known clusters or cluster candidates, ordered by significance. Galactic and equatorial coordinates are given in degrees. The joint S/N is indicated, as well as the SZ component of this S/N. Finally, the joint threshold $q_{\rm J}$, the difference between the S/N and the threshold, and the significance are shown. The last column indicates whether the candidate is also a candidate for $P_{\rm FA}=2.04 \cdot 10^{-7}$ and $P_{\rm FA}=1.90 \cdot 10^{-8}$ (* indicates that it is also a candidate for $P_{\rm FA}=2.04 \cdot 10^{-7}$, ** indicates that it is a candidate for both $P_{\rm FA}=2.04 \cdot 10^{-7}$ and $P_{\rm FA}=1.90 \cdot 10^{-8}$).}
	\label{table:candidates}
	\centering 
	\begin{tabular}{c | c c c c c c c c c c}
		\hline
		\noalign{\smallskip}
		Id. &  G. lon. & G. lat. & RA J2000 & Dec J2000 & (S/N)$_{\rm J}$ & (S/N)$_{\rm SZ}$ & $q_{\rm J}$ & (S/N)$_{\rm J}$-$q_{\rm J}$ & Significance & Notes \\
		&  [$\degr$] & [$\degr$] & [$\degr$] & [$\degr$] & & & & & & \\
		\noalign{\smallskip}
		\hline
		\noalign{\smallskip}
    1 & 262.127 & -30.865 &  86.910 & -54.310 &   13.72 &    3.28 &    4.70 &    9.02 &   10.62 & * *  \\ 
    2 & 272.610 & -28.890 &  91.930 & -63.244 &    9.46 &    4.23 &    4.87 &    4.59 &    7.62 & * *  \\ 
    3 & 356.016 & -51.958 & 330.063 & -43.514 &    9.05 &    3.08 &    6.00 &    3.04 &    6.57 & * *  \\ 
    4 & 269.926 & -33.562 &  81.955 & -60.928 &    7.70 &    4.06 &    4.84 &    2.86 &    6.44 & * *  \\ 
    5 & 266.726 & -34.077 &  81.239 & -58.245 &    7.58 &    3.23 &    4.92 &    2.65 &    6.30 & * *  \\ 
    6 & 299.961 & -53.485 &  16.829 & -63.551 &    7.45 &    3.04 &    5.15 &    2.30 &    6.06 & * *  \\ 
    7 & 282.665 & -54.841 &  35.371 & -58.601 &    6.77 &    3.94 &    4.77 &    2.00 &    5.75 & * *  \\ 
    8 & 265.689 & -27.575 &  93.313 & -57.045 &    7.12 &    3.48 &    4.71 &    2.42 &    5.67 & * *  \\ 
    9 & 270.400 & -44.745 &  60.078 & -58.641 &    6.93 &    3.01 &    5.00 &    1.93 &    5.65 & * *  \\ 
    10 & 254.861 & -20.784 & 100.312 & -45.874 &    6.20 &    3.74 &    4.84 &    1.36 &    5.47 & * *  \\ 
    11 & 327.280 & -75.170 &   4.893 & -40.415 &    7.36 &    4.03 &    5.54 &    1.82 &    5.43 & * *  \\ 
    12 & 255.846 & -41.583 &  69.662 & -49.106 &    5.56 &    3.58 &    4.53 &    1.02 &    5.35 & *   \\ 
    13 & 346.731 & -57.456 & 339.853 & -46.882 &    9.40 &    3.52 &    6.77 &    2.63 &    5.26 & * *  \\ 
    14 & 290.504 & -71.045 &  18.586 & -45.527 &    7.07 &    3.02 &    5.64 &    1.43 &    5.25 & *   \\ 
    15 & 249.394 & -34.217 &  80.293 & -43.930 &    5.58 &    3.56 &    4.84 &    0.74 &    5.19 & *   \\ 
    16 & 357.238 & -34.672 & 306.147 & -43.350 &    5.79 &    3.51 &    4.87 &    0.91 &    5.14 & *   \\ 
    17 & 283.847 & -65.365 &  25.079 & -49.914 &    6.65 &    3.52 &    5.86 &    0.79 &    4.93 &    \\ 
    18 &   0.622 & -48.152 & 324.445 & -41.074 &    6.72 &    3.11 &    5.98 &    0.74 &    4.89 &    \\ 
    19 & 352.487 & -33.184 & 303.471 & -47.119 &    5.69 &    3.73 &    5.14 &    0.55 &    4.86 &    \\ 
    20 & 335.748 & -37.097 & 310.996 & -60.647 &    5.56 &    3.36 &    5.03 &    0.53 &    4.84 &      \\
    21 & 260.921 & -35.301 &  79.333 & -53.436 &    5.54 &    3.03 &    4.98 &    0.57 &    4.84 &      \\
    22 & 265.024 & -30.437 &  87.988 & -56.765 &    5.11 &    3.45 &    4.74 &    0.37 &    4.74 &      \\
    23 & 356.548 & -41.337 & 315.336 & -44.232 &    4.89 &    3.84 &    4.69 &    0.20 &    4.70 &      \\
    24 & 277.913 & -65.311 &  28.381 & -48.691 &    5.34 &    3.97 &    5.06 &    0.28 &    4.68 &      \\
    25 & 274.955 & -58.096 &  37.411 & -53.368 &    5.72 &    3.12 &    5.43 &    0.29 &    4.65 &      \\
    26 & 282.535 & -53.687 &  36.837 & -59.477 &    5.34 &    3.13 &    5.16 &    0.18 &    4.62 &      \\
    27 & 258.667 & -34.133 &  81.229 & -51.574 &    4.98 &    3.45 &    4.85 &    0.13 &    4.58 &      \\
    28 & 260.348 & -20.324 & 103.375 & -50.613 &    4.94 &    4.43 &    4.87 &    0.07 &    4.53 &      \\
    29 & 252.269 & -23.928 &  95.065 & -44.475 &    4.96 &    3.64 &    4.88 &    0.08 &    4.53 &      \\
    30 & 264.135 & -35.097 &  79.543 & -56.066 &    4.63 &    3.41 &    4.61 &    0.03 &    4.51 &      \\
    31 & 254.233 & -25.041 &  94.169 & -46.485 &    4.84 &    3.19 &    4.82 &    0.01 &    4.51 &      \\
    32 & 328.390 & -59.170 & 351.147 & -53.453 &    5.77 &    3.46 &    5.65 &    0.13 &    4.43 &      \\
  	\end{tabular}
  \end{table*}
  
  \begin{figure*}[]
  	\centering
  	\subfigure[]{\includegraphics[width=.65\columnwidth]{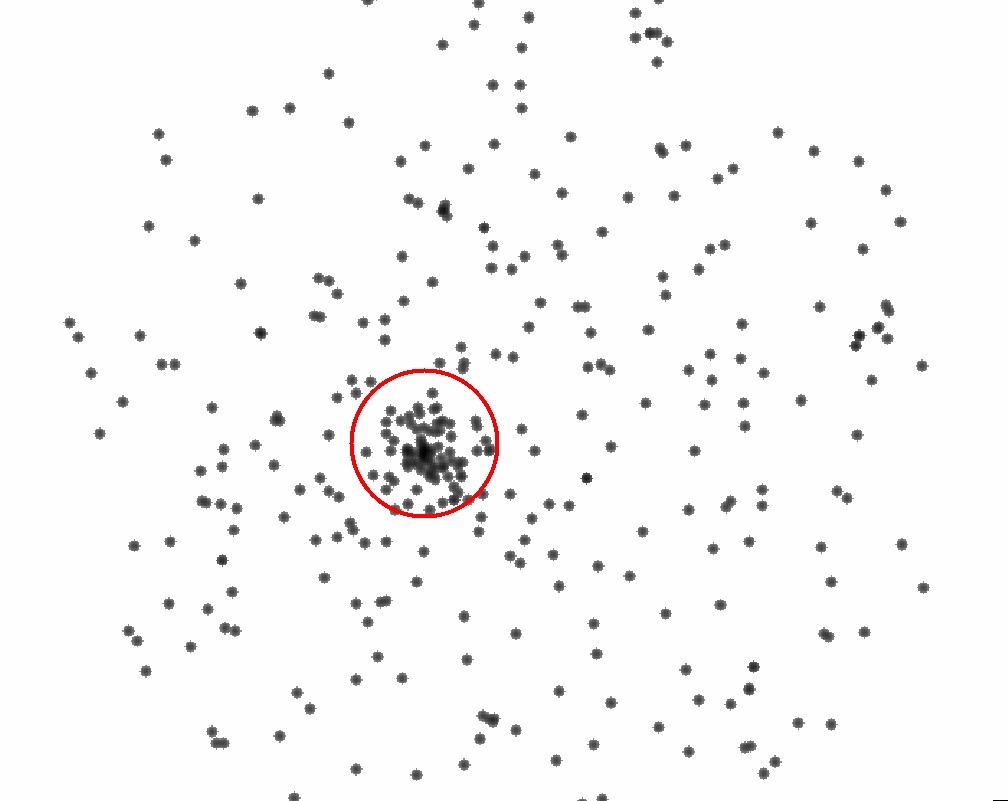}\label{fig:swift1}}
  	\subfigure[]{\includegraphics[width=.65\columnwidth]{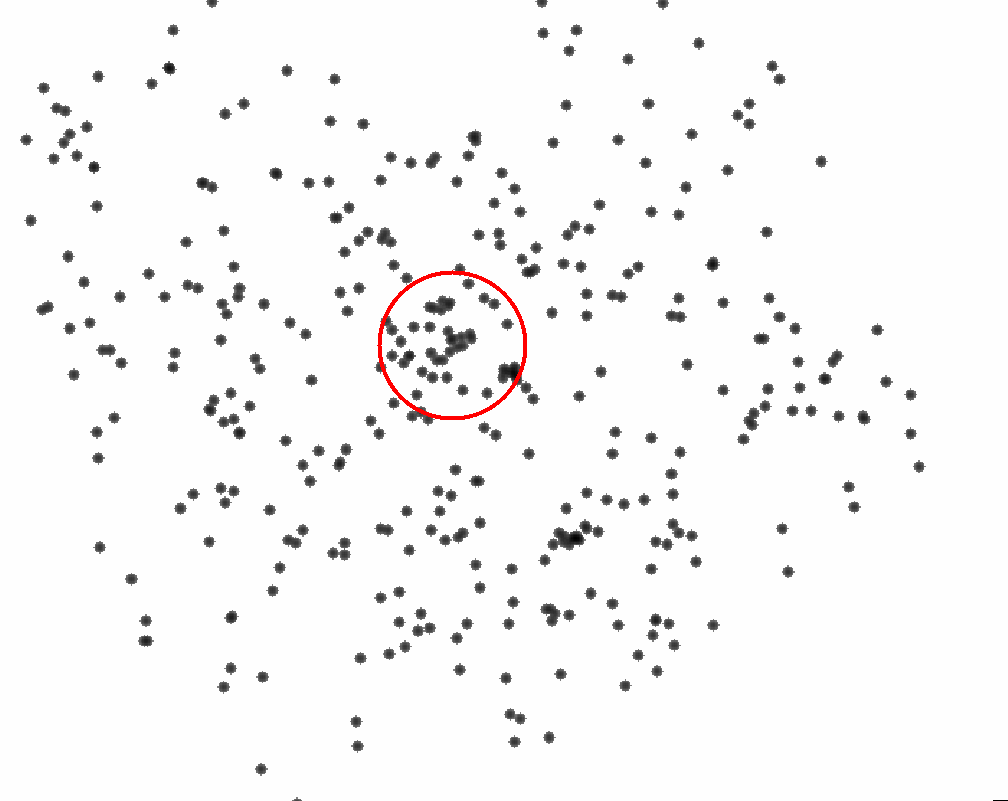}\label{fig:swift4}}
  	\subfigure[]{\includegraphics[width=.65\columnwidth]{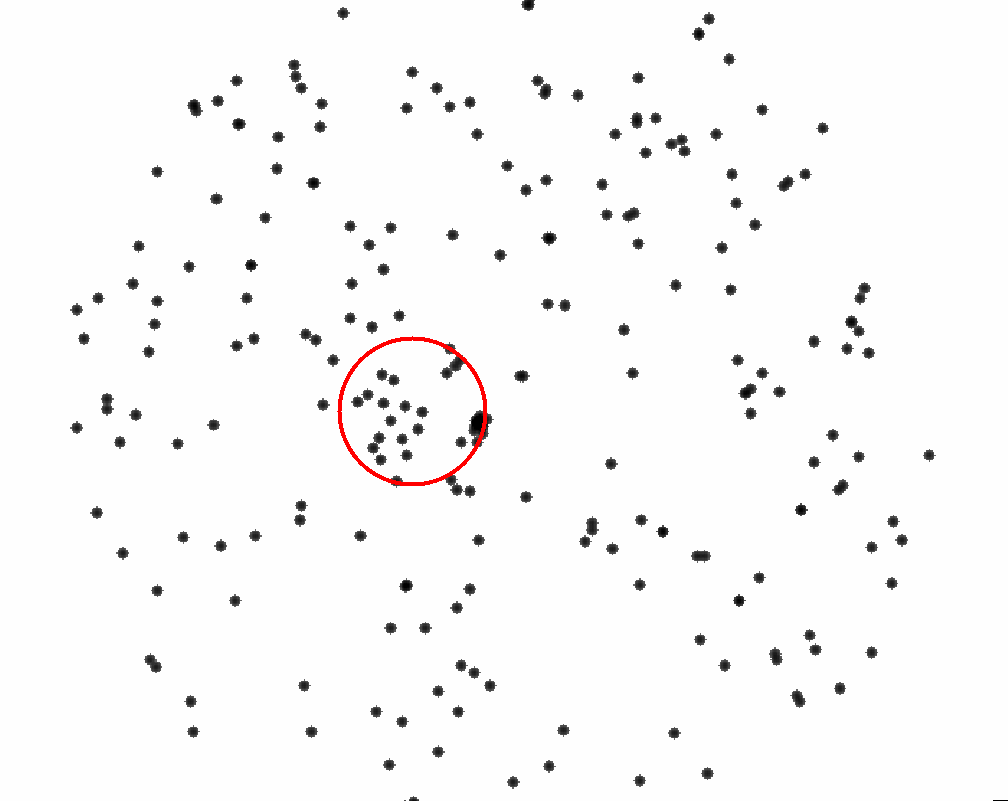}\label{fig:swift12}}
  	\caption{\textit{Swift} observations of candidates 1 (a), 8 (b) and 12 (c). A red circle with 2 arcmin radius centered at the candidate position was added for visual reference.}
  	\label{fig:swift}
  \end{figure*}
  
We searched for archival X-ray observations covering these 32 positions. Three of the candidates (1, 8 and 12) were observed by \textit{Swift} \citep{Burrows2005} and two more (14 and 32) by XMM-Newton. In the three \textit{Swift} observations, there is evidence that the candidates are real clusters, since we can see an extended emission, as shown in Fig. \ref{fig:swift}. On the contrary, the two XMM-Newton observations show that the candidates are false detections: candidate 14 is a point source, while candidate 32 is just a noise fluctuation. As shown in the Table \ref{table:candidates}, the joint S/N of candidate 32 is just above the threshold. 

We also looked around these 32 positions for other known galaxy clusters or groups in the NED\footnote{http://nedwww.ipac.caltech.edu/} and SIMBAD\footnote{http://simbad.u-strasbg.fr/simbad} databases. Table \ref{table:candidates_matchings} shows all the clusters and groups found closer than 10 arcmin to our candidates. Most of the joint candidates do not have any NED or SIMBAD cluster close to them. For 13 candidates, we found some close-by objects, but in most of the cases, they do not seem to be associated with the candidate, since their separation is too big. Only candidates 12 and 21 might be associated with clusters: the SPT cluster SPT-CL J0438-4907 and the optical cluster LCS-CL J051723-5325.5, respectively. 

We clarify that we did not match candidate 12 with SPT-CL J0438-4907 before (Sect. \ref{ssec:crossmatch}) because this cluster is not included in the published SPT catalogue of \cite{Bleem2015} since its significance is lower than 4.5, which was the limit for the published catalogue. However, it is detected at lower significance and confirmed with optical observations in \cite{Saro2015}. The presence of this cluster at only 0.7 arcmin from candidate 12, together with the \textit{Swift} observation presented in Fig. \ref{fig:swift12} are strong indicators that this candidate is a real cluster. Moreover, the mass obtained from the joint extraction assuming the redshift of the SPT cluster is $M_{500}=3.04 \cdot 10^{14} M_{\odot}$, very close to the mass published by \cite{Saro2015} for the SPT cluster ($M_{500}=(3.13 \pm 0.81) \cdot 10^{14} M_{\odot}$).

Regarding candidate 21, the estimated mass assuming the redshift of the nearby optical cluster is $M_{500}=3.38 \cdot 10^{14} M_{\odot}$. If we apply the richness-mass relation of \cite{Rozo2014}, we get an estimated richness of $\lambda_e=63.5$. According to \cite{Bleem2015b}, the optical cluster LCS-CL J051723-5325.5 has a richness of $\lambda=29.2$, which differs from  $\lambda_e$ by 2.92 $\sigma_{\rm{ln} \lambda}$. Given the large scatter of the richness-mass relation, it is reasonable to associate candidate 21 with cluster LCS-CL J051723-5325.5.

\begin{table}
	\caption{Galaxy clusters and galaxy groups found close to the 32 candidates of Table \ref{table:candidates}. The search was done in NED and SIMBAD databases.}
	\label{table:candidates_matchings}
	\centering 
	\small
	\begin{tabular}{c | c c c c}

		\hline
		\noalign{\smallskip}
		Id. &  Name  & Type & Redshift & Separation\\
		    &        &      &          & [arcmin]\\
		\noalign{\smallskip}
		\hline
		\noalign{\smallskip}
		1  & Str 0547-543		     & Cluster &       &  3.2 \\
		3  & NGC 7166                & Group   & 0.0077&  7.1 \\
		   & [CHM2007] HDC 1172      & Group   &       &  4.6 \\
		   & LGG 449                 & Group   &       &  7.1 \\
		   & NOGG H 1003             & Group   &       &  7.5 \\
		7  & [LH2011] 3692           & Group   &       &  7.0 \\
	    11 & [RZZ99] ESP 121         & Group   &       &  5.0 \\
		   & Str 0018-407            & Cluster &       &  5.0 \\
		   & EDCC 435                & Cluster & 0.15  &  8.6 \\
        12 & SPT-CL J0438-4907       & Cluster &  0.24 &  0.7 \\
		14 & LCLG -45 038            & Group   & 0.0897&  7.6 \\
		18 & APMCC 688               & Group   & 0.065 &  3.6 \\
		   & LCLG -42 164            & Group   & 0.065 &  9.2 \\
		20 & APM CC 607              & Cluster &       &  2.3 \\
		   & Str 2040-608            & Cluster &       &  2.8 \\
		21 & LCS-CL J051723-5325.5   & Cluster &  0.37 &  0.8 \\
		   & LCS-CL J051759-5326.4   & Cluster &  0.39 &  5.9 \\
		   & LCS-CL J051813-5327.2   & Cluster &  0.62 &  8.0 \\
		23 & APMCC 621               & Cluster & 0.143 &  4.9 \\
		   & [DEM94] 205740.8-442233 & Cluster &       &  5.8  \\
		   & QW 146                  & Cluster &       &  6.8  \\
		27 & LCS-CL J052516-5134.1   & Cluster &  0.22 &  3.3 \\
		   & LCS-CL J052502-5134.1   & Cluster &  0.24 &  9.8 \\
		30 & SCSO J051755-555727     & Cluster &  0.66 &  6.9 \\
		32 & SCSO J232529-532420     & Cluster &  0.74 &  8.5 \\

	\end{tabular}
\end{table}

\section{Conclusions}\label{sec:conclusions}

In this paper we have proposed a galaxy cluster detection method based on matched multifrequency filters (MMF) that combines X-ray and SZ observations. This method builds on the previously proposed joint X-ray-SZ extraction method and allows to blindly detect clusters, that is finding new clusters without knowing their position, size or redshift, by searching on SZ and X-ray maps simultaneously. It can be seen as an evolution of the MMF3 detection method, one of the MMF methods used to detect clusters from \textit{Planck} observations, that incorporates X-ray observations to improve the detection performance. 

The main challenge to solve was to obtain a high purity, since the addition of the X-ray information increases the cluster detection probability, but also the number of false detections, produced by AGNs and Poisson noise. To deal with the Poisson noise correctly, we proposed to use an adaptive S/N threshold to keep or discard detections depending on the noise characteristics of the region. To discard AGN detections, we propose an additional classification according to the SZ part of the S/N. 

The proposed method is tested using data from the RASS and \textit{Planck} surveys and evaluated by comparing the detection results with existing cluster catalogs in the area of the sky covered by the SPT survey. We have shown that, thanks to the addition of the X-ray information, the method is able to achieve simultaneously better purity, better detection efficiency and better position accuracy than its predecessor, the MMF3 detection method.

We have also shown that, if the redshift of a cluster is known by any other means, the joint detection allows to obtain a good estimation for its mass. Some bias may appear in the presence of a cool core (overestimated mass) or if there is an offset between the X-ray and SZ peaks (underestimated mass).

Finally, we have produced a catalogue of candidates in the SPT region composed of 225 objects, with 32 new objects that are not included in other SZ or X-ray cluster catalogues. We have found, using \textit{Swift} observations, that three of these new objects are probably real clusters. This supports the fact that the proposed method can be used to find new clusters. 

In future work we will run the joint detection on all the sky using \emph{Planck} and RASS maps and provide the last and deepest all-sky cluster catalogue before the e-ROSITA mission.

\begin{acknowledgements}
This research is based on observations obtained with \textit{Planck} (http://www.esa.int/Planck), an ESA science mission with instruments and contributions directly funded by ESA Member States, NASA, and Canada. This research has made use of the ROSAT all-sky survey data which have been processed at MPE. The authors acknowledge the use of the HEALPix package \citep{Gorski2005}. The authors would like to thank Iacopo Bartalucci for his help in the preparation of the \textit{Swift} images. The research leading to these results has received funding from the European Research Council under the European Union's Seventh Framework Programme (FP7/2007-2013) / ERC grant agreement n$^{\circ}$ 340519. 
\end{acknowledgements}

\bibliographystyle{aa} 
\bibliography{mybiblio}

\begin{appendix}
	
\section{Matching of simulation results with real maps parameters}\label{app:conversion}
	
As explained in Sect. \ref{ssec:jointthreshold}, the X-ray noise maps used in the Monte Carlo simulations were simulated as homogeneous Poisson random fields, characterized by a given mean value $\lambda$ (in counts). To express these count maps into $\Delta T/T$ units, as done with real RASS count maps, we need to assume an exposure time map and a $N_{\rm H}$ map, and then apply the conversion procedure described in Appendix B of \cite{Tarrio2016}, which can be summarized as follows:
\begin{equation}\label{eq:conversion_counts_deltaTT}
M[\Delta T/T] = \frac{M[\rm counts]}{t_{\rm exp}[\rm s]} \cdot c(N_{\rm H}) \cdot \left[ \frac{F_{\rm X}}{Y_{500}} \right]_{z_{\rm ref}}^{-1} \cdot \frac{g(\nu_{\rm ref})}{d_{\rm pix}^2}
\end{equation}
In this expression $c(N_{\rm H})$ represents the factor that converts the countrate into X-ray flux and it depends on the $ N_{\rm H} $ map; the expected $ F_{\rm X}/Y_{500} $ relation is used to convert the X-ray flux into equivalent $ Y_{500} $ integrated flux and depends on the reference redshift; $d_{\rm pix}^2$ is the HEALPix pixel area and $g(\nu_{\rm ref})$ is the factor that converts from $y$ units into $ \Delta T/T_{\rm CMB} $ units, which depends on the reference frequency assumed for the map (1000 GHz in our case).

For the Monte Carlo simulations of this paper, we assumed a constant exposure time $t_{\rm exp} = 400$ s and a constant $N_{\rm H}=2 \cdot 10^{20}$ cm$^{-2}$ (average values in the SPT region). If other values were used, the resulting X-ray maps in $\Delta T/T$ units would differ from the ones obtained with these values by just a constant factor $a$. This allows to convert some of the simulation results obtained for the reference $t_{\rm exp}$ and $N_{\rm H}$ values (in particular $\sigma_{\theta_{\rm s}}^{\rm x}$ and S/N) into the results corresponding to any other value of $t_{\rm exp}$ and/or $N_{\rm H}$. In the following, we explain how this conversion is done.

Let $M_{\rm x}$ and $M'_{\rm x}$ be two X-ray maps in $\Delta T/T$ units, calculated from the same count map using different values of $t_{\rm exp}$ and $N_{\rm H}$. From eq. \ref{eq:conversion_counts_deltaTT}, we have the following relation between the two maps:
\begin{equation}\label{eq:Mreal_Msimu}
M'_{\rm x}= \frac{M_{\rm x}}{a}
\end{equation}
where
\begin{equation}\label{eq:a}
a = \frac{t_{\rm exp}'}{t_{\rm exp}} \cdot \frac{c(N_{\rm H})}{c(N_{\rm H}')}
\end{equation}

Let $\mathbf{M} = \left[\mathbf{N}_{\rm sz}, N_{\rm x}\right]^{\rm T}$ and $\mathbf{M}' = \left[\mathbf{N}_{\rm sz}, N'_{\rm x}\right]^{\rm T} = \left[\mathbf{N}_{\rm sz}, N_{\rm x}/a\right]^{\rm T}$ be two multifrequency noise maps whose SZ components are equal and whose X-ray components differ by a constant factor $a$. 

Considering that the noise in the X-ray map and the SZ maps is uncorrelated, we can write the noise power spectrum of $\mathbf{M}$ as:
\begin{equation}\label{eq:P_joint}
\mathbf{P} = \left[ \begin{array}{c c}
\mathbf{P}_{\rm sz} & \mathbf{0}_{N_\nu \times 1} \\
\mathbf{0}_{1 \times N_\nu} & P_{\rm x}\\
\end{array}\right] 
\end{equation}
where $\mathbf{P}_{\rm sz}$ is the noise power spectrum of the SZ maps $\mathbf{N}_{\rm sz}$, $P_{\rm x}$ is the noise power spectrum of the X-ray map $N_{\rm x}$, and $\mathbf{0}_{n \times m}$ denotes a vector with $n$ rows and $m$ columns whose elements are all equal to 0. The noise power spectrum of $\mathbf{M'}$ can be decomposed into $\mathbf{P}_{\rm sz}$ and $P'_{\rm x}$ in an analogous way. Using the definition of the noise power spectrum (see Sect. \ref{ssec:jointalgorithm}), it is immediate to show that $P'_{\rm x} = a^{-2} P_{\rm x}$.

Using the definition of the variance of the filtered maps (eq. \ref{eq:sigma_sz}) and applying \ref{eq:P_joint}, we can also decompose the variance of the filtered maps into an SZ and an X-ray component as follows:
\begin{equation}\label{eq:sigmasz_sigmaxr}
\sigma_{\theta_{\rm s}}^{-2} = \mathbf{F}_{\rm sz}^{\rm T} \mathbf{P}_{\rm sz}^{-1} \mathbf{F}_{\rm sz} +  {F}_{\rm x}^{\rm T} {P}_{\rm x}^{-1} F_{\rm x}  = \sigma_{{\rm sz}}^{-2} + \sigma_{{\rm x}}^{-2}
\end{equation}
where $\mathbf{F}_{\rm sz}$ and ${F}_{\rm x}$ are the SZ and X-ray components of $\mathbf{F}_{\theta_{\rm s}}$ (eq. \ref{eq:F_joint}). From this expression it is easy to show that:
\begin{equation}\label{eq:sigmareal_sigmasimu}
\sigma'_{\rm x} = \frac{\sigma_{\rm x}}{a}
\end{equation}

If we filter $\mathbf{M}$ and $\mathbf{M}'$ with the proposed joint filter (eq. \ref{eq:filter_sz}), the S/N of the filtered maps will be given by:
\begin{equation}\label{eq:SNR}
S/N= \sigma_{\theta_{\rm s}} \left[\mathbf{F}_{\rm sz}^{*} \mathbf{P}_{\rm sz}^{*-1} \mathbf{N}_{\rm sz} +  {F}_{\rm x}^{*} {P}_{\rm x}^{*-1} N_{\rm x} \right]
\end{equation}
and 
\begin{equation}\label{eq:SNR_prima}
S/N' = \sigma'_{\theta_{\rm s}} \left[\mathbf{F}_{\rm sz}^{*} \mathbf{P}_{\rm sz}^{*-1} \mathbf{N}_{\rm sz} +  a{F}_{\rm x}^{*} {P}_{\rm x}^{*-1} N_{\rm x} \right]
\end{equation}
From these two expressions, and taking into account \ref{eq:sigmasz_sigmaxr} and \ref{eq:sigmareal_sigmasimu}, we can obtain the relation between the S/N of the two noise maps as:
\begin{equation}\label{eq:SNRreal_SNRsimu}	S/N'= S/N \cdot \sqrt{\frac{1+\omega}{1+a^2 \omega}} \cdot \frac{1+a W}{1+W}
\end{equation}
where
\begin{equation}\label{eq:omega}
\omega = \left(\frac{\sigma_{\rm sz}}{\sigma_{\rm x}} \right)^2
\end{equation}
and
\begin{equation}\label{eq:W}	
W = \frac{{F}_{\rm x}^{*} {P}_{\rm x}^{*-1} N_{\rm x}}{\mathbf{F}_{\rm sz}^{*} \mathbf{P}_{\rm sz}^{*-1} \mathbf{N}_{\rm sz}}
\end{equation}

\section{Significance estimation}\label{app:significance}
As explained in Sect. \ref{ssec:mass_estimation}, the significance of a detection is calculated with the aid of the S/N maps obtained from the numerical simulations described in Sect. \ref{ssec:jointthreshold}. In particular, it is obtained by measuring the number of pixels in the simulations that have a S/N higher than the S/N of the detection. Given that the number of simulated pixels is finite, when the joint S/N of the detection is very high, there may be no pixels satisfying this condition. In these cases, it is not possible to calculate the significance directly. Since for each combination of  mean Poisson level $\lambda$  and  mean Gaussian level $\sigma_{217}$ we created 450 random realizations of the noise maps, and since each map has 700x700 pixels, the minimum false alarm probability that we can determine is $4.5\cdot 10^{-9}$. This means that when we do not find any pixel above a given S/N, we can only affirm that the significance will be higher than 5.75.

\begin{figure}[]
	\centering
	\includegraphics[width=0.99\columnwidth]{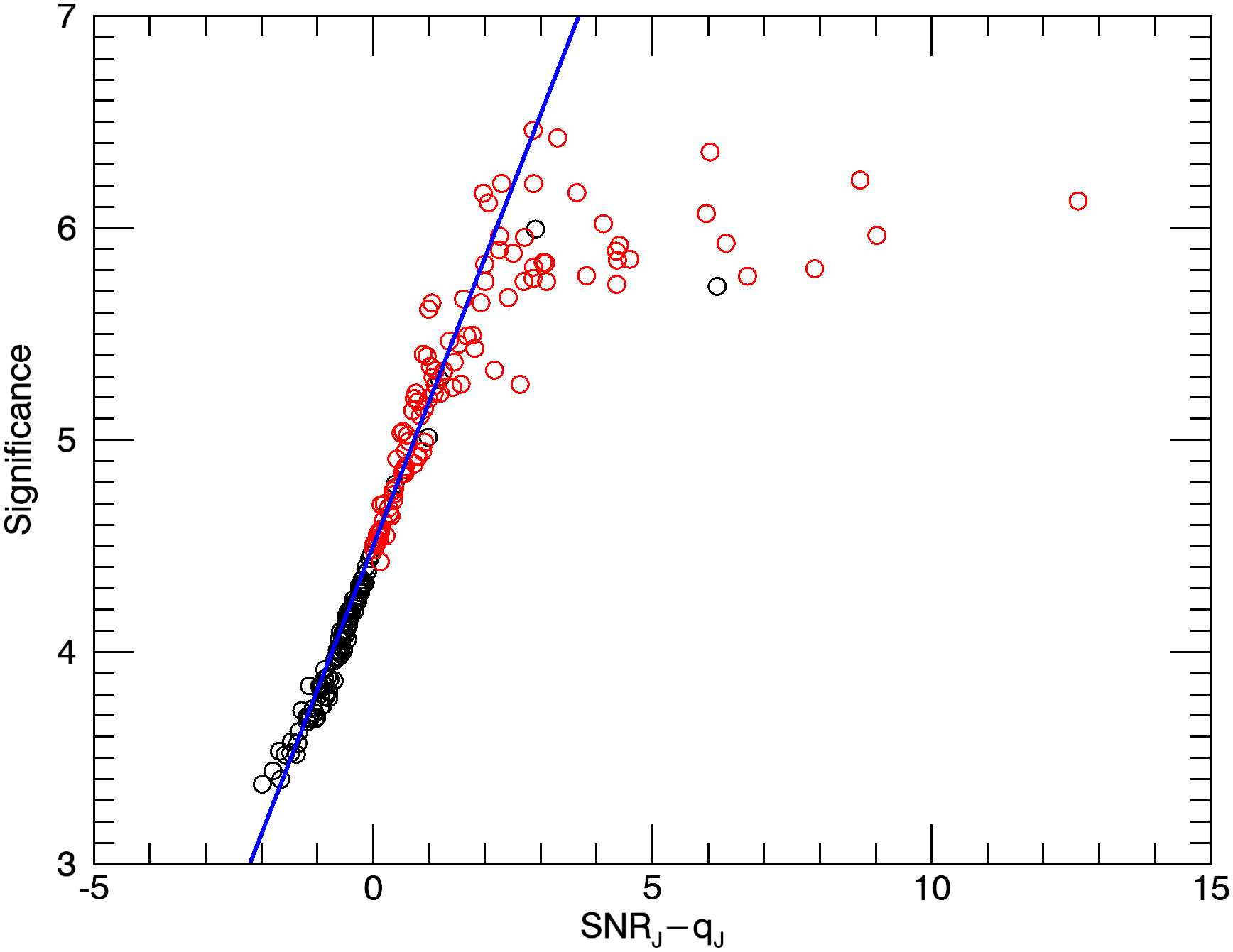}
	\caption{Significance as a function of (S/N)$_{\rm J}-q_{\rm J}$ for the second pass detections corresponding to $P_{\rm FA}=3.40 \cdot 10^{-6}$, with (S/N)$_{\rm SZ}>3$ and outside the SZ mask. Red circles indicate the detections that pass all the other cuts ($t_{\rm exp}>100$ s and (S/N)$_{\rm J}>q_{\rm J}$), while black circles indicate the detections that do not pass them. The solid blue line is the best linear fit with a fixed intercept of 4.5.}
	\label{fig:significance}
\end{figure}

To find a way to estimate the significance in these cases, we analyzed the second pass detections corresponding to $P_{\rm FA}=3.40 \cdot 10^{-6}$. We calculated the significance of the detections with (S/N)$_{\rm SZ}>3$ outside the SZ mask and  we found a good correlation between the significance and the difference between the joint S/N and the threshold $q_{\rm J}$, as shown in Fig. \ref{fig:significance}. Therefore, we decided to use a linear extrapolation to estimate the values of significance as a function of (S/N)$_{\rm J}-q_{\rm J}$ for the detections whose calculated significance is greater than 5.75. By definition, the significance corresponding to (S/N)$_{\rm J}=q_{\rm J}$ is 4.5 for $P_{\rm FA}=3.40 \cdot 10^{-6}$. Thus, the linear extrapolation was found by fitting a line with a fixed intercept of 4.5 to the points in Fig. \ref{fig:significance}, excluding the outliers.   
The final expression we obtained is:
\begin{equation}\label{eq:significance}	
\rm{significance} = 4.5 + 0.68\cdot((\rm{S/N})_{\rm J}-q_{\rm J}).
\end{equation}

\section{List of candidates}\label{app:candidates}
	
	\begin{sidewaystable*}
		\caption{List of candidates for $P_{\rm FA}=3.40 \cdot 10^{-6}$, ordered according to their significance. Galactic and equatorial coordinates are given in degrees in Cols. 2 to 5. Column 6 indicates the joint S/N, while Col. 7 shows the SZ component of this S/N. The joint threshold $q_{\rm J}$ and the difference between the S/N and the threshold are shown in Cols. 8 and 9. Column 10 indicates the significance, calculated as explained in Sect. \ref{sec:evaluation}. Columns 11 to 14 show the names of the objects from PSZ2, SPT, MCXC and Abell catalogues that match our candidates. Column 15 indicates the redshift of the matching cluster, if it is available (if several are available, we show preferentially the SZ redshift, then the MCXC redshift, and finally the Abell redshift). The last column shows our mass estimation using the SZ redshift or the MCXC redshift (the first one is shown if the two are available).}
		\label{table:allcandidates1}
		\centering 
		\tiny
		\begin{tabular}{c | c c c c c c c c c c c c c c c}
			\hline
			\noalign{\smallskip}
			Id. &  G. lon.   & G. lat.   & RA J2000  & Dec J2000 & (S/N)$_{\rm J}$ & (S/N)$_{\rm SZ}$ & $q_{\rm J}$ & (S/N)$_{\rm J}$-$q_{\rm J}$ & Significance  &  &  &  &  & z & $M_{500}$\\
			&  [$\degr$] & [$\degr$] & [$\degr$] & [$\degr$] &                 &                  &             &                             & & PSZ2 & SPT & MCXC & Abell &  & [$10^{14} M_{\odot}$]\\
			\noalign{\smallskip}
			\hline
			\noalign{\smallskip}
  1 & 252.918 & -56.071 &  49.505 & -44.232 &  124.37 &    9.77 &    5.13 &  119.24 &   85.47 & PSZ2 G252.99-56.09 &  & RXC J0317.9-4414  & ABELL 3112 & 0.08 &  6.89 \\ 
  2 & 272.108 & -40.136 &  67.857 & -61.429 &   92.35 &   32.68 &    4.58 &   87.77 &   64.10 & PSZ2 G272.08-40.16 & SPT-CLJ0431-6126 & RXC J0431.4-6126  & ABELL 3266 & 0.06 &  4.86 \\ 
  3 & 265.042 & -48.932 &  55.733 & -53.630 &   83.91 &   17.71 &    4.73 &   79.18 &   58.27 & PSZ2 G265.02-48.96 &  & RXC J0342.8-5338  & ABELL 3158 & 0.06 &  4.43 \\ 
  4 & 348.350 & -64.811 & 348.488 & -42.727 &   66.96 &    3.74 &    5.15 &   61.81 &   46.47 & PSZ2 G348.46-64.83 & SPT-CLJ2313-4243 & RXC J2313.9-4244  & ACOS 1101  & 0.06 &  4.78 \\ 
  5 & 340.852 & -33.392 & 303.128 & -56.840 &   55.74 &   21.76 &    4.65 &   51.09 &   39.20 & PSZ2 G340.88-33.36 & SPT-CLJ2012-5649 & RXC J2012.5-5649  & ABELL 3667 & 0.06 &  4.97 \\ 
  6 & 349.492 & -59.929 & 342.173 & -44.527 &   55.29 &   21.36 &    5.51 &   49.78 &   38.30 & PSZ2 G349.46-59.95 & SPT-CLJ2248-4431 & RXC J2248.7-4431  & ACOS 1063  & 0.35 & 14.52 \\ 
  7 & 332.212 & -46.369 & 330.472 & -59.961 &   51.71 &   19.52 &    4.78 &   46.92 &   36.36 & PSZ2 G332.23-46.37 & SPT-CLJ2201-5956 & RXC J2201.9-5956  & ABELL 3827 & 0.10 &  5.67 \\ 
  8 & 266.038 & -21.238 & 104.652 & -55.956 &   48.79 &   28.65 &    4.98 &   43.81 &   34.25 & PSZ2 G266.04-21.25 & SPT-CLJ0658-5556 & RXC J0658.5-5556  &  & 0.30 & 12.46 \\ 
  9 & 263.681 & -22.551 & 101.369 & -54.229 &   46.11 &   22.82 &    4.91 &   41.19 &   32.47 & PSZ2 G263.68-22.55 & SPT-CLJ0645-5413 & RXC J0645.4-5413  & ABELL 3404 & 0.16 &  7.49 \\ 
  10 & 321.963 & -47.975 & 342.492 & -64.418 &   42.97 &   13.80 &    4.82 &   38.15 &   30.40 & PSZ2 G321.98-47.96 & SPT-CLJ2249-6426 & RXC J2249.9-6425  & ABELL 3921 & 0.09 &  4.59 \\ 
  11 & 271.197 & -30.970 &  87.315 & -62.087 &   40.24 &   14.30 &    4.93 &   35.31 &   28.48 & PSZ2 G271.18-30.95 & SPT-CLJ0549-6205 &  &  & 0.37 &  7.50 \\ 
  12 & 246.413 & -30.288 &  85.030 & -40.838 &   39.70 &    4.43 &    4.79 &   34.91 &   28.20 & PSZ2 G246.36-30.27 &  & RXC J0540.1-4050  & ACOS 540   & 0.04 &  1.86 \\ 
  13 & 263.130 & -23.428 &  99.699 & -53.966 &   39.12 &   13.50 &    4.75 &   34.37 &   27.84 & PSZ2 G263.14-23.41 & SPT-CLJ0638-5358 & RXC J0638.7-5358  & ACOS 592   & 0.23 &  8.61 \\ 
  14 & 335.592 & -46.464 & 328.547 & -57.851 &   36.98 &   13.24 &    4.81 &   32.17 &   26.35 & PSZ2 G335.58-46.44 &  & RXC J2154.1-5751  & ABELL 3822 & 0.08 &  3.61 \\ 
  15 & 270.988 & -31.757 &  85.642 & -61.902 &   35.85 &    4.57 &    4.90 &   30.95 &   25.51 & PSZ2 G270.95-31.78 &  &  &  & 0.14 &  2.28 \\ 
  16 & 269.311 & -49.893 &  52.142 & -55.710 &   35.12 &    9.63 &    4.88 &   30.24 &   25.03 & PSZ2 G269.31-49.87 & SPT-CLJ0328-5541 & RXC J0328.6-5542  & ABELL 3126 & 0.09 &  3.46 \\ 
  17 & 336.594 & -55.435 & 341.577 & -52.727 &   33.57 &   13.74 &    4.96 &   28.62 &   23.93 & PSZ2 G336.60-55.43 &  & RXC J2246.3-5243  & ABELL 3911 & 0.10 &  3.66 \\ 
  18 & 249.887 & -21.650 &  97.204 & -41.723 &   33.44 &    5.97 &    4.89 &   28.55 &   23.89 & PSZ2 G249.87-21.65 & SPT-CLJ0628-4143 & RXC J0628.8-4143  & ABELL 3396 & 0.18 &  6.58 \\ 
  19 & 262.372 & -25.143 &  96.602 & -53.690 &   31.42 &   10.38 &    4.60 &   26.82 &   22.71 & PSZ2 G262.36-25.15 &  & RXC J0626.3-5341  & ABELL 3391 & 0.05 &  2.41 \\ 
  20 & 259.969 & -63.445 &  38.079 & -44.350 &   32.04 &   12.57 &    5.26 &   26.78 &   22.68 & PSZ2 G259.98-63.43 & SPT-CLJ0232-4421 & RXC J0232.2-4420  &  & 0.28 &  7.82 \\ 
  21 & 255.316 & -56.290 &  48.595 & -45.420 &   31.59 &    4.60 &    5.19 &   26.40 &   22.43 & PSZ2 G255.32-56.27 &  & RXC J0314.3-4525  & ABELL 3104 & 0.07 &  2.62 \\ 
  22 & 256.487 & -65.712 &  36.470 & -41.903 &   29.49 &   11.21 &    5.24 &   24.26 &   20.97 & PSZ2 G256.53-65.70 & SPT-CLJ0225-4155 & RXC J0225.9-4154  & ABELL 3017 & 0.22 &  6.33 \\ 
  23 & 262.254 & -35.383 &  79.148 & -54.520 &   28.33 &   24.13 &    4.88 &   23.45 &   20.42 & PSZ2 G262.27-35.38 & SPT-CLJ0516-5430 & RXC J0516.6-5430  & ACOS 520   & 0.30 &  7.42 \\ 
  24 & 337.997 & -45.692 & 326.009 & -56.630 &   28.13 &    3.30 &    5.03 &   23.10 &   20.19 &  &  & RXC J2143.9-5637  &  & 0.08 &  3.62 \\ 
  25 & 246.506 & -26.043 &  90.489 & -39.971 &   24.82 &   10.53 &    4.56 &   20.26 &   18.26 & PSZ2 G246.50-26.09 &  & RXC J0601.7-3959  &  & 0.05 &  1.70 \\ 
  26 & 273.539 & -30.283 &  88.852 & -64.099 &   24.30 &    7.92 &    4.93 &   19.37 &   17.65 & PSZ2 G273.54-30.28 & SPT-CLJ0555-6406 &  &  & 0.35 &  5.79 \\ 
  27 & 264.820 & -51.135 &  52.478 & -52.602 &   23.44 &    7.39 &    4.72 &   18.72 &   17.21 & PSZ2 G264.60-51.07 & SPT-CLJ0330-5228 & RXC J0330.0-5235  & ABELL 3128 & 0.44 &  7.47 \\ 
  28 & 356.079 & -49.531 & 326.747 & -43.895 &   23.48 &    4.51 &    4.96 &   18.52 &   17.08 & PSZ2 G356.04-49.50 &  & RXC J2146.9-4354  & ABELL 3809 & 0.06 &  1.90 \\ 
  29 & 263.289 & -25.265 &  96.695 & -54.527 &   22.63 &    9.68 &    4.55 &   18.08 &   16.77 & PSZ2 G263.19-25.19 &  & RXC J0627.2-5428  &  & 0.05 &  2.02 \\ 
  30 & 271.527 & -56.581 &  41.363 & -53.030 &   23.04 &   11.17 &    5.02 &   18.02 &   16.73 & PSZ2 G271.53-56.57 &  &  & ACOS 295   & 0.30 &  7.27 \\ 
  31 & 272.091 & -29.042 &  91.534 & -62.800 &   22.41 &    6.21 &    4.71 &   17.69 &   16.52 & PSZ2 G272.08-29.06 &  &  & ACOS 567   & 0.10 &  2.19 \\ 
  32 & 285.508 & -62.255 &  26.260 & -53.022 &   22.41 &    9.13 &    4.97 &   17.43 &   16.34 & PSZ2 G285.52-62.23 & SPT-CLJ0145-5301 & RXC J0145.0-5300  & ABELL 2941 & 0.12 &  3.46 \\ 
  33 & 249.914 & -39.851 &  72.466 & -44.685 &   22.51 &    7.71 &    5.09 &   17.42 &   16.33 & PSZ2 G249.91-39.86 &  & RXC J0449.9-4440  & ABELL 3292 & 0.15 &  4.39 \\ 
  34 & 324.518 & -44.971 & 334.513 & -65.195 &   21.12 &    8.47 &    4.62 &   16.50 &   15.70 & PSZ2 G324.54-44.97 & SPT-CLJ2217-6509 & RXC J2218.0-6511  &  & 0.10 &  3.56 \\ 
  35 & 250.889 & -36.270 &  77.568 & -45.324 &   20.94 &   12.43 &    4.95 &   15.99 &   15.36 & PSZ2 G250.89-36.24 & SPT-CLJ0510-4519 & RXC J0510.2-4519  & ABELL 3322 & 0.20 &  5.15 \\ 
  36 & 274.753 & -32.179 &  84.354 & -65.072 &   20.59 &    4.58 &    5.03 &   15.56 &   15.07 & PSZ2 G274.73-32.20 & SPT-CLJ0537-6504 &  &  & 0.20 &  3.23 \\ 
  37 & 250.720 & -30.352 &  85.851 & -44.508 &   20.22 &    3.08 &    4.79 &   15.44 &   14.98 &  &  & RXC J0543.4-4430  &  & 0.16 &  4.29 \\ 
  38 & 247.159 & -23.314 &  94.123 & -39.806 &   20.18 &    8.21 &    4.78 &   15.40 &   14.95 & PSZ2 G247.19-23.31 &  & RXC J0616.5-3948  & ACOS 579   & 0.15 &  4.14 \\ 
  39 & 253.480 & -33.724 &  81.467 & -47.254 &   20.20 &    9.44 &    4.82 &   15.39 &   14.95 & PSZ2 G253.48-33.72 & SPT-CLJ0525-4715 & RXC J0525.8-4715  & ABELL 3343 & 0.19 &  4.83 \\ 
  40 & 336.953 & -45.747 & 326.608 & -57.284 &   20.20 &    6.52 &    4.83 &   15.37 &   14.94 & PSZ2 G336.95-45.75 &  & RXC J2146.3-5717  & ABELL 3806 & 0.08 &  2.50 \\ 
  41 & 322.634 & -49.140 & 343.512 & -63.256 &   20.35 &    7.86 &    4.99 &   15.36 &   14.93 & PSZ2 G322.63-49.15 & SPT-CLJ2254-6314 & RXC J2254.0-6315  &  & 0.21 &  5.30 \\ 
  42 & 293.098 & -70.858 &  17.477 & -45.916 &   19.89 &    6.54 &    4.71 &   15.18 &   14.80 & PSZ2 G293.12-70.85 &  & RXC J0110.0-4555  & ABELL 2877 & 0.02 &  0.91 \\ 
  43 & 277.729 & -51.733 &  43.613 & -58.952 &   20.23 &   15.04 &    5.09 &   15.14 &   14.78 & PSZ2 G277.76-51.74 & SPT-CLJ0254-5857 &  &  & 0.44 &  8.58 \\ 
  44 & 352.851 & -49.322 & 326.976 & -45.999 &   19.33 &    4.04 &    5.04 &   14.29 &   14.20 &  &  & RXC J2147.9-4600  & ACOS 974   & 0.06 &  1.62 \\ 
  45 &   0.403 & -41.864 & 316.089 & -41.354 &   19.08 &    9.61 &    4.79 &   14.28 &   14.20 & PSZ2 G000.40-41.86 &  & RXC J2104.3-4120  & ABELL 3739 & 0.17 &  5.25 \\ 
  46 & 275.968 & -41.473 &  62.877 & -63.680 &   18.87 &    3.70 &    4.92 &   13.95 &   13.97 &  & SPT-CLJ0411-6340 &  & ABELL 3230 & 0.14 &  3.46 \\ 
  47 & 257.328 & -22.198 &  99.324 & -48.485 &   18.43 &   10.77 &    4.70 &   13.73 &   13.83 & PSZ2 G257.32-22.19 & SPT-CLJ0637-4829 & RXC J0637.3-4828  & ABELL 3399 & 0.20 &  4.72 \\ 
  48 & 348.315 & -66.454 & 350.402 & -41.892 &   18.62 &    3.19 &    4.98 &   13.64 &   13.76 &  &  & RXC J2321.5-4153  & ABELL 3998 & 0.09 &  3.21 \\ 
  49 & 339.612 & -69.321 & 356.169 & -42.739 &   18.34 &    8.83 &    5.49 &   12.84 &   13.22 & PSZ2 G339.63-69.34 & SPT-CLJ2344-4243 &  &  & 0.60 & 11.36 \\ 
  50 & 256.571 & -68.894 &  32.834 & -40.283 &   18.38 &    3.56 &    5.55 &   12.84 &   13.22 &  &  & RXC J0211.4-4017  & ABELL 2984 & 0.10 &  2.31 \\ 

  		\end{tabular}
	\end{sidewaystable*}
	
	\begin{sidewaystable*}
		\caption{Continuation of Table \ref{table:allcandidates1}.}
		\label{table:allcandidates2}
		\centering 
		\tiny
		\begin{tabular}{c | c c c c c c c c c c c c c c c}
			\hline
			\noalign{\smallskip}
			Id. &  G. lon.   & G. lat.   & RA J2000  & Dec J2000 & (S/N)$_{\rm J}$ & (S/N)$_{\rm SZ}$ & $q_{\rm J}$ & (S/N)$_{\rm J}$-$q_{\rm J}$ & Significance  &  &  &  &  & z & $M_{500}$\\
			&  [$\degr$] & [$\degr$] & [$\degr$] & [$\degr$] &                 &                  &             &                             & & PSZ2 & SPT & MCXC & Abell &  & [$10^{14} M_{\odot}$]\\		
			\noalign{\smallskip}
			\hline
			\noalign{\smallskip}
   51 & 245.675 & -33.761 &  80.365 & -40.813 &   17.40 &    5.53 &    4.77 &   12.62 &   13.07 & PSZ2 G245.70-33.75 &  & RXC J0521.4-4049  & ABELL 3336 & 0.08 &  2.18 \\ 
   52 & 291.343 & -55.329 &  26.309 & -60.572 &   16.77 &    6.40 &    4.86 &   11.91 &   12.59 & PSZ2 G291.34-55.31 & SPT-CLJ0145-6033 & RXC J0145.2-6033  &  & 0.18 &  4.58 \\ 
   53 & 278.217 & -41.515 &  61.078 & -65.175 &   16.66 &    5.76 &    4.92 &   11.75 &   12.48 & PSZ2 G278.33-41.53 & SPT-CLJ0404-6510 &  & ABELL 3216 & 0.12 &  2.87 \\ 
   54 & 255.663 & -25.317 &  94.221 & -47.808 &   16.40 &    6.86 &    4.76 &   11.64 &   12.40 & PSZ2 G255.64-25.30 &  & RXC J0616.8-4748  &  & 0.12 &  2.89 \\ 
   55 & 298.024 & -67.760 &  15.703 & -49.259 &   17.15 &   12.71 &    5.56 &   11.59 &   12.37 & PSZ2 G297.97-67.74 & SPT-CLJ0102-4915 &  &  & 0.87 & 11.05 \\ 
   56 & 342.531 & -50.944 & 332.329 & -51.829 &   16.57 &    4.14 &    5.15 &   11.42 &   12.25 & PSZ2 G342.51-50.97 &  & RXC J2209.3-5148  & ABELL 3836 & 0.11 &  2.48 \\ 
   57 & 345.825 & -34.284 & 304.692 & -52.711 &   16.05 &    4.21 &    4.70 &   11.35 &   12.21 & PSZ2 G345.82-34.29 &  & RXC J2018.7-5242  & ACOS 861   & 0.05 & 2.01 \\ 
 58 & 245.487 & -53.615 &  54.070 & -40.631 &   16.25 &    8.23 &    4.90 &   11.34 &   12.20 & PSZ2 G245.47-53.62 &  & RXC J0336.3-4037  & ABELL 3140 & 0.17 & 4.75 \\ 
 59 & 247.558 & -56.069 &  50.582 & -41.347 &   14.55 &    4.37 &    4.79 &    9.76 &   11.13 & PSZ2 G247.56-56.06 &  & RXC J0322.3-4121  & ABELL 3122 & 0.06 & 1.60 \\ 
 60 & 336.016 & -51.369 & 336.131 & -55.260 &   14.56 &    4.05 &    4.94 &    9.62 &   11.03 & PSZ2 G336.01-51.27 &  & RXC J2224.4-5515  &  & 0.08 & 1.84 \\ 
 61 & 254.713 & -30.502 &  86.388 & -47.934 &   13.87 &    5.40 &    4.64 &    9.22 &   10.76 & PSZ2 G254.73-30.52 &  & RXC J0545.5-4756  & ABELL 3363 & 0.13 & 2.38 \\ 
 62 & 262.127 & -30.865 &  86.910 & -54.310 &   13.72 &    3.28 &    4.70 &    9.02 &   10.62 &  &  &  &  &  &  \\ 
 63 & 262.730 & -40.971 &  69.523 & -54.313 &   14.46 &   10.10 &    5.45 &    9.01 &   10.62 & PSZ2 G262.73-40.92 & SPT-CLJ0438-5419 &  &  & 0.42 & 7.07 \\ 
 64 & 312.556 & -66.427 &   6.857 & -50.250 &   13.58 &    6.11 &    4.86 &    8.72 &   10.42 & PSZ2 G312.63-66.40 & SPT-CLJ0027-5015 & RXC J0027.3-5015  & ABELL 2777 & 0.14 & 3.48 \\ 
 65 & 331.971 & -45.773 & 329.592 & -60.398 &   13.36 &    5.37 &    4.69 &    8.66 &   10.38 & PSZ2 G331.96-45.74 &  & RXC J2158.4-6023  & ABELL 3825 & 0.08 & 1.90 \\ 
 66 & 277.045 & -41.037 &  63.006 & -64.598 &   13.76 &    4.07 &    5.10 &    8.66 &   10.38 &  &  &  & ABELL 3231 & 0.06 &  \\ 
 67 & 271.303 & -36.123 &  76.374 & -61.753 &   13.30 &    5.81 &    4.94 &    8.37 &   10.18 & PSZ2 G271.28-36.11 & SPT-CLJ0505-6145 &  &  & 0.25 & 3.95 \\ 
 68 & 359.072 & -32.114 & 302.999 & -41.483 &   13.21 &    7.04 &    4.90 &    8.30 &   10.14 & PSZ2 G359.07-32.12 & SPT-CLJ2012-4130 & RXC J2012.0-4129  & ABELL 3668 & 0.15 & 3.59 \\ 
 69 & 309.397 & -72.852 &  10.208 & -44.143 &   13.92 &    6.07 &    5.64 &    8.28 &   10.12 & PSZ2 G309.43-72.86 & SPT-CLJ0040-4407 &  &  & 0.35 & 6.58 \\ 
 70 &   0.757 & -35.699 & 307.971 & -40.613 &   13.11 &    6.50 &    4.92 &    8.19 &   10.06 & PSZ2 G000.77-35.69 & SPT-CLJ2031-4037 & RXC J2031.8-4037  &  & 0.34 & 6.53 \\ 
 71 & 255.541 & -35.698 &  78.659 & -49.058 &   12.88 &    5.85 &    4.69 &    8.19 &   10.06 & PSZ2 G255.52-35.66 &  & RXC J0514.6-4903  & ABELL 3330 & 0.09 & 1.90 \\ 
 72 & 280.272 & -53.834 &  38.666 & -58.516 &   13.56 &    5.95 &    5.47 &    8.09 &    9.99 & PSZ2 G280.23-53.84 & SPT-CLJ0234-5831 &  &  & 0.41 & 6.56 \\ 
 73 & 267.196 & -34.967 &  79.498 & -58.576 &   12.58 &    4.39 &    4.68 &    7.91 &    9.87 &  &  &  & ABELL 3334 & 0.10 &  \\ 
 74 & 358.353 & -47.312 & 323.476 & -42.659 &   13.72 &    5.19 &    5.94 &    7.79 &    9.79 & PSZ2 G358.34-47.31 & SPT-CLJ2134-4238 &  & ABELL 3783 & 0.20 & 3.53 \\ 
 75 & 276.802 & -59.815 &  34.313 & -52.741 &   12.70 &    5.11 &    5.03 &    7.67 &    9.71 & PSZ2 G276.75-59.82 & SPT-CLJ0217-5245 & RXC J0217.2-5244  &  & 0.34 & 5.14 \\ 
 76 & 265.129 & -59.508 &  40.895 & -48.568 &   12.82 &    6.38 &    5.17 &    7.65 &    9.69 & PSZ2 G265.10-59.50 & SPT-CLJ0243-4833 &  &  & 0.50 & 6.31 \\ 
 77 & 270.660 & -35.672 &  77.479 & -61.308 &   12.82 &    8.01 &    5.23 &    7.59 &    9.65 & PSZ2 G270.63-35.67 & SPT-CLJ0509-6118 &  &  & 0.31 & 5.38 \\ 
 78 & 254.531 & -27.306 &  91.021 & -47.241 &   12.27 &    7.26 &    4.78 &    7.50 &    9.59 & PSZ2 G254.51-27.32 & SPT-CLJ0603-4714 &  &  & 0.27 & 4.78 \\ 
 79 & 342.335 & -34.935 & 305.851 & -55.577 &   12.11 &    9.15 &    4.87 &    7.24 &    9.42 & PSZ2 G342.33-34.93 & SPT-CLJ2023-5535 & RXC J2023.4-5535  &  & 0.23 & 5.22 \\ 
 80 & 270.952 & -58.797 &  38.928 & -51.363 &   12.30 &    8.63 &    5.31 &    6.99 &    9.24 & PSZ2 G270.93-58.78 & SPT-CLJ0235-5121 &  &  & 0.28 & 5.24 \\ 
 81 &   1.844 & -46.923 & 322.772 & -40.310 &   12.88 &    4.51 &    5.97 &    6.92 &    9.20 &  & SPT-CLJ2131-4019 &  &  & 0.45 & 6.72 \\ 
 82 & 347.229 & -52.463 & 332.847 & -48.563 &   12.89 &    4.37 &    6.00 &    6.88 &    9.17 & PSZ2 G347.27-52.46 & SPT-CLJ2211-4833 &  & ABELL 3841 & 0.24 & 4.02 \\ 
 83 & 348.940 & -67.400 & 351.306 & -41.187 &   12.66 &    7.68 &    5.83 &    6.83 &    9.14 & PSZ2 G348.90-67.37 & SPT-CLJ2325-4111 &  & ACOS 1121  & 0.36 & 7.10 \\ 
 84 & 245.996 & -51.753 &  56.451 & -41.200 &   11.62 &    3.30 &    4.92 &    6.70 &    9.05 & PSZ1 G246.01-51.76 &  & RXC J0345.7-4112  & ACOS 384   & 0.06 & 1.69 \\ 
 85 & 259.191 & -19.122 & 104.589 & -49.191 &   11.19 &    5.33 &    4.82 &    6.38 &    8.83 & PSZ2 G259.22-19.10 &  &  & ABELL 3406 & 0.16 & 3.39 \\ 
 86 & 255.625 & -46.169 &  62.807 & -48.319 &   11.08 &    9.28 &    4.76 &    6.32 &    8.79 & PSZ2 G255.60-46.18 & SPT-CLJ0411-4819 &  &  & 0.42 & 6.34 \\ 
 87 & 340.431 & -52.163 & 335.034 & -52.490 &   11.29 &    5.52 &    5.00 &    6.29 &    8.77 & PSZ2 G340.46-52.20 &  & RXC J2220.1-5228  & ABELL 3864 & 0.10 & 1.81 \\ 
 88 & 250.275 & -17.306 & 102.809 & -40.622 &   11.14 &    5.72 &    4.88 &    6.26 &    8.75 & PSZ2 G250.29-17.29 & SPT-CLJ0651-4037 &  &  & 0.27 & 4.57 \\ 
 89 & 252.457 & -25.533 &  92.937 & -45.042 &   10.76 &    4.05 &    4.73 &    6.03 &    8.60 &  &  &  & ACOS 574   & 0.01 &  \\ 
 90 & 266.550 & -27.326 &  93.971 & -57.766 &   10.67 &    7.09 &    4.70 &    5.97 &    8.55 & PSZ2 G266.54-27.31 & SPT-CLJ0615-5746 &  &  & 0.97 & 6.74 \\ 
 91 & 260.463 & -54.460 &  49.448 & -48.820 &   10.91 &    6.07 &    4.97 &    5.94 &    8.53 & PSZ2 G260.52-54.50 & SPT-CLJ0317-4849 &  & ABELL 3113 & 0.16 & 2.43 \\ 
 92 & 329.599 & -54.687 & 344.995 & -56.285 &   10.89 &    3.75 &    5.02 &    5.87 &    8.49 & PSZ2 G329.53-54.73 & SPT-CLJ2259-5617 &  & ABELL 3950 & 0.15 & 2.88 \\ 
 93 & 254.072 & -58.458 &  46.081 & -44.021 &   11.62 &    5.61 &    5.81 &    5.80 &    8.44 & PSZ2 G254.08-58.45 & SPT-CLJ0304-4401 &  &  & 0.46 & 5.92 \\ 
 94 & 279.498 & -44.872 &  53.238 & -64.211 &   10.19 &    5.11 &    4.86 &    5.34 &    8.12 & PSZ2 G279.51-44.85 &  &  & ACOS 362   & 0.08 & 1.58 \\ 
 95 & 346.823 & -45.341 & 322.386 & -50.819 &   10.36 &    4.30 &    5.03 &    5.33 &    8.12 & PSZ2 G346.86-45.38 &  & RXC J2129.8-5048  & ABELL 3771 & 0.08 & 1.70 \\ 
 96 & 335.934 & -41.349 & 319.209 & -59.492 &    9.94 &    5.12 &    4.78 &    5.16 &    8.00 &  &  & RXC J2116.8-5929  & ACOS 927   & 0.06 & 1.44 \\ 
 97 & 330.555 & -40.546 & 320.482 & -63.588 &    9.84 &    4.50 &    4.69 &    5.15 &    8.00 & PSZ2 G330.53-40.56 & SPT-CLJ2121-6335 &  & ACOS 937   & 0.22 & 4.88 \\ 
 98 & 292.079 & -71.965 &  17.569 & -44.764 &   10.76 &    4.07 &    5.63 &    5.13 &    7.98 &  & SPT-CLJ0110-4445 &  &  & 0.38 & 5.62 \\ 
 99 & 295.629 & -51.955 &  23.372 & -64.573 &    9.78 &    8.48 &    4.83 &    4.95 &    7.86 & PSZ2 G295.63-51.96 & SPT-CLJ0133-6434 &  &  & 0.33 & 4.99 \\ 
 100 & 258.317 & -37.882 &  75.219 & -51.263 &    9.49 &    4.00 &    4.78 &    4.71 &    7.70 &  & SPT-CLJ0500-5116 &  & ABELL 3303 & 0.11 & 1.95 \\ 
		\end{tabular}
	\end{sidewaystable*}
	
	\begin{sidewaystable*}
		\caption{Continuation of Table \ref{table:allcandidates1}.}
		\label{table:allcandidates3}
		\centering 
		\tiny
		\begin{tabular}{c | c c c c c c c c c c c c c c c}
			\hline
			\noalign{\smallskip}
			Id. &  G. lon.   & G. lat.   & RA J2000  & Dec J2000 & (S/N)$_{\rm J}$ & (S/N)$_{\rm SZ}$ & $q_{\rm J}$ & (S/N)$_{\rm J}$-$q_{\rm J}$ & Significance  &  &  &  &  & z & $M_{500}$\\
			&  [$\degr$] & [$\degr$] & [$\degr$] & [$\degr$] &                 &                  &             &                             & & PSZ2 & SPT & MCXC & Abell &  & [$10^{14} M_{\odot}$]\\
			\noalign{\smallskip}
			\hline
			\noalign{\smallskip}
 101 & 269.739 & -64.391 &  33.219 & -47.147 &    9.45 &    4.67 &    4.79 &    4.66 &    7.66 & PSZ2 G269.82-64.35 &  & RXC J0212.8-4707  & ABELL 2988 & 0.12 & 1.78 \\ 
 102 & 272.610 & -28.890 &  91.930 & -63.244 &    9.46 &    4.23 &    4.87 &    4.59 &    7.62 &  &  &  &  &  &  \\ 
 103 & 292.968 & -65.767 &  19.319 & -50.858 &    9.68 &    6.49 &    5.18 &    4.50 &    7.56 & PSZ2 G293.01-65.78 &  &  & ABELL 2893 &  &  \\ 
 104 & 260.280 & -26.081 &  94.442 & -52.035 &    9.20 &    3.58 &    4.80 &    4.41 &    7.49 &  &  &  & ABELL 3385 &  &  \\ 
 105 & 314.240 & -55.337 & 359.720 & -60.623 &    9.02 &    5.49 &    4.65 &    4.37 &    7.47 & PSZ2 G314.26-55.35 &  & RXC J2359.3-6042  & ABELL 4067 & 0.10 & 2.32 \\ 
 106 & 251.713 & -41.693 &  69.812 & -46.017 &    8.94 &    3.99 &    4.59 &    4.35 &    7.46 &  & SPT-CLJ0439-4600 &  &  & 0.34 & 5.36 \\ 
 107 & 257.241 & -46.829 &  61.490 & -49.279 &    8.99 &    4.48 &    4.87 &    4.12 &    7.30 &  & SPT-CLJ0405-4916 &  &  & 0.32 & 5.13 \\ 
 108 & 251.438 & -37.938 &  75.214 & -45.836 &    9.53 &    3.13 &    5.52 &    4.01 &    7.23 &  & SPT-CLJ0500-4551 &  &  &  &  \\ 
 109 & 333.905 & -43.600 & 324.438 & -60.119 &    9.03 &    7.49 &    5.11 &    3.93 &    7.17 & PSZ2 G333.89-43.60 & SPT-CLJ2138-6008 &  &  & 0.32 & 5.00 \\ 
 110 & 260.626 & -28.926 &  89.924 & -52.810 &    8.56 &    7.07 &    4.74 &    3.82 &    7.10 & PSZ2 G260.63-28.94 & SPT-CLJ0559-5249 &  &  & 0.60 & 5.91 \\ 
 111 & 324.286 & -73.835 &   5.023 & -41.968 &    8.90 &    3.50 &    5.17 &    3.73 &    7.03 &  &  &  & ABELL 2763 &  &  \\ 
 112 & 244.117 & -53.593 &  54.249 & -39.829 &    8.40 &    6.26 &    4.76 &    3.65 &    6.98 & PSZ2 G244.11-53.59 &  & RXC J0337.0-3949  & ABELL 3142 & 0.10 & 1.93 \\ 
113 & 269.508 & -47.131 &  56.470 & -57.092 &    8.08 &    4.14 &    4.50 &    3.58 &    6.93 & PSZ2 G269.36-47.20 &  & RXC J0346.1-5702  & ABELL 3164 & 0.06 & 0.75 \\ 
114 & 347.588 & -35.374 & 306.490 & -51.277 &    8.19 &    5.57 &    4.78 &    3.41 &    6.82 & PSZ2 G347.58-35.35 & SPT-CLJ2025-5117 &  & ACOS 871   & 0.22 & 3.99 \\ 
115 & 303.027 & -68.574 &  12.807 & -48.554 &    8.35 &    3.41 &    5.05 &    3.30 &    6.74 & PSZ2 G303.03-68.49 & SPT-CLJ0051-4834 & RXC J0051.1-4833  & ABELL 2830 & 0.19 & 3.09 \\ 
116 & 341.173 & -36.116 & 308.056 & -56.454 &    8.04 &    5.92 &    4.80 &    3.24 &    6.70 & PSZ2 G341.19-36.12 & SPT-CLJ2032-5627 & RXC J2032.1-5627  & ABELL 3685 & 0.28 & 4.80 \\ 
117 & 334.025 & -49.837 & 334.988 & -57.142 &    8.37 &    4.03 &    5.28 &    3.09 &    6.60 &  & SPT-CLJ2219-5708 &  & ABELL 3860 & 0.33 & 4.67 \\ 
118 & 263.029 & -56.198 &  46.079 & -49.352 &    8.22 &    4.75 &    5.13 &    3.09 &    6.60 & PSZ2 G263.03-56.19 & SPT-CLJ0304-4921 &  &  & 0.39 & 4.28 \\ 
119 & 285.899 & -74.959 &  18.675 & -41.391 &    8.34 &    6.13 &    5.27 &    3.08 &    6.59 & PSZ2 G285.87-74.93 & SPT-CLJ0114-4123 &  &  & 0.21 & 2.75 \\ 
120 & 356.016 & -51.958 & 330.063 & -43.514 &    9.05 &    3.08 &    6.00 &    3.04 &    6.57 &  &  &  &  &  &  \\ 
121 & 291.577 & -51.264 &  29.413 & -64.380 &    7.85 &    3.03 &    4.84 &    3.01 &    6.54 &  &  &  & ACOS 210   &  &  \\ 
122 & 248.030 & -26.392 &  90.488 & -41.383 &    7.89 &    3.87 &    4.92 &    2.97 &    6.51 & PSZ2 G247.99-26.37 & SPT-CLJ0601-4122 &  &  & 0.23 & 3.08 \\ 
123 & 345.335 & -60.017 & 343.644 & -46.335 &    9.07 &    5.44 &    6.13 &    2.94 &    6.50 & PSZ2 G345.32-59.97 & SPT-CLJ2254-4620 &  & ABELL 3937 & 0.27 & 3.85 \\ 
124 & 260.905 & -62.219 &  39.226 & -45.362 &    7.95 &    5.40 &    5.05 &    2.90 &    6.47 & PSZ2 G260.85-62.19 &  &  & ABELL 3036 & 0.19 & 2.24 \\ 
125 & 309.077 & -72.506 &  10.274 & -44.498 &    8.11 &    3.50 &    5.24 &    2.87 &    6.45 &  & SPT-CLJ0041-4428 &  & ACOS 67    & 0.33 & 4.49 \\ 
126 & 318.946 & -66.813 &   3.308 & -49.110 &    8.39 &    3.58 &    5.52 &    2.87 &    6.45 &  & SPT-CLJ0013-4906 &  &  & 0.41 & 5.90 \\ 
127 & 250.693 & -27.274 &  90.059 & -43.897 &    7.61 &    3.92 &    4.75 &    2.87 &    6.45 &  & SPT-CLJ0600-4353 &  &  & 0.36 & 4.13 \\ 
128 & 269.926 & -33.562 &  81.955 & -60.928 &    7.70 &    4.06 &    4.84 &    2.86 &    6.44 &  &  &  &  &  &  \\ 
129 & 264.378 & -35.048 &  79.615 & -56.268 &    7.65 &    3.32 &    4.83 &    2.82 &    6.41 &  &  &  & ACOS 522   &  &  \\ 
130 & 273.632 & -68.444 &  27.637 & -45.174 &    7.87 &    5.14 &    5.07 &    2.80 &    6.40 & PSZ2 G273.69-68.38 & SPT-CLJ0150-4511 &  &  & 0.32 & 3.86 \\ 
131 & 254.686 & -34.395 &  80.582 & -48.300 &    7.73 &    3.94 &    4.93 &    2.80 &    6.40 & PSZ2 G254.66-34.45 & SPT-CLJ0522-4818 &  & ABELL 3338 & 0.30 & 3.72 \\ 
132 & 265.211 & -24.979 &  97.822 & -56.170 &    7.30 &    3.96 &    4.59 &    2.71 &    6.34 & PSZ2 G265.21-24.83 &  & RXC J0631.3-5610  &  & 0.05 & 0.89 \\ 
133 & 266.726 & -34.077 &  81.239 & -58.245 &    7.58 &    3.23 &    4.92 &    2.65 &    6.30 &  &  &  &  &  &  \\ 
134 & 259.625 & -51.080 &  54.540 & -49.609 &    7.88 &    4.22 &    5.25 &    2.63 &    6.28 & PSZ2 G259.58-51.07 &  &  &  &  &  \\ 
135 & 295.552 & -60.100 &  19.547 & -56.644 &    7.48 &    3.14 &    4.97 &    2.51 &    6.20 &  & SPT-CLJ0118-5638 &  & ABELL 2897 & 0.21 & 3.38 \\ 
136 & 261.307 & -36.488 &  77.318 & -53.709 &    7.42 &    5.23 &    4.91 &    2.51 &    6.20 & PSZ2 G261.28-36.47 & SPT-CLJ0509-5342 &  &  & 0.46 & 4.53 \\ 
137 & 299.961 & -53.485 &  16.829 & -63.551 &    7.45 &    3.04 &    5.15 &    2.30 &    6.06 &  &  &  &  &  &  \\ 
138 & 280.717 & -52.374 &  40.161 & -59.786 &    7.34 &    4.62 &    5.08 &    2.26 &    6.04 &  & SPT-CLJ0240-5946 &  &  & 0.40 & 4.34 \\ 
139 & 256.067 & -37.447 &  75.983 & -49.495 &    7.49 &    3.87 &    5.23 &    2.26 &    6.03 &  & SPT-CLJ0504-4929 &  & ABELL 3311 & 0.20 & 2.69 \\ 
140 & 281.296 & -46.908 &  47.813 & -63.917 &    7.07 &    5.85 &    4.82 &    2.26 &    6.03 & PSZ2 G281.26-46.90 & SPT-CLJ0311-6354 &  &  & 0.28 & 3.61 \\ 
141 & 268.370 & -64.487 &  33.712 & -46.652 &    7.29 &    3.30 &    5.13 &    2.16 &    5.97 &  & SPT-CLJ0214-4638 &  &  & 0.27 & 3.36 \\ 
142 & 340.088 & -52.584 & 335.806 & -52.480 &    7.33 &    3.84 &    5.24 &    2.09 &    5.92 &  & SPT-CLJ2223-5227 &  & ABELL 3872 & 0.27 & 3.23 \\ 
143 & 250.615 & -25.048 &  93.026 & -43.309 &    6.85 &    4.22 &    4.79 &    2.06 &    5.90 & PSZ2 G250.59-25.03 & SPT-CLJ0612-4317 &  &  & 0.55 & 4.91 \\ 
144 & 357.745 & -41.742 & 315.890 & -43.334 &    6.73 &    3.53 &    4.68 &    2.05 &    5.89 & PSZ2 G357.75-41.77 &  & RXC J2103.4-4319  & ABELL 3736 & 0.05 & 1.02 \\ 
145 & 254.991 & -41.239 &  70.265 & -48.496 &    6.58 &    4.66 &    4.58 &    2.00 &    5.86 &  &  &  & ABELL 3283 &  &  \\ 
146 & 296.229 & -52.033 &  22.493 & -64.590 &    6.80 &    6.00 &    4.82 &    1.98 &    5.84 & PSZ2 G296.27-52.05 & SPT-CLJ0129-6432 &  &  & 0.33 & 4.06 \\ 
147 & 351.816 & -54.778 & 334.758 & -45.238 &    7.26 &    5.00 &    5.29 &    1.97 &    5.84 & PSZ2 G351.76-54.71 & SPT-CLJ2218-4519 &  &  & 0.61 & 5.28 \\ 
148 & 252.960 & -41.382 &  70.199 & -46.964 &    6.55 &    4.43 &    4.58 &    1.97 &    5.84 & PSZ2 G252.95-41.35 & SPT-CLJ0440-4657 &  &  & 0.35 & 4.27 \\ 
149 & 292.252 & -53.065 &  26.997 & -62.875 &    6.46 &    3.85 &    4.54 &    1.92 &    5.80 &  &  &  & ACOS 194   &  &  \\ 
150 & 268.551 & -28.149 &  92.830 & -59.627 &    7.41 &    4.32 &    4.71 &    2.70 &    5.75 & PSZ2 G268.51-28.14 & SPT-CLJ0611-5938 &  &  & 0.46 & 3.94 \\ 
		\end{tabular}
	\end{sidewaystable*}
	
	\begin{sidewaystable*}
		\caption{Continuation of Table \ref{table:allcandidates1}.}
		\label{table:allcandidates4}
		\centering 
		\tiny
		\begin{tabular}{c | c c c c c c c c c c c c c c c}
			\hline
			\noalign{\smallskip}
			Id. &  G. lon.   & G. lat.   & RA J2000  & Dec J2000 & (S/N)$_{\rm J}$ & (S/N)$_{\rm SZ}$ & $q_{\rm J}$ & (S/N)$_{\rm J}$-$q_{\rm J}$ & Significance  &  &  &  &  & z & $M_{500}$\\
			&  [$\degr$] & [$\degr$] & [$\degr$] & [$\degr$] &                 &                  &             &                             & & PSZ2 & SPT & MCXC & Abell &  & [$10^{14} M_{\odot}$]\\
			\noalign{\smallskip}
			\hline
			\noalign{\smallskip}
151 & 263.223 & -33.187 &  82.956 & -55.344 &    7.88 &    6.80 &    4.78 &    3.10 &    5.75 & PSZ2 G263.24-33.18 &  &  &  & 0.30 & 3.83 \\ 
152 & 282.665 & -54.841 &  35.371 & -58.601 &    6.77 &    3.94 &    4.77 &    2.00 &    5.75 &  &  &  &  &  &  \\ 
153 & 345.317 & -62.251 & 346.407 & -45.204 &    9.63 &    3.57 &    5.26 &    4.36 &    5.73 &  &  & RXC J2305.5-4513  & ABELL 3970 & 0.13 & 2.13 \\ 
154 & 265.689 & -27.575 &  93.313 & -57.045 &    7.12 &    3.48 &    4.71 &    2.42 &    5.67 &  &  &  &  &  &  \\ 
155 & 257.118 & -27.233 &  91.795 & -49.483 &    6.20 &    3.06 &    4.59 &    1.62 &    5.66 &  &  & RXC J0607.0-4928  & ABELL 3380 & 0.06 & 0.87 \\ 
156 & 270.400 & -44.745 &  60.078 & -58.641 &    6.93 &    3.01 &    5.00 &    1.93 &    5.65 &  &  &  &  &  &  \\ 
157 & 330.754 & -32.806 & 302.919 & -65.315 &    5.77 &    3.69 &    4.72 &    1.05 &    5.65 &  &  &  & ACOS 856   &  &  \\ 
158 & 250.409 & -41.393 &  70.287 & -45.052 &    5.61 &    4.04 &    4.61 &    0.99 &    5.62 & PSZ2 G250.43-41.42 &  &  & ABELL 3284 & 0.15 & 2.02 \\ 
159 & 329.901 & -74.561 &   3.711 & -40.593 &    7.31 &    5.34 &    5.53 &    1.78 &    5.49 &  & SPT-CLJ0014-4036 &  &  & 0.55 & 5.67 \\ 
160 & 258.554 & -22.618 &  99.187 & -49.690 &    6.48 &    4.00 &    4.80 &    1.68 &    5.49 &  & SPT-CLJ0636-4942 &  &  & 0.35 & 3.89 \\ 
161 & 254.861 & -20.784 & 100.312 & -45.874 &    6.20 &    3.74 &    4.84 &    1.36 &    5.47 &  &  &  &  &  &  \\ 
162 & 260.733 & -31.256 &  86.092 & -53.151 &    6.26 &    4.57 &    4.73 &    1.53 &    5.45 & PSZ2 G260.76-31.27 &  &  &  & 0.25 & 2.95 \\ 
163 & 327.280 & -75.170 &   4.893 & -40.415 &    7.36 &    4.03 &    5.54 &    1.82 &    5.43 &  &  &  &  &  &  \\ 
164 & 356.230 & -43.104 & 317.814 & -44.422 &    5.59 &    4.46 &    4.70 &    0.90 &    5.40 & PSZ2 G356.21-43.11 &  &  &  &  &  \\ 
165 & 247.682 & -25.090 &  92.055 & -40.751 &    5.88 &    4.00 &    4.91 &    0.96 &    5.39 & PSZ2 G247.69-25.06 &  &  &  & 0.34 & 3.35 \\ 
166 & 252.248 & -50.642 &  57.037 & -45.249 &    6.62 &    5.24 &    5.17 &    1.45 &    5.37 & PSZ2 G252.23-50.62 & SPT-CLJ0348-4515 &  &  & 0.36 & 4.12 \\ 
167 & 255.846 & -41.583 &  69.662 & -49.106 &    5.56 &    3.58 &    4.53 &    1.02 &    5.35 &  &  &  &  &  &  \\ 
168 & 347.751 & -59.916 & 342.696 & -45.317 &    8.50 &    3.28 &    6.32 &    2.17 &    5.33 &  &  & RXC J2250.8-4521  & ACOS 1067  & 0.05 & 0.82 \\ 
169 & 252.121 & -34.155 &  80.699 & -46.168 &    6.04 &    3.08 &    4.92 &    1.11 &    5.33 &  &  &  & ACOS 526   &  &  \\ 
170 & 295.198 & -57.013 &  21.188 & -59.613 &    6.00 &    5.23 &    4.74 &    1.26 &    5.33 & PSZ2 G295.19-56.99 & SPT-CLJ0124-5937 &  & ACOS 157   & 0.22 & 3.03 \\ 
171 & 319.166 & -55.033 & 354.344 & -59.699 &    6.02 &    4.51 &    4.94 &    1.07 &    5.29 &  & SPT-CLJ2337-5942 &  &  & 0.77 & 5.75 \\ 
172 & 246.470 & -38.671 &  73.948 & -41.976 &    7.03 &    3.00 &    5.45 &    1.57 &    5.26 &  & SPT-CLJ0455-4159 &  &  & 0.31 & 3.38 \\ 
173 & 346.731 & -57.456 & 339.853 & -46.882 &    9.40 &    3.52 &    6.77 &    2.63 &    5.26 &  &  &  &  &  &  \\ 
174 & 258.418 & -38.557 &  74.136 & -51.305 &    5.89 &    4.48 &    4.78 &    1.11 &    5.26 & PSZ2 G258.33-38.54 & SPT-CLJ0456-5116 &  &  & 0.56 & 4.58 \\ 
175 & 290.504 & -71.045 &  18.586 & -45.527 &    7.07 &    3.02 &    5.64 &    1.43 &    5.25 &  &  &  &  &  &  \\ 
176 & 248.782 & -29.959 &  85.981 & -42.789 &    5.66 &    3.68 &    4.89 &    0.76 &    5.22 &  & SPT-CLJ0543-4250 &  &  & 0.59 & 4.49 \\ 
177 & 277.000 & -51.584 &  44.412 & -58.716 &    6.48 &    3.58 &    5.28 &    1.20 &    5.22 &  & SPT-CLJ0257-5842 &  &  & 0.44 & 4.51 \\ 
178 & 245.510 & -29.222 &  86.190 & -39.854 &    6.27 &    3.18 &    5.18 &    1.09 &    5.22 &  & SPT-CLJ0544-3950 &  &  & 0.45 & 4.25 \\ 
179 & 249.394 & -34.217 &  80.293 & -43.930 &    5.58 &    3.56 &    4.84 &    0.74 &    5.19 &  &  &  &  &  &  \\ 
180 & 262.045 & -32.976 &  83.288 & -54.353 &    5.84 &    4.60 &    4.85 &    0.99 &    5.19 & PSZ2 G262.01-32.98 &  &  &  & 0.23 & 2.72 \\ 
181 & 269.014 & -55.228 &  44.560 & -52.727 &    5.62 &    3.09 &    4.82 &    0.80 &    5.18 &  &  &  & ABELL 3074 &  &  \\ 
182 & 357.238 & -34.672 & 306.147 & -43.350 &    5.79 &    3.51 &    4.87 &    0.91 &    5.14 &  &  &  &  &  &  \\ 
183 & 323.083 & -63.990 & 358.996 & -50.913 &    5.07 &    5.08 &    4.37 &    0.71 &    5.14 & PSZ2 G323.08-63.98 & SPT-CLJ2355-5055 &  &  & 0.32 & 4.22 \\ 
184 & 345.215 & -32.903 & 302.377 & -53.179 &    5.87 &    3.47 &    5.02 &    0.84 &    5.11 &  &  &  & ABELL 3665 & 0.24 &  \\ 
185 & 251.548 & -42.792 &  68.244 & -45.837 &    5.09 &    4.28 &    4.55 &    0.54 &    5.04 & PSZ2 G251.55-42.78 &  &  & ABELL 3270 &  &  \\ 
186 & 318.019 & -58.895 & 358.740 & -56.548 &    5.06 &    4.58 &    4.57 &    0.50 &    5.03 & PSZ2 G318.05-58.88 & SPT-CLJ2354-5633 &  &  & 0.53 & 4.87 \\ 
187 & 268.411 & -43.638 &  63.012 & -57.710 &    5.46 &    3.37 &    4.85 &    0.61 &    5.02 & PSZ2 G268.34-43.64 & SPT-CLJ0412-5743 &  &  & 0.34 & 3.34 \\ 
188 & 267.374 & -46.219 &  59.056 & -56.117 &    5.42 &    4.14 &    4.79 &    0.63 &    4.99 & PSZ2 G267.30-46.19 &  &  &  &  &  \\ 
189 & 279.016 & -56.371 &  36.612 & -56.124 &    6.64 &    3.19 &    5.71 &    0.92 &    4.99 &  &  &  & ABELL 3019 &  &  \\ 
190 & 268.091 & -59.722 &  39.274 & -49.647 &    5.73 &    3.42 &    5.15 &    0.58 &    4.95 &  & SPT-CLJ0236-4938 &  &  & 0.33 & 3.43 \\ 
191 & 333.125 & -35.416 & 308.253 & -63.019 &    6.05 &    3.70 &    5.16 &    0.89 &    4.95 &  &  &  & ABELL 3687 & 0.08 &  \\ 
192 & 283.847 & -65.365 &  25.079 & -49.914 &    6.65 &    3.52 &    5.86 &    0.79 &    4.93 &  &  &  &  &  &  \\ 
193 & 305.262 & -71.942 &  11.835 & -45.167 &    6.35 &    3.12 &    5.54 &    0.81 &    4.92 &  & SPT-CLJ0047-4506 &  &  & 0.39 & 4.56 \\ 
194 & 324.836 & -46.138 & 336.297 & -64.282 &    4.98 &    3.13 &    4.56 &    0.42 &    4.91 &  &  &  & ACOS 1022  & 0.09 &  \\ 
195 &   0.622 & -48.152 & 324.445 & -41.074 &    6.72 &    3.11 &    5.98 &    0.74 &    4.89 &  &  &  &  &  &  \\ 
196 &   1.499 & -35.778 & 308.172 & -40.026 &    5.86 &    3.71 &    5.28 &    0.57 &    4.87 &  &  &  & ABELL 3688 &  &  \\ 
197 & 272.628 & -24.433 & 101.650 & -62.645 &    5.47 &    3.83 &    4.91 &    0.57 &    4.86 &  & SPT-CLJ0646-6236 &  &  & 1.20 & 4.66 \\ 
198 & 352.487 & -33.184 & 303.471 & -47.119 &    5.69 &    3.73 &    5.14 &    0.55 &    4.86 &  &  &  &  &  &  \\ 
199 & 335.748 & -37.097 & 310.996 & -60.647 &    5.56 &    3.36 &    5.03 &    0.53 &    4.84 &  &  &  &  &  &  \\ 
200 & 260.921 & -35.301 &  79.333 & -53.436 &    5.54 &    3.03 &    4.98 &    0.57 &    4.84 &  &  &  &  &  &  \\ 
\end{tabular}
	\end{sidewaystable*}

		\begin{sidewaystable*}
			\caption{Continuation of Table \ref{table:allcandidates1}.}
			\label{table:allcandidates5}
			\centering 
			\tiny
			\begin{tabular}{c | c c c c c c c c c c c c c c c}
				\hline
				\noalign{\smallskip}
				Id. &  G. lon.   & G. lat.   & RA J2000  & Dec J2000 & (S/N)$_{\rm J}$ & (S/N)$_{\rm SZ}$ & $q_{\rm J}$ & (S/N)$_{\rm J}$-$q_{\rm J}$ & Significance  &  &  &  &  & z & $M_{500}$\\
				&  [$\degr$] & [$\degr$] & [$\degr$] & [$\degr$] &                 &                  &             &                             & & PSZ2 & SPT & MCXC & Abell &  & [$10^{14} M_{\odot}$]\\
				\noalign{\smallskip}
				\hline
				\noalign{\smallskip}
				201 & 255.932 & -52.549 &  53.538 & -46.977 &    5.76 &    3.70 &    5.20 &    0.56 &    4.84 &  & SPT-CLJ0334-4659 &  &  & 0.49 & 4.23 \\ 
				202 & 305.616 & -54.819 &   9.538 & -62.240 &    5.33 &    4.11 &    4.94 &    0.39 &    4.77 & PSZ2 G305.59-54.80 &  &  &  &  &  \\ 
				203 & 292.378 & -57.126 &  24.015 & -59.079 &    5.20 &    3.69 &    4.84 &    0.35 &    4.76 & PSZ2 G292.40-57.11 & SPT-CLJ0135-5904 &  &  & 0.49 & 4.57 \\ 
				204 & 265.024 & -30.437 &  87.988 & -56.765 &    5.11 &    3.45 &    4.74 &    0.37 &    4.74 &  &  &  &  &  &  \\ 
				205 & 284.523 & -47.939 &  42.558 & -64.722 &    5.35 &    4.02 &    4.99 &    0.36 &    4.71 &  &  &  & ABELL 3061 &  &  \\ 
				206 & 356.548 & -41.337 & 315.336 & -44.232 &    4.89 &    3.84 &    4.69 &    0.20 &    4.70 &  &  &  &  &  &  \\ 
				207 & 310.626 & -63.554 &   7.143 & -53.231 &    4.46 &    4.46 &    4.32 &    0.14 &    4.69 &  &  &  & ABELL 2779 &  &  \\ 
				208 & 277.913 & -65.311 &  28.381 & -48.691 &    5.34 &    3.97 &    5.06 &    0.28 &    4.68 &  &  &  &  &  &  \\ 
				209 & 274.955 & -58.096 &  37.411 & -53.368 &    5.72 &    3.12 &    5.43 &    0.29 &    4.65 &  &  &  &  &  &  \\ 
				210 & 327.637 & -54.292 & 345.963 & -57.369 &    5.51 &    4.67 &    5.22 &    0.29 &    4.65 & PSZ2 G327.66-54.26 &  &  &  &  &  \\ 
				211 & 274.185 & -53.288 &  44.073 & -56.304 &    5.89 &    4.36 &    5.57 &    0.32 &    4.64 &  & SPT-CLJ0256-5617 &  &  & 0.58 & 5.05 \\ 
				212 & 282.535 & -53.687 &  36.837 & -59.477 &    5.34 &    3.13 &    5.16 &    0.18 &    4.62 &  &  &  &  &  &  \\ 
				213 & 258.667 & -34.133 &  81.229 & -51.574 &    4.98 &    3.45 &    4.85 &    0.13 &    4.58 &  &  &  &  &  &  \\ 
				214 & 253.736 & -47.436 &  61.300 & -46.822 &    4.96 &    3.54 &    4.82 &    0.13 &    4.57 &  & SPT-CLJ0405-4648 &  &  & 0.39 & 3.88 \\ 
				215 & 261.447 & -30.962 &  86.667 & -53.736 &    4.91 &    3.12 &    4.79 &    0.12 &    4.56 &  & SPT-CLJ0546-5345 &  &  & 1.07 & 4.00 \\ 
				216 & 245.673 & -50.813 &  57.723 & -41.115 &    5.06 &    3.44 &    4.99 &    0.08 &    4.55 &  & SPT-CLJ0351-4109 &  &  & 0.68 & 5.16 \\ 
				217 & 326.406 & -68.015 &   0.288 & -46.755 &    5.89 &    3.02 &    5.66 &    0.23 &    4.55 &  &  &  & ABELL 4075 &  &  \\ 
				218 & 274.682 & -64.879 &  30.412 & -48.217 &    5.25 &    3.51 &    5.13 &    0.12 &    4.54 &  &  &  & ACOS 218   & 0.09 &  \\ 
				219 & 260.348 & -20.324 & 103.375 & -50.613 &    4.94 &    4.43 &    4.87 &    0.07 &    4.53 &  &  &  &  &  &  \\ 
				220 & 252.269 & -23.928 &  95.065 & -44.475 &    4.96 &    3.64 &    4.88 &    0.08 &    4.53 &  &  &  &  &  &  \\ 
				221 & 319.266 & -48.599 & 346.674 & -65.073 &    5.14 &    4.48 &    5.06 &    0.08 &    4.52 & PSZ2 G319.19-48.59 & SPT-CLJ2306-6505 &  &  & 0.53 & 4.34 \\ 
				222 & 264.135 & -35.097 &  79.543 & -56.066 &    4.63 &    3.41 &    4.61 &    0.03 &    4.51 &  &  &  &  &  &  \\ 
				223 & 254.233 & -25.041 &  94.169 & -46.485 &    4.84 &    3.19 &    4.82 &    0.01 &    4.51 &  &  &  &  &  &  \\ 
				224 & 325.000 & -49.228 & 341.209 & -62.122 &    5.38 &    4.60 &    5.36 &    0.01 &    4.48 & PSZ2 G324.99-49.26 & SPT-CLJ2245-6206 &  &  & 0.58 & 4.70 \\ 
				225 & 328.390 & -59.170 & 351.147 & -53.453 &    5.77 &    3.46 &    5.65 &    0.13 &    4.43 &  &  &  &  &  &  \\ 	\end{tabular}
		\end{sidewaystable*}

\end{appendix}

\end{document}